\documentclass[a4paper,11pt]{article}
\pdfoutput=1 

\usepackage{jcappub}
\usepackage[utf8]{inputenc}
\usepackage{graphicx,color,amssymb,amsmath,mathtools,cases,bm,float,tocbibind,subfigure,upgreek,longtable,geometry,blindtext,hyperref}
\usepackage[format=hang,font=small,textfont=it]{caption}
\allowdisplaybreaks[2]

\title{CMB polarization analysis on circular scans}
\author[a,b,c,1]{Jia-Rui Li,\note{Corresponding author}} 
\author[a,b,c,]{Chunlong Li,} 
\author[a,b,c,]{Jie Jiang,} 
\author[a,b,c,1]{Yi-Fu Cai,} 
\author[d,e,a,1]{Jacques Delabrouille,} 
\author[f,g]{Deliang Wu,}
\author[f]{Hong Li}
\affiliation[a]{Department of Astronomy, School of Physical Sciences, University of Science and Technology of China, Hefei, Anhui 230026, China}
\affiliation[b]{CAS Key Laboratory for Researches in Galaxies and Cosmology, University of Science and Technology of China, Hefei, Anhui 230026, China}
\affiliation[c]{School of Astronomy and Space Science, University of Science and Technology of China, Hefei, Anhui 230026, China}
\affiliation[d]{Laboratoire Astroparticule et Cosmologie (APC), CNRS/IN2P3, Universit\'e Paris Diderot, 75205 Paris Cedex 13, France}
\affiliation[e]{IRFU, CEA, Universit\'e Paris Saclay, 91191 Gif-sur-Yvette, France}
\affiliation[f]{Key Laboratory of Particle Astrophysics, Institute of High Energy Physics, Chinese Academy of Sciences, Beijing 100086, China}
\affiliation[g]{University of Chinese Academy of Sciences, Beijing 100086, China}
\date{}

\emailAdd{jr981025@mail.ustc.edu.cn}
\emailAdd{yifucai@ustc.edu.cn}
\emailAdd{delabrouille@apc.in2p3.fr}

\abstract{
Most cosmic microwave background experiments observe the sky along circular or near-circular scans on the celestial sphere. For such experiments, we show that simple linear systems connect the Fourier spectra of temperature and polarization time-ordered data to the harmonic spectra of $T$, $E$ and $B$ on the sphere. We show how this can be used to estimate those spectra directly from data streams. In addition, the inversion of the linear system that connects Fourier spectra to angular power spectra offers a natural way to down-weight those modes of observation most contaminated by low-frequency noise, ground pickup, or fluctuations of atmospheric emission on large angular scale. This can be of interest for the analysis of future CMB data sets, as an alternative or in complement to other approaches that involve map-making as a first analysis step.
}

\begin{document}


\maketitle


\section{Introduction}

The cosmic microwave background (CMB), relic radiation emitted at a redshift $z\simeq 1080$, when light nuclei and electrons in the primordial plasma first combined to produce neutral atoms, carries a wealth of information about the global properties and history of the Universe we live in. Over the past 30 years, several generations of experiments have observed CMB temperature and polarization spatial fluctuations imprinted at the time of last scattering by the seeds of the large scale structure observable in the low-redshift. Recently, the Planck space mission, launched in May 2009 by ESA \citep{2011A&A...536A...1P}, has measured with unprecedented accuracy the temperature and polarization power spectra of the CMB fluctuations \citep{2019arXiv190712875P}, establishing the current best-fit $\Lambda$CDM cosmological scenario and measuring its six main parameters with precision ranging from fractions of a per-cent to a few per-cent \citep{2018arXiv180706209P}.
In spite of this success, complemented by a vigorous ongoing experimental programme involving CMB observations from ground-based observatories
\citep{2014JCAP...10..007N, 2018JCAP...09..005K, 2018PhRvL.121v1301B, 2019arXiv190800480D, 2018JLTP..193.1066N, 2019arXiv191002608A, 2019Univ....5...42M, 2019arXiv191005748S, 2019JCAP...02..056A} and from stratospheric balloons \citep{2018ApJS..239....7E, 2018JLTP..193.1112G, 2012SPIE.8452E..3FD, 2018SPIE10708E..06P}, only a fraction of the information available in CMB polarization has been collected so far. Available CMB polarization maps are either limited by instrumental sensitivity, or cover only small fractions of the observable sky.

Scientific motivations for improving on existing CMB polarization observations are strong. CMB polarization patterns on the sky can be decomposed into even and odd parity components, the so-called $E$ and $B$ modes. $E$ modes are primarily generated at last scattering ($z\simeq 1080$), by the plasma motions induced by primordial scalar fluctuations. Their precise measurement would contribute to drastically improve constraints on the cosmological parameters \citep{2018JCAP...04..017D}. Lower amplitude $B$ modes are generated on small angular scales by tiny distortions of the primary E-mode polarization pattern, due to gravitational lensing along the photon path across the large scale structures present in the lower redshift Universe \citep{2006PhR...429....1L}. Together with that of $E$ modes, their observation would allow us to map the distribution of dark matter over a large fraction of the Hubble volume \citep{2018JCAP...04..018C}. In addition, early-universe tensor perturbations of the metric are expected to generate both primordial $E$ modes and primordial $B$ modes on large angular scale. Large scale primordial $B$ modes, which still escape detection so far, are of particular interest: they potentially bear the most easily detectable signature of the energy scale of cosmic inflation, many models of which predict primordial CMB polarization $B$ modes at a level that could be reached with next-decade CMB experiments \citep{2016arXiv161002743A, 2018JCAP...04..016F, 2019BAAS...51c.338S}.

CMB experiments usually measure angular fluctuations of CMB temperature and polarization by scanning the sky along circular or nearly circular scans on the sky. Such scans arise because of the fast rotation of the instrument around a fixed axis to modulate the CMB signal impinging the detectors. The Planck space mission instruments, for instance, scanned the sky by  rotating the spacecraft at $\simeq$1~RPM around a spin axis fixed with respect to the spacecraft, $85^\circ$ away from the line of sight of the focal plane center. The spin axis direction on the sky was displaced by a few arcminutes every $\sim$40 scans to slowly cover the whole sky \cite{2010A&A...520A...1T,2011A&A...536A...1P}. Scanning strategies for many proposed future space missions involve such rotations. Similarly, most ground-based observatories scan the sky along parts of circular scans to keep the line-of-sight elevation constant. Ground-based experiments such as GroundBird and STRIP plan on a scanning strategy that consists of 360$^\circ$ circular scans \citep{2020JLTP..200..384L, 1967JMP.....8.2155G, 2018SPIE10708E..1GF, 2018SPIE10708E..2FI}. It is also the case for the future SWIPE stratospheric balloon \citep{2020arXiv200811049T}.

The usage of the circular scans as intermediate steps for CMB data analysis has been discussed in the context of the preparation of the analysis of Planck mission data \citep{1998astro.ph..4180D,1998MNRAS.298..445D,2002MNRAS.331..975V,2002MNRAS.331..994C}. In particular, it has been shown in \citep{1998MNRAS.298..445D} that the Fourier spectrum $\Gamma_m$ of CMB temperature anisotropies along circular scans can be easily connected to the full-sky CMB temperature angular power spectrum $C_\ell^{TT}$. In view of upcoming polarization CMB experiments, it is of interest to investigate whether similar relations exist for polarization signals. This would allow us to also calculate the Fourier spectra of $E$ and $B$ modes in CMB time-ordered data, as well as the Fourier cross spectra between $T$, $E$ and $B$, and potentially to use such data as intermediate steps in the analysis of future polarization data, with intermediate data products that would be directly connected both to the theoretical predictions and to the geometry of the observations.

An approximate method to recover the $C_\ell^{TT}$ power spectrum from the $T$ mode $\Gamma_m$ coefficients has been proposed by \cite{2003MNRAS.343..552A}. In that work, the authors invert the transformation matrix for $\Theta=90^\circ$, and re-scale the abscissa of the $\Gamma_m$ for $\Theta<90^\circ$ to match the corresponding $\Theta=90^\circ$ one. Here, we investigate a different method, based on the direct inversion of the linear system that connects $\Gamma_m$ to a binned version of the angular power spectrum $C_\ell$. We also extend previous work to include the analysis of polarization spectra.

The paper is organised as follows.
In Section 2 we describe the scanning on circular scans and the connection between the Fourier spectra on the rings and the $T$, $E$, $B$ spectra on the sphere. Section 3 discusses the inversion of the system in practice when we do not have enough $m$-modes to measure the $C_\ell$ for all $\ell$. The effectiveness of the method to calculate angular power spectra is investigated with numerical simulations in Section 4. Section 5 extends this work to the case where the scans are only approximately circular, before we conclude in Section 6.

\section{Relations between Fourier and Spherical Harmonic power spectra}


\subsection{Polarization measurements on circular scans}


\begin{figure}[hbt]
\centering
\includegraphics[width=0.9\textwidth]{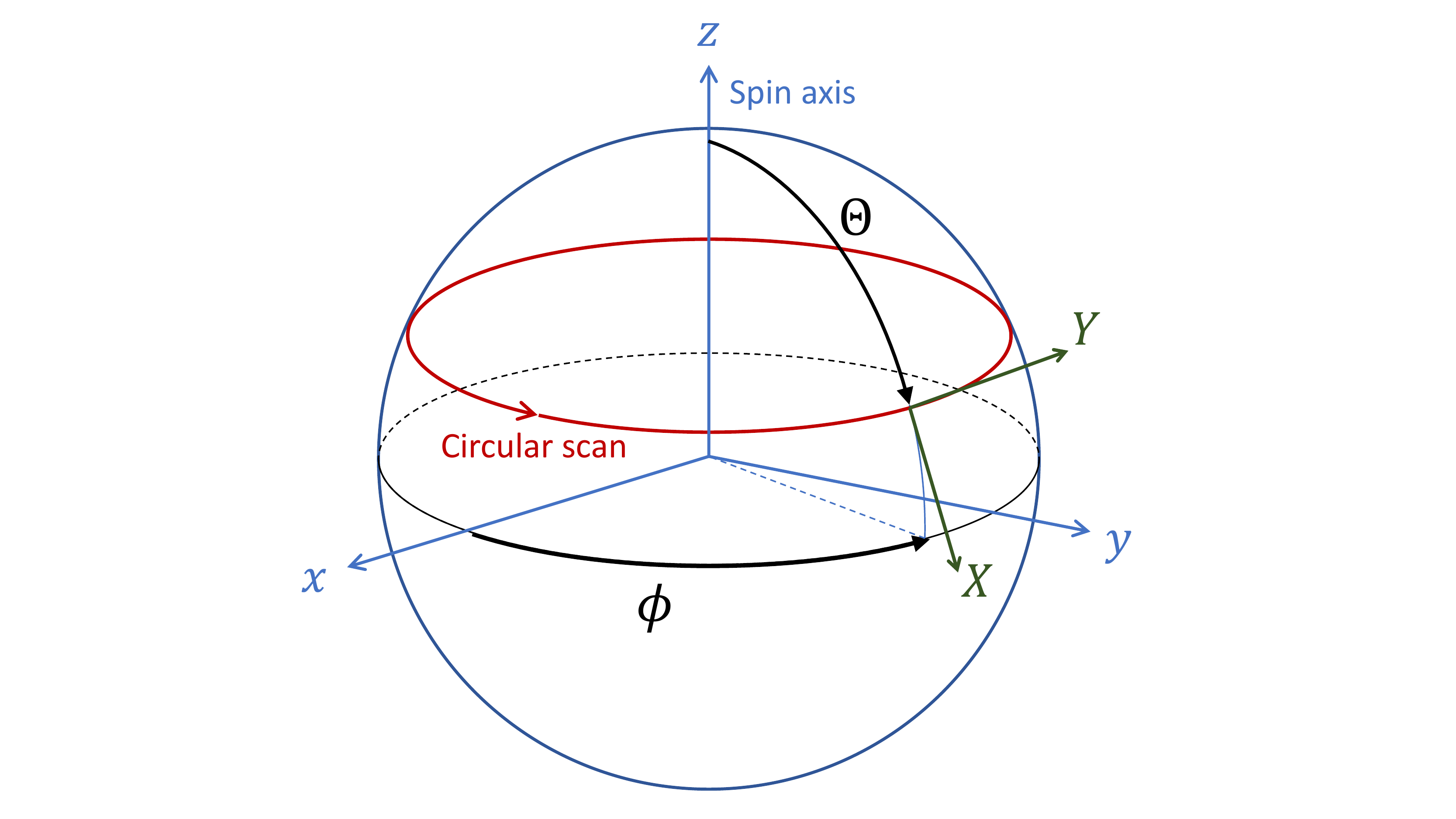}
\caption{\label{fig:scan} Circular scan with opening angle $\Theta$. The $X$ and $Y$ axes are references for the measurement of linear polarization Stokes parameters $Q$ and $U$. At time $t$, the detector of interest points towards a point with angular coordinates $\theta = \Theta$ and $\phi=\phi(t)$, where the time dependence of $\phi$ defines the scanning strategy along the circular scan.}
\end{figure}

Consider an experiment that continuously scans the sky along a single given ring on the sky. The opening angle of the ring, denoted as $\Theta$, is the angular radius of the ring on the sphere. Without loss of generality, we can choose spherical coordinates for the harmonic analysis of the CMB sky in such a way that the North Pole is at the center of the ring (see Fig.~\ref{fig:scan}). As the experiment scans the ring, the colatitude $\theta$ of the observation remains constant ($\theta=\Theta$), while the time-dependence of the longitude $\phi(t)$ defines the scanning strategy of the experiment along the ring.

The time-stream signal $x(t)$ observed by a single, ideal, perfectly polarized detector (polarimeter) with a scanning strategy $\phi(t)$ is:
\begin{eqnarray}
    x(t) & = & s(\Theta,\phi) + n(t) \\
         & = & T(\Theta,\phi) + Q(\Theta,\phi)\cos{2\psi} + U(\Theta,\phi)\sin{2\psi} + n(t),
\end{eqnarray}
where $T$, $Q$, $U$ are linear polarization Stokes parameters, $\psi$ is the orientation angle between the polarimeter and the polarization reference axis, and $n(t)$ is detector noise.
For a particular ring-shaped scan, it is convenient to measure $Q$ and $U$ in a reference frame for which the $X$-axis is perpendicular to the scan, away from the North Pole, and the $Y$-axis along the scan, towards the East. The ring-shaped circular scan and the notations used here are illustrated in Fig.~\ref{fig:scan}. The angle $\psi$ is measured from vector $X$, towards $Y$.

Using data from a set of polarization detectors to scan the same ring with different polarization angles $\psi$ evenly spread in $[0,\pi]$, it is possible to measure independently $T$, $Q$ and $U$ in each of the observed pixels \citep{1999A&AS..135..579C}. In the following, we assume that the detector data streams have been preprocessed to get measurements of $T(\Theta,\phi)$, $(Q+iU)(\Theta,\phi)$ and $(Q-iU)(\Theta,\phi)$ in the coordinate system described in Fig.~\ref{fig:scan}. Those ring-shaped data sets are the main data sets considered next for harmonic analysis.

\subsection{Fourier analysis of polarization}


We note $\alpha_{j,m}$ ($j=1, 2, 3$) as the coefficients of the Fourier decomposition of $T(\Theta,\phi)$, $(Q+iU)(\Theta,\phi)$ and $(Q-iU)(\Theta,\phi)$ respectively,
i.e.
\begin{equation}
\alpha_{1,m} = \frac{1}{2\pi}\int_{0}^{2\pi} {\rm d} \phi \, T \, {\rm e}^{-{\rm i}m\phi},
\end{equation}
\begin{equation}
\alpha_{2,m} = \frac{1}{2\pi}\int_{0}^{2\pi} {\rm d} \phi \,  (Q+iU) \, {\rm e}^{-{\rm i}m\phi},
\end{equation}
and
\begin{equation}
\alpha_{3,m} = \frac{1}{2\pi}\int_{0}^{2\pi} {\rm d} \phi \,  (Q-iU) \, {\rm e}^{-{\rm i}m\phi},
\end{equation}
where the dependence of $T$, $Q$ and $U$ on the angles $(\Theta,\phi)$ has been omitted in the notations for simplicity.

With the (spin-weighted) spherical harmonic expansion coefficients of the Stokes parameters, the multivariate ring power spectrum $\Gamma^{jk}_m\equiv\langle \alpha_{j,m}^*\alpha_{k,m}\rangle$ \cite{1998MNRAS.298..445D} is connected to the ensemble average temperature and polarization power spectra by a linear equation:
\begin{equation}
\label{relation between gamma and C}
\Gamma_m^{jk}=\sum_{\ell=|m|}^{+\infty}K^\ell_m(\theta,j,k,X,Y) \cdot C^{XY}_\ell
\end{equation}
where $j,k$ range from 1 to 3 and $X,Y$ stand here for $T$, $E$ and $B$.
The concrete form of $K^\ell_m(\theta,j,k,X,Y)$ is derived in appendix \ref{app:math}. We have:
\begin{align}
\Gamma_m^{11}=&\sum \mathcal P^1_{\ell m}\mathcal P^1_{\ell m}C_\ell^{TT} \nonumber \\
\Gamma_m^{12}=&\sum \mathcal P^1_{\ell m}\mathcal P^2_{\ell m}\big(-C_\ell^{TE}-iC_\ell^{TB}\big) \nonumber \\
\Gamma_m^{21}=&\sum \mathcal P^2_{\ell m}\mathcal P^1_{\ell m}\big(-C_\ell^{TE}+iC_\ell^{TB}\big) \nonumber \\
\Gamma_m^{13}=&\sum \mathcal P^1_{\ell m}\mathcal P^3_{\ell m}\big(-C_\ell^{TE}+iC_\ell^{TB}\big) \nonumber \\
\Gamma_m^{31}=&\sum \mathcal P^3_{\ell m}\mathcal P^1_{\ell m}\big(-C_\ell^{TE}-iC_\ell^{TB}\big) \nonumber \\
\Gamma_m^{22}=&\sum \mathcal P^2_{\ell m}\mathcal P^2_{\ell m}\big(C_\ell^{EE}+C_\ell^{BB}\big) \nonumber \\
\Gamma_m^{23}=&\sum \mathcal P^2_{\ell m}\mathcal P^3_{\ell m}\big(C_\ell^{EE}-C_\ell^{BB}-2iC_\ell^{EB}\big) \nonumber \\
\Gamma_m^{32}=&\sum \mathcal P^3_{\ell m}\mathcal P^2_{\ell m}\big(C_\ell^{EE}-C_\ell^{BB}+2iC_\ell^{EB}\big) \nonumber \\
\Gamma_m^{33}=&\sum \mathcal P^3_{\ell m}\mathcal P^3_{\ell m}\big(C_\ell^{EE}+C_\ell^{BB}\big),
\label{equation: relation between Gamma and C}
\end{align}
where all sums range from $\ell=|m|$ to $+\infty$, and where $\mathcal P^1_{\ell m}(\theta)$,
$\mathcal P^2_{\ell m}(\theta)$ and
$\mathcal P^3_{\ell m}(\theta)$ are defined by
\begin{equation}
Y_{\ell m}(\theta,\phi)\equiv\prescript{}{0}{Y}_{\ell m}(\theta,\phi)=\mathcal P^1_{\ell m}(\theta)e^{im\phi}
\label{definition of P1 function}
\end{equation}
\begin{equation}
\prescript{}{2}{Y}_{\ell m}(\theta,\phi)=\mathcal P^2_{\ell m}(\theta)e^{im\phi}
\label{definition of P2 function}
\end{equation}
\begin{equation}
\prescript{}{-2}{Y}_{\ell m}(\theta,\phi)=\mathcal P^3_{\ell m}(\theta)e^{im\phi},
\label{eq:definition of P3 functions}
\end{equation}
and where $Y_{\ell m}$ is Spherical Harmonics while $\prescript{}{0}{Y}_{\ell m}$, $\prescript{}{2}{Y}_{\ell m}$ and $\prescript{}{-2}{Y}_{\ell m}$ are spin-weighted Spherical Harmonics whose definition can be found in the Appendix. We have
\begin{equation}
\mathcal P_{\ell m}^1(\theta)=\sqrt{\frac{2\ell +1}{4\pi}\frac{(\ell-m)!}{(\ell+m)!}}P_{\ell m}(\cos\theta),
\end{equation}
where $P_{\ell m}$ is an Associated Legendre Polynomial.\footnote{There are two different definitions of Associated Legendre Polynomials: Hobson's notation and Ferrer's notation. Our definition of spin-weighted Spherical Harmonics $\prescript{}{s}{Y}_{\ell m}$ is consistent with the Hobson's notation. The relationship between these two notations is : $P_{\ell m}({\rm Ferrer})\equiv (-1)^mP_{\ell m}({\rm Hobson})$.}
We note the following properties
\begin{itemize}
    \item Diagonal terms $\Gamma_m^{ii}$ are always real;
    \item Off-diagonal terms $\Gamma_m^{ij}$ and $\Gamma_m^{ji}$ ($i \neq j$) are complex conjugates, and are real and equal if either $C_\ell^{TB}$ or $C_\ell^{EB}$ vanishes;
    \item Appropriate linear combinations of the various $\Gamma_m^{ij}$ can be formed to isolate a weighted sum of $C_\ell^{TT}$ with no contribution from polarization, and of $(C_\ell^{EE}+C_\ell^{BB})$, or $(C_\ell^{EE}-C_\ell^{BB})$ with no contribution from intensity; There is, however, no direct way to isolate a weighted sum of $C_\ell^{EE}$ or a weighted sum of $C_\ell^{BB}$ independently of each other.
\end{itemize}

\section{Inversion of the system -- Spectral estimation}


\subsection{Approximations and practical calculations}

\begin{figure}[thbp]
	\centering
	\subfigure{\includegraphics[scale=0.37]{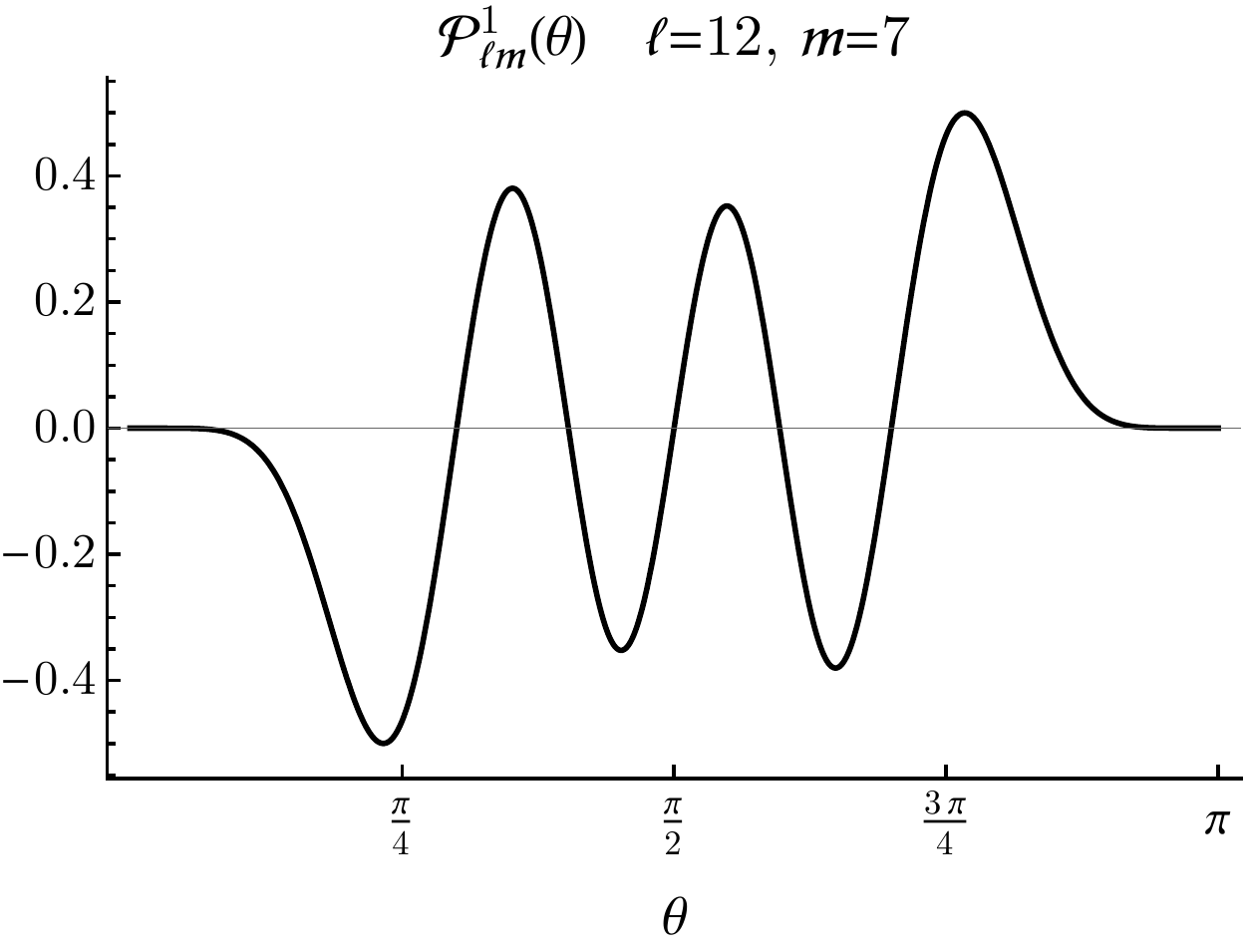}}
	\subfigure{\includegraphics[scale=0.37]{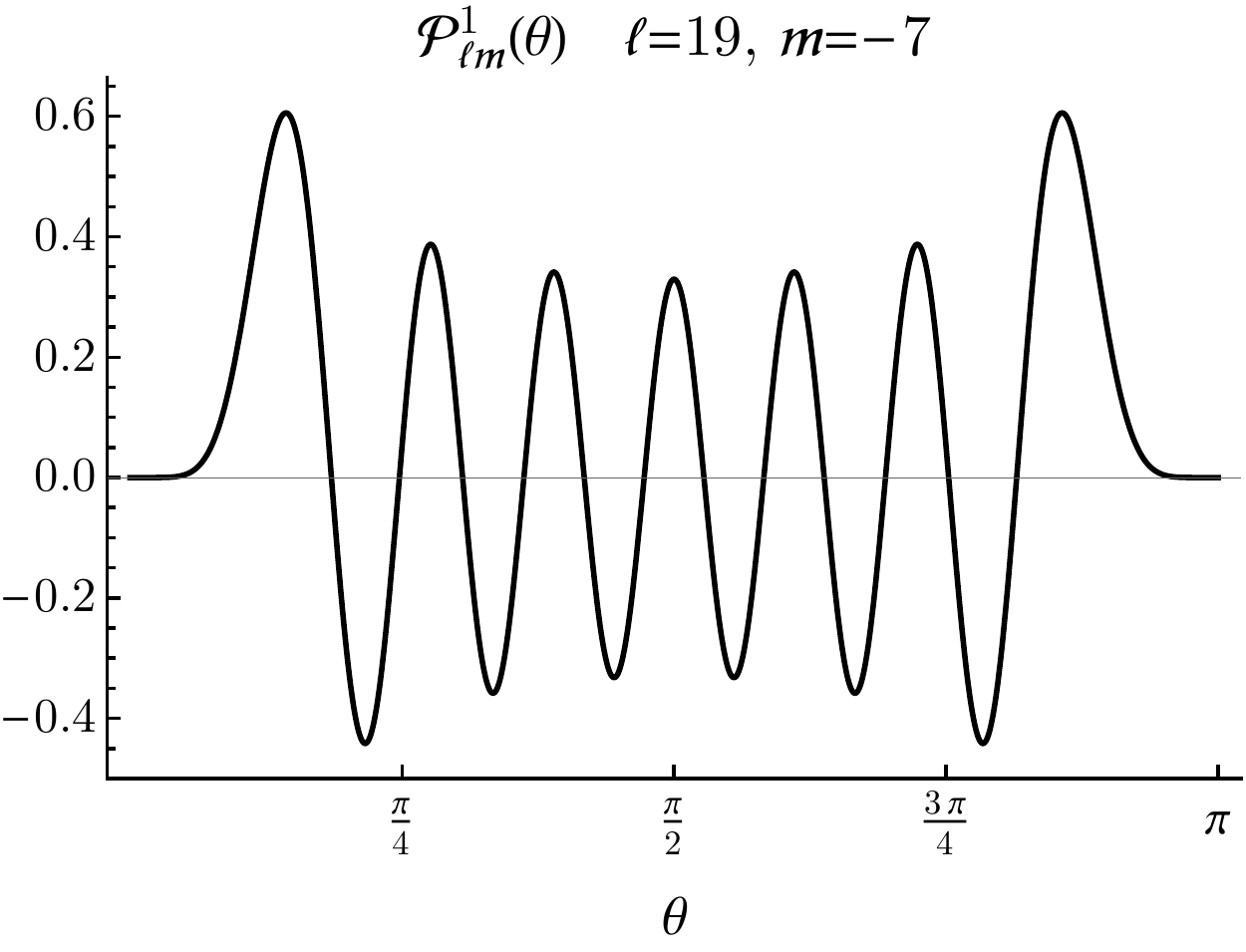}}
	\subfigure{\includegraphics[scale=0.37]{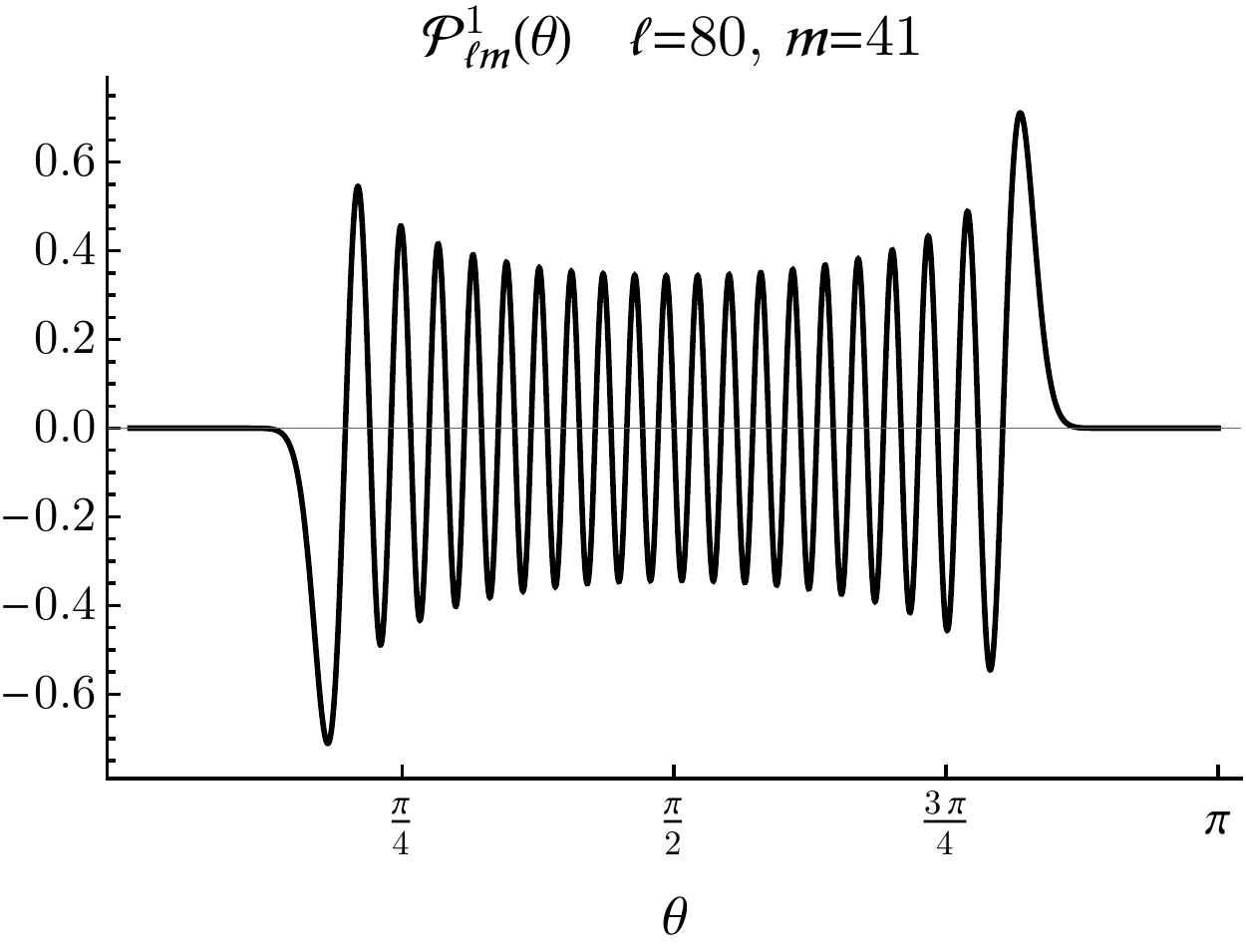}}
	
	\subfigure{\includegraphics[scale=0.37]{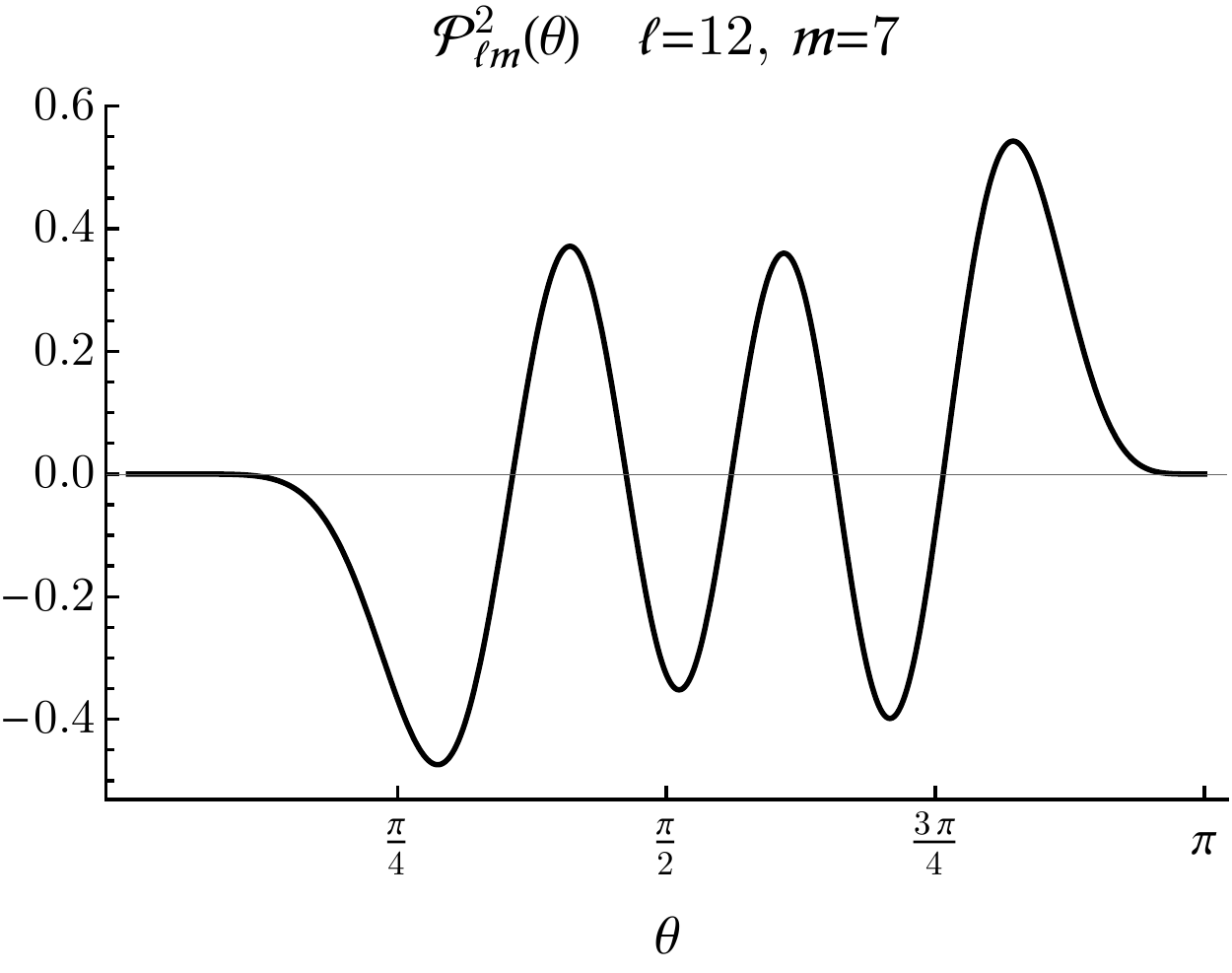}}
	\subfigure{\includegraphics[scale=0.37]{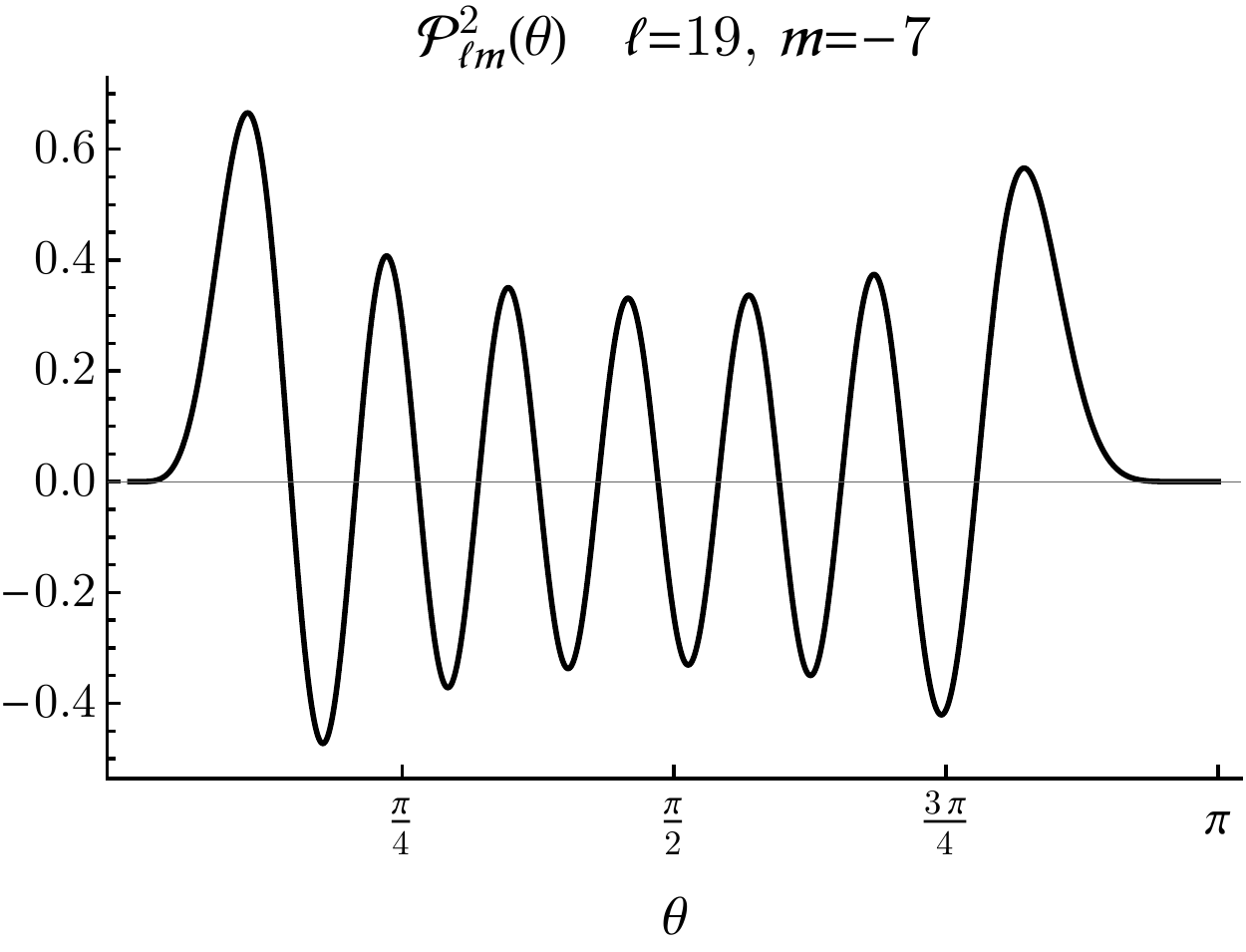}}
	\subfigure{\includegraphics[scale=0.37]{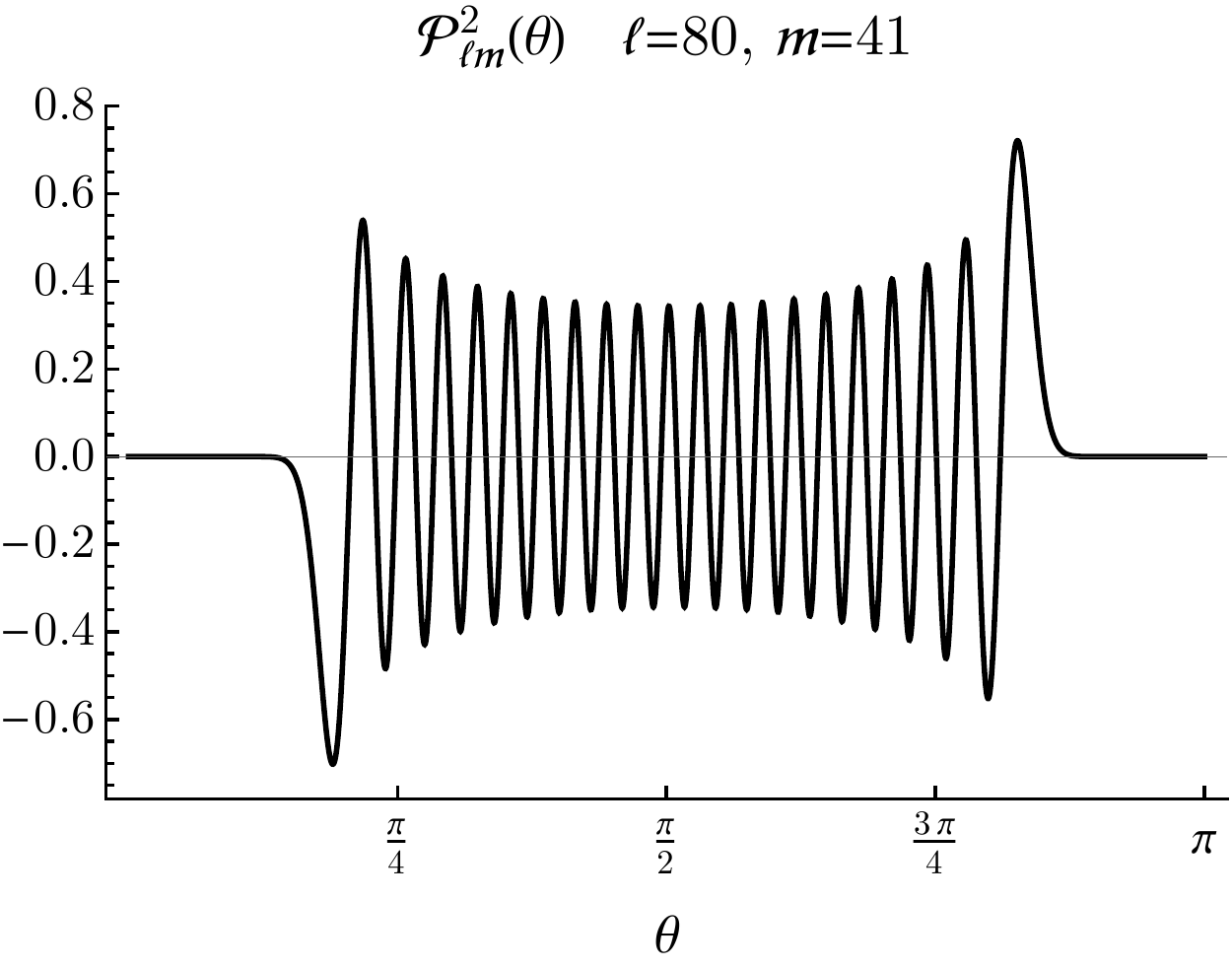}}
	
	\subfigure{\includegraphics[scale=0.37]{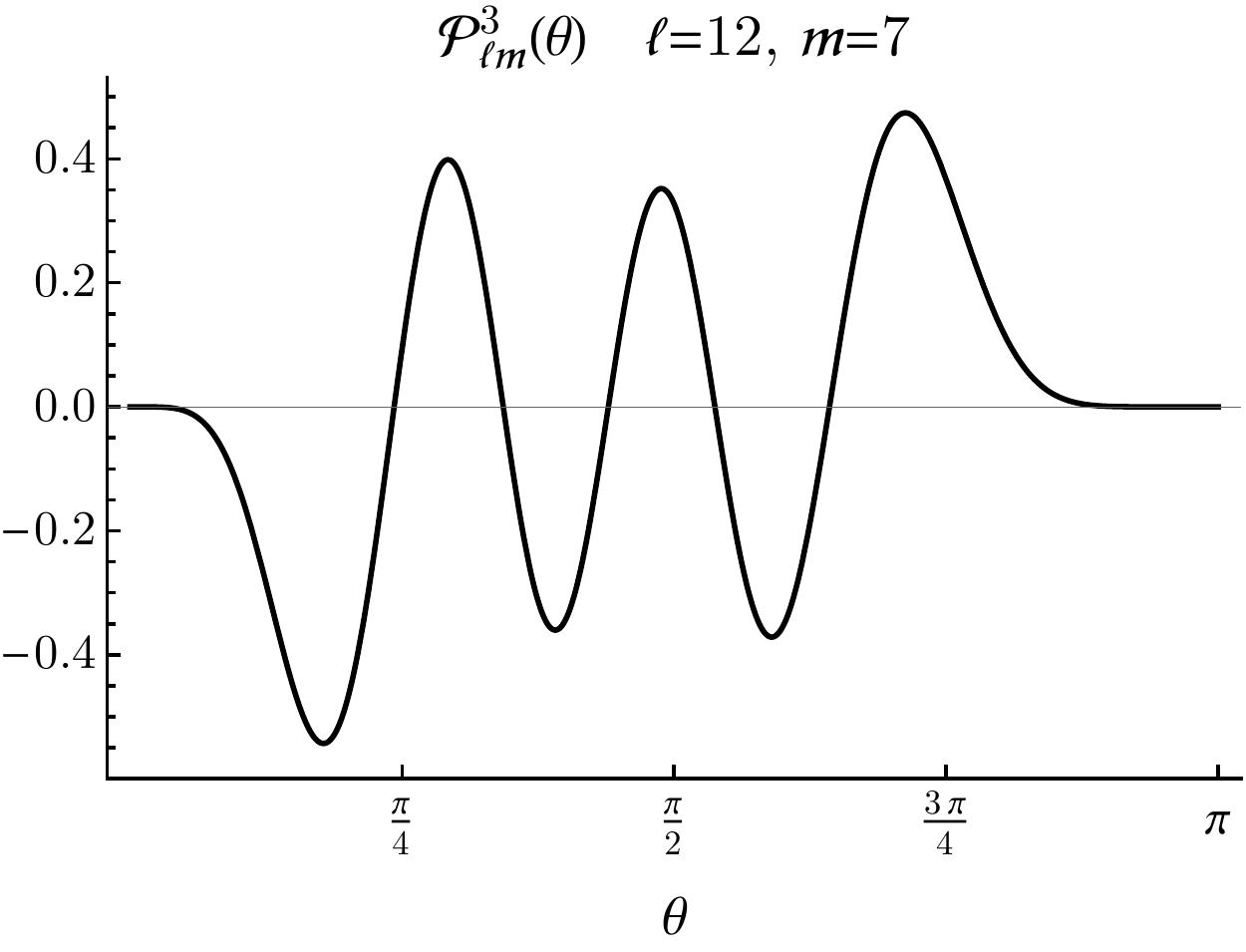}}
	\subfigure{\includegraphics[scale=0.37]{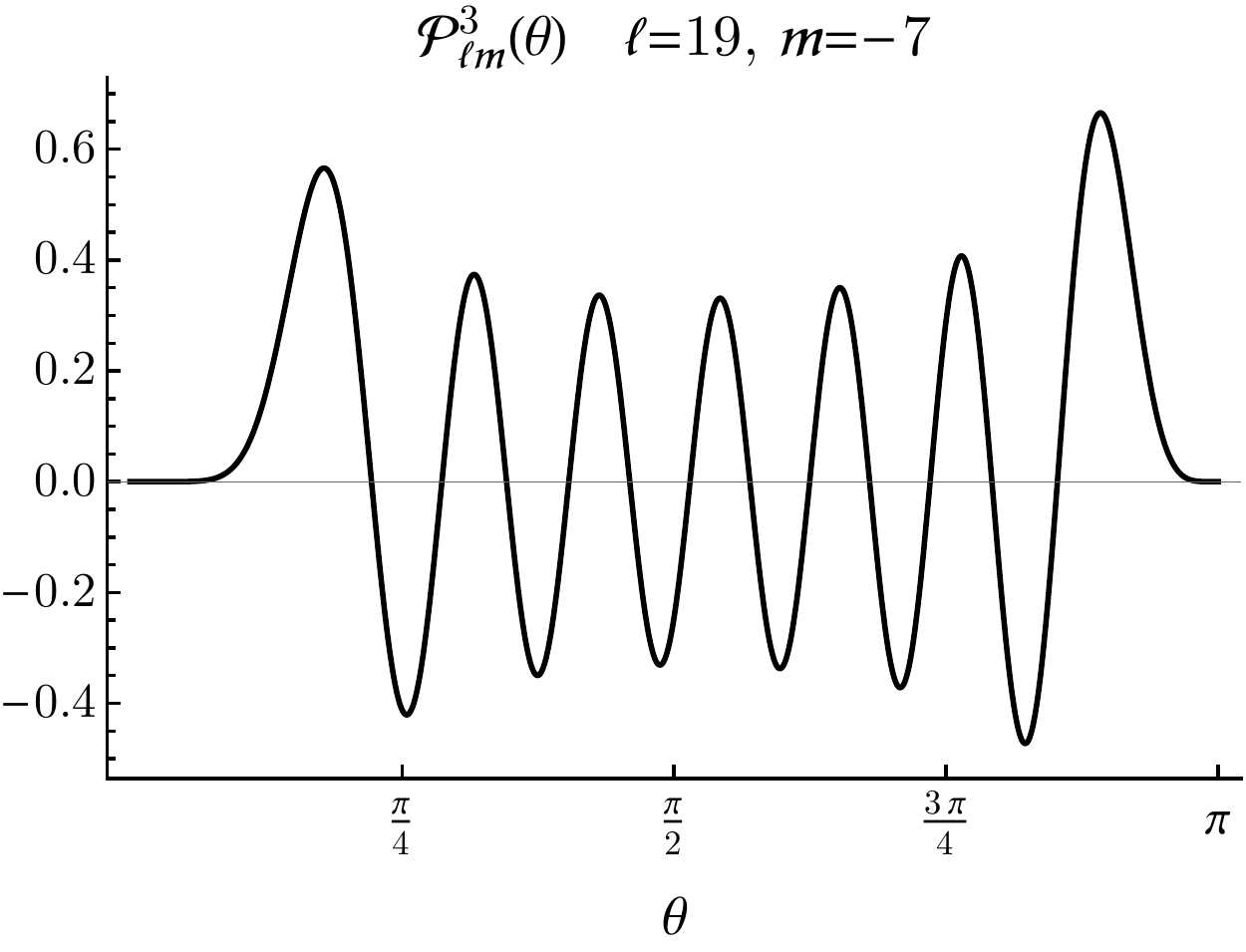}}
	\subfigure{\includegraphics[scale=0.37]{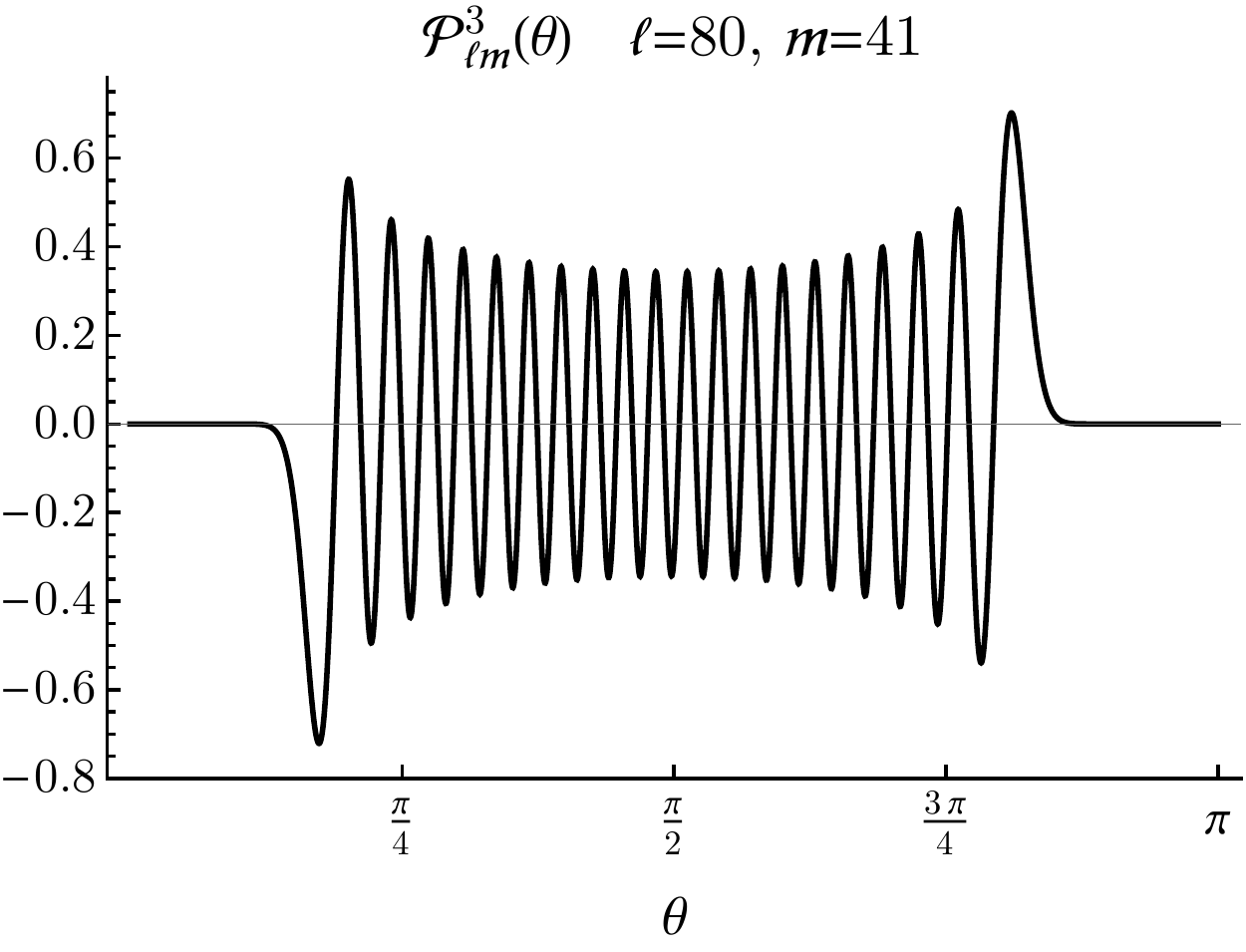}}
	\caption{Curves of the functions $\mathcal{P}^k_{\ell m}, k=1,2,3$. When $\theta$ is close to $0$ or $\pi$, the value of these functions is close to zero. With the increase of $\ell$, they exhibit a strong oscillatory behavior (Right panels). }
\label{plot_function_P}
\end{figure}

In practical observations of CMB temperature and polarization, the angular resolution of the observation is limited by the beam of the experiment. In addition, the CMB power spectrum at high $\ell$ is exponentially damped by photon diffusion \cite{1968ApJ...151..459S}. Hence, when $\ell$ is sufficiently large, the value of $C_\ell^{XY}$ becomes close to zero, and we can truncate the sum in formula (\ref{relation between gamma and C}) at a maximum value $L$. Thus we re-write Eq.~\eqref{relation between gamma and C} as:
\begin{equation}
\Gamma_m^{jk}=\sum_{\ell=|m|}^{L}K^\ell_m(\theta,j,k,X,Y)\cdot C^{XY}_\ell.
\label{approx relation between gamma and C}
\end{equation}
The value of $L$ can be inferred from the smallest scale that is being considered on the sky map. For a map with $N_{\rm pix}$ independent pixels, we get a number of harmonic modes equal to the number of sky pixels when
\begin{equation}
\sum_{\ell=0}^L (2\ell+1) \simeq N_{\rm pix}.
\end{equation}
which results in
\begin{equation}
L+1 \simeq \sqrt{N_{\rm pix}}.
\end{equation}
On a ring-shaped scan of angular length $2\pi \sin\Theta$, the maximum value $M$ of $m$ to be used for practical calculations can be set to
\begin{equation}
M \simeq L\sin\Theta.
\end{equation}
Details and practical implications of this are discussed in \citep{2003MNRAS.343..552A}.
We then reformulate Eq.~\eqref{approx relation between gamma and C} in matrix notation as:
\begin{equation}
\bm{\Gamma^{jk}}=\bm{K}.\bm{C^{XY}}
\label{eq:Gamma-C-linear-system}
\end{equation}
where $\bm{\Gamma^{jk}}=\begin{pmatrix}\Gamma^{jk}_{2}&\cdots&\Gamma^{jk}_{M}\end{pmatrix}^T$, for $j,k=1,2,3$, and $\bm{C^{XY}}=\begin{pmatrix}C^{XY}_{2}&\cdots&C^{XY}_{L}\end{pmatrix}^T$, for $X,Y=T,E,B$.\footnote{See appendix for detailed discussions about the minimum value for $m$ and $\ell$.}
All matrices are now finite-dimensional, and can be computed numerically.

\begin{figure}[tbp]
	\centering
	\subfigure{\includegraphics[scale=0.46]{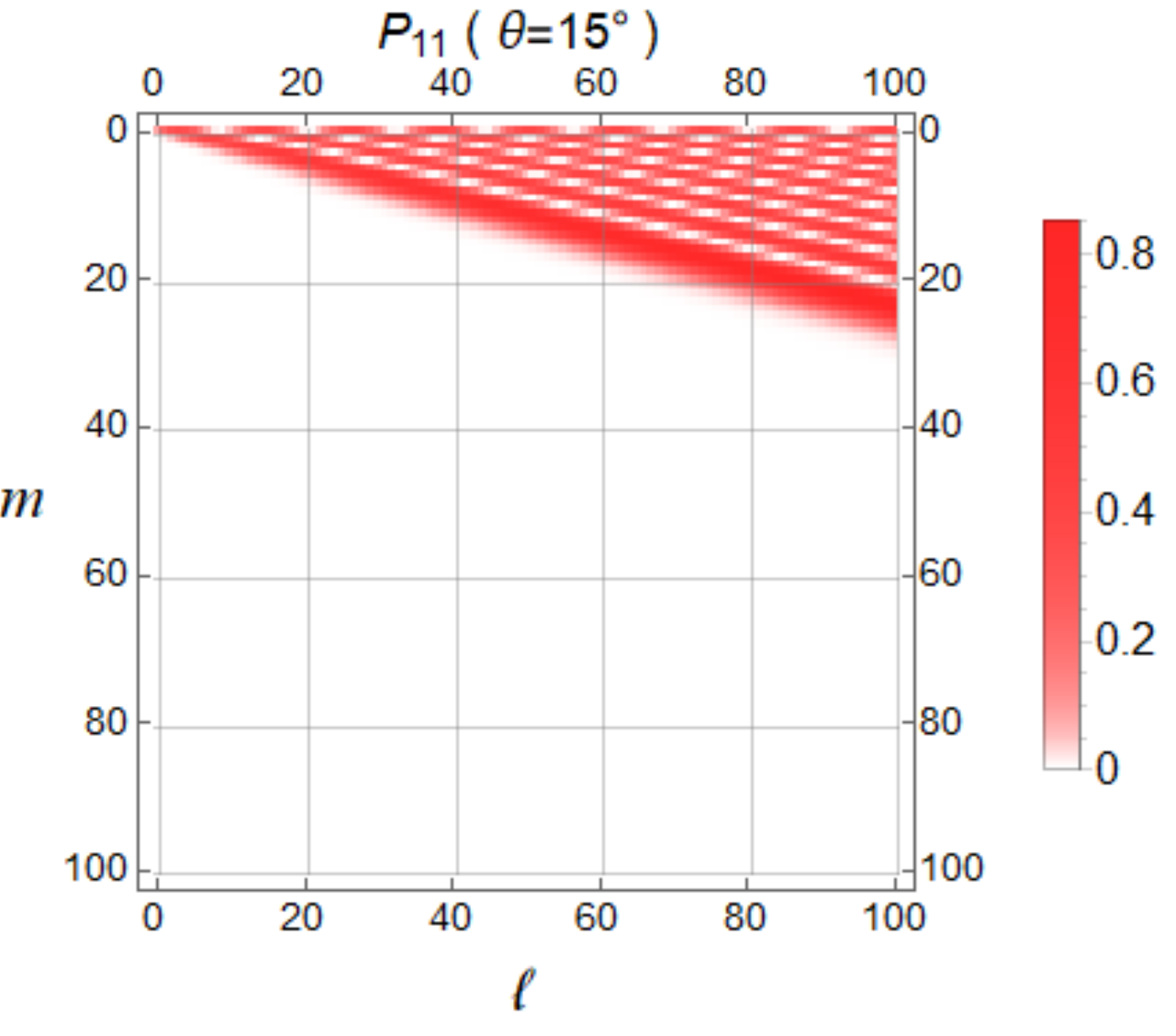}}\hspace{0.5cm}
	\subfigure{\includegraphics[scale=0.46]{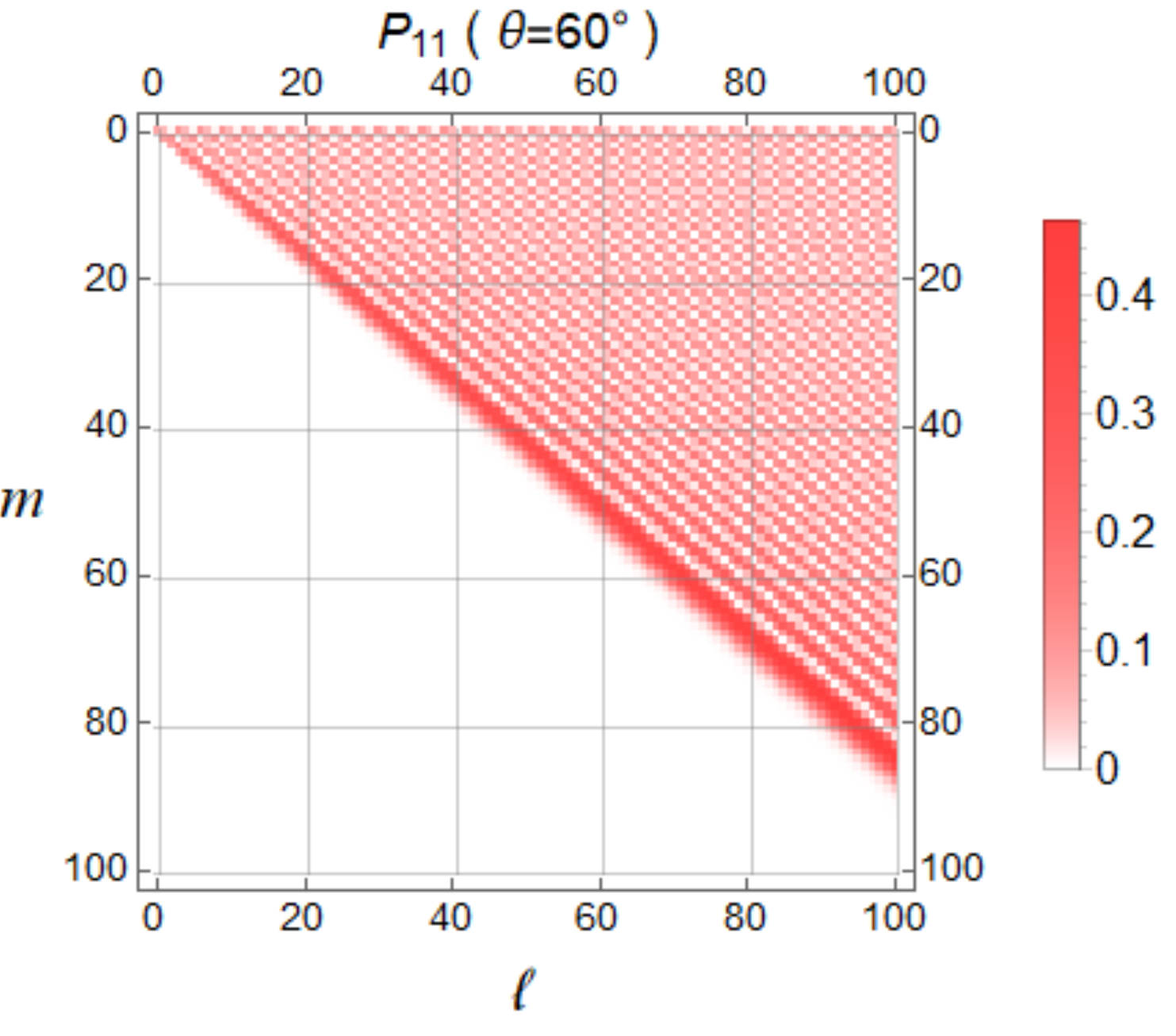}}
	\subfigure{\includegraphics[scale=0.46]{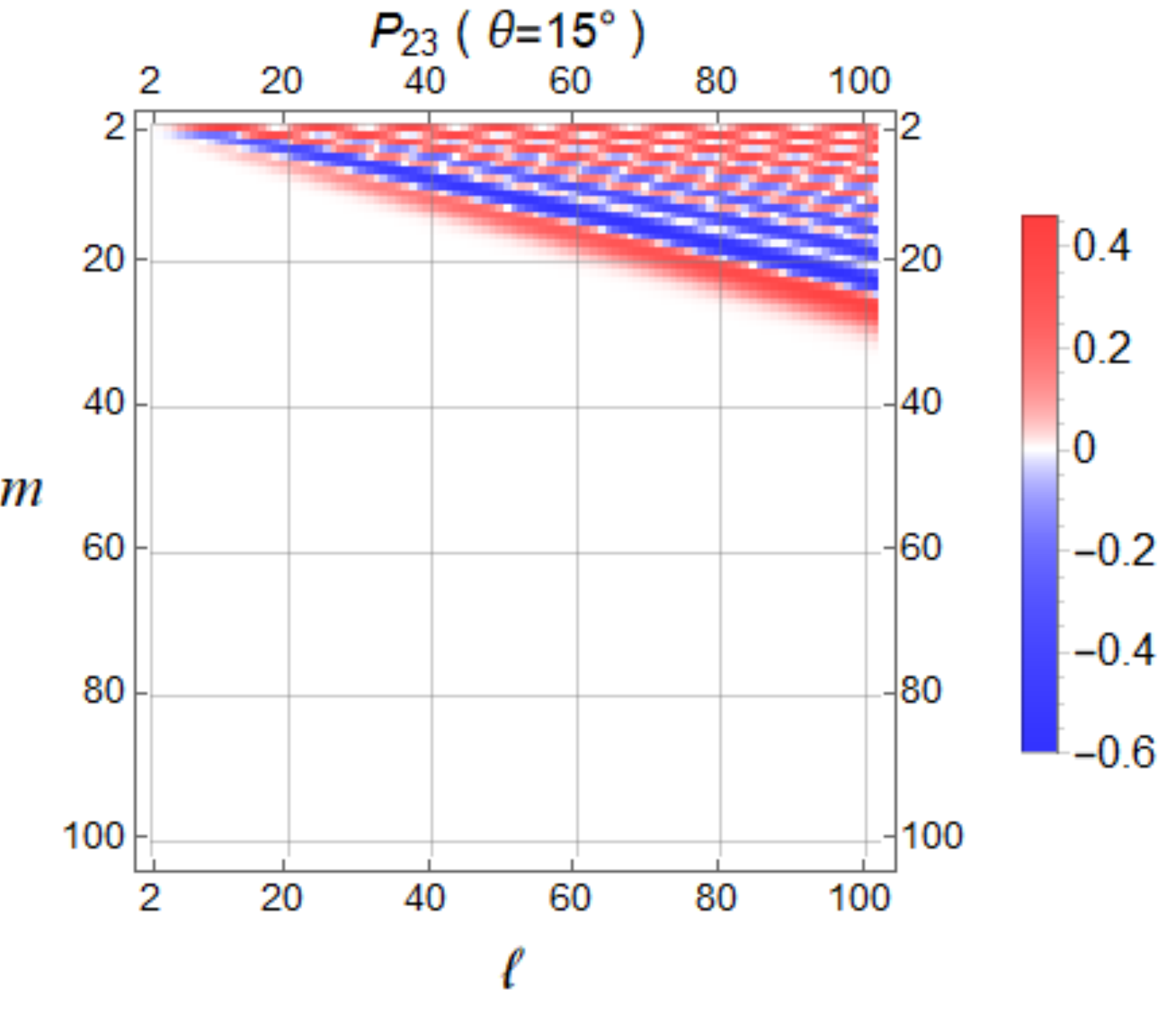}}\hspace{0.4cm}
	\subfigure{\includegraphics[scale=0.46]{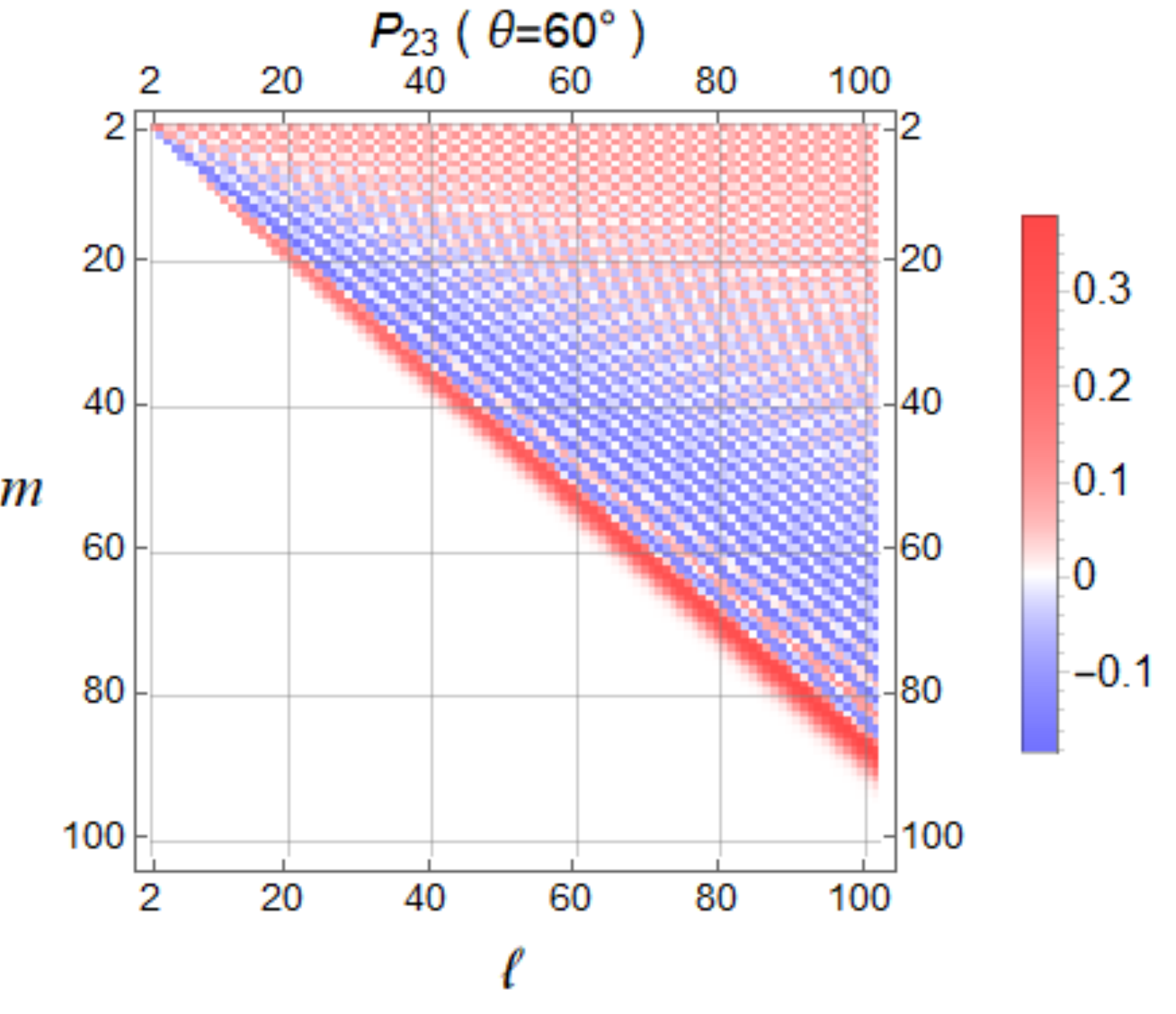}}
	\caption{Visual representation of the first elements of matrix \texorpdfstring{$\bm{P_{11}}$}{} (top two panels) and \texorpdfstring{$\bm{P_{23}}$}{} (bottom two panels), for two different ring opening angles. These plots display zero values (or those very close to zero) as white, with negative and positive values in various shades of red and blue respectively. The left panels correspond to a small ring ($\theta=15^\circ$) and the right panels  to a large one ($\theta=60^\circ$). The ratio of the short side to the long side of the colored triangles is about $\sin\theta$.}
\label{plot_matrix_P}
\end{figure}

For notation convenience, we define a new matrix, $\bm{P_{jk}}$ as follows:
\begin{equation}
\bm{P_{jk}}\equiv
\begin{pmatrix}
\mathcal P^j_{22}(\theta)\mathcal P^k_{22}(\theta)&\mathcal P^j_{32}(\theta)\mathcal P^k_{32}(\theta)&\cdots&\mathcal P^j_{L2}(\theta)\mathcal P^k_{L2}(\theta)\\
0&\mathcal P^j_{33}(\theta)\mathcal P^k_{33}(\theta)&\cdots&\mathcal P^j_{L3}(\theta)\mathcal P^k_{L3}(\theta)\\
\vdots&&\ddots&\vdots\\
0&0&&\mathcal P^j_{LM}(\theta)\mathcal P^k_{LM}(\theta)
\end{pmatrix}
\end{equation}
In Appendix~\ref{app:math}, we show how the matrix $\bm{K}$ can be decomposed into several $\bm{P_{jk}}$ matrices.
Fig.~\ref{plot_function_P} and Fig.~\ref{plot_matrix_P} show the visualization of some examples of $\mathcal{P}^k_{\ell m}$ functions  and $\bm{P_{jk}}$ matrices respectively.

\subsection{Inverting the system}

Given a multivariate power spectrum of the ring data, $\bm{\Gamma^{jk}_m}$, which can be directly computed from the ring-projection of the time-ordered data, it is interesting to invert the linear system of Eq.~\eqref{eq:Gamma-C-linear-system} to get an estimate of the harmonic power spectra $\bm{C^{XY}_\ell}$. However, since $M \simeq L\sin\theta$, and hence $M \leq L$, the matrix $\bm{K}$ is in general not square, and hence in general not invertible. For $\Theta \neq \pi/2$, the system is degenerate.
To lift this degeneracy, one can modify the linear system to express it in terms of band-averaged $C_\ell$, i.e. we can bin the vector $\bm{C_\ell}$, and modify matrix $\bm{K}$ to avoid the singularity:\footnote{Considering that $D_\ell \equiv \ell (\ell+1) C_\ell/2\pi$ is shown more often than $C_\ell$ in practice, we can choose to bin $D_\ell$ instead of $C_\ell$. This amounts to replacing the 1's in the bin matrix of Eq.~\eqref{eq:binmatrix} by $2\pi/(\ell (\ell+1))$.} 

\begin{equation}
\bm{C_\ell}\simeq
\begin{pmatrix}
1\\
\vdots\\
1\\
&1\\
&\vdots\\
&1\\
&&\ddots\\
&&&1\\
&&&\vdots\\
&&&1
\end{pmatrix}\bm{\bar{C}}=\bm{B}\bm{\bar{C}},
\label{eq:binmatrix}
\end{equation}
where $\bm{\bar{C}}$ denotes a binned version of the harmonic power spectrum $\bm{C}$.
In the $i$-th column, the number of non-zero elements of the ``bin matrix" $\bm{B}$ is $b_i$, which is named as ``bin-step". In order to eliminate the singularity, increasing the value of bin-steps in geometrical progression usually works. For example,
\begin{equation}
b_i= {\rm Round}\big(b_1\times q^{i-1}\big)
\end{equation}
where ${\rm Round}(x)$ gives the integer closest to $x$. 
As is usual in CMB observations, the value of the first term $b_1$ and common ratio $q$ can be chosen as a function of the geometry of the observations and the noise level of the experiment.
%
Then in the binned approximation, we can re-write,
\begin{equation}
\bm{\Gamma_m}\simeq\bm{K}\bm{B}\bm{\bar{C}}
\end{equation}
and solve for an estimate of $\bm{C_\ell}$ using
\begin{equation}
\bm{C_\ell}\simeq\bm{B}\big(\bm{K}\bm{B}\big)^{-1}\bm{\Gamma_m}.
\end{equation}
The asymptotically equal symbol occurs here since we ``flattened" the power spectrum $C_\ell$ by binning. 

We note that in general $\big(\bm{K}\bm{B}\big)$ is not a square matrix either. For strong binning there are more values of $\bm{\Gamma_m}$ than different values of $\bm{\bar{C}}$. We can then use the left pseudo inverse $\big(\bm{K}\bm{B}\big)^+$ of
$\big(\bm{K}\bm{B}\big)$ to perform the inversion, with
\begin{equation}
\bm{M}^+ = \left( \bm{M}^T \bm{M} \right)^{-1} \bm{M}^T.
\end{equation}
It is also possible to take into account specifics of the observations and replace the above pseudo-inverse by a weighted version
\begin{equation}
\bm{M}^+ = \left( \bm{M}^T \bm{W} \bm{M} \right)^{-1} \bm{M}^T \bm{W},
\label{eq:weighted-inversion}
\end{equation}
where $\bm{W}$ is a weighting matrix (for instance, the inverse of the noise covariance of the observations, or a filter which down-weights or cuts out measurements contaminated by strong systematics).

\subsection{Discarding some modes}

We note that by reason of the triangular form of $\bm{P_{jk}}$ matrices, it is possible to restrict the system to high values of $\ell$ and $m$. This can be useful for practical purposes, when the Fourier spectrum of rings for the lowest $m$ values is poorly measured, and/or when the observed sky patch does not allow for a precise measurement of the lowest $\ell$ modes.

In practical observations indeed, it often happens that the lowest $m$-modes of the ring-shaped scans are poorly measured. This can happen by reason of fluctuations of atmospheric emission on large scales, ground-pickup in the sidelobes of the instrumental optical response, or low frequency noise in the readout electronics.

It is standard in CMB data analysis to filter-out these contaminated modes before map-making from the instrumental data streams. After re-projection of the filtered data streams onto sky maps, this results in an anisotropic filtering of the CMB maps, which must be taken into account by Monte-Carlo simulations for the analysis of the co-added maps.

As an alternative, when the CMB $C_\ell$ are directly computed from the observed $\Gamma_m$, one can remove the contaminated modes from the linear system, keeping only the equations corresponding to $m \geq m_{\rm min}$. This also removes from the unknowns all the $C_\ell$ for which $\ell < m_{\rm min}$, with no impact on the capacity to measure all the remaining $C_\ell$, without the need to use computer-intensive simulations to evaluate the transfer function of the pipeline.
We make use of this flexibility in numerical tests of system inversion in the following sections.


\subsection{Connection with the spectra of time streams}

CMB observations are actually collected in the form of time streams. We note that in the case where rings are scanned repeatedly at constant angular speed $\omega$, there is a direct connection between the angle $\phi$ along the ring, and the time variable ($\phi=\omega t + \phi_0$). This yields a direct connection between Fourier spectra of observation time streams and harmonic spectra $C_\ell$.

\section{Numerical simulations}


In this section, we present numerical computations of $\bm{\Gamma_m}$ from $\bm{C_\ell}$ for a standard cosmological scenario, followed by an inversion to recover the (binned) input $\bm{C_\ell}$. We also demonstrate the computation of the harmonic power spectra $\bm{C_\ell}$ from empirical Fourier spectra of rings obtained by scanning simulated CMB maps along circles with various angular diameters.

\subsection{Fourier coefficients \texorpdfstring{$\Gamma_m^{jk}$}{} for \texorpdfstring{$\Lambda$}{}CDM models}

\begin{figure}
    \centering
    \includegraphics{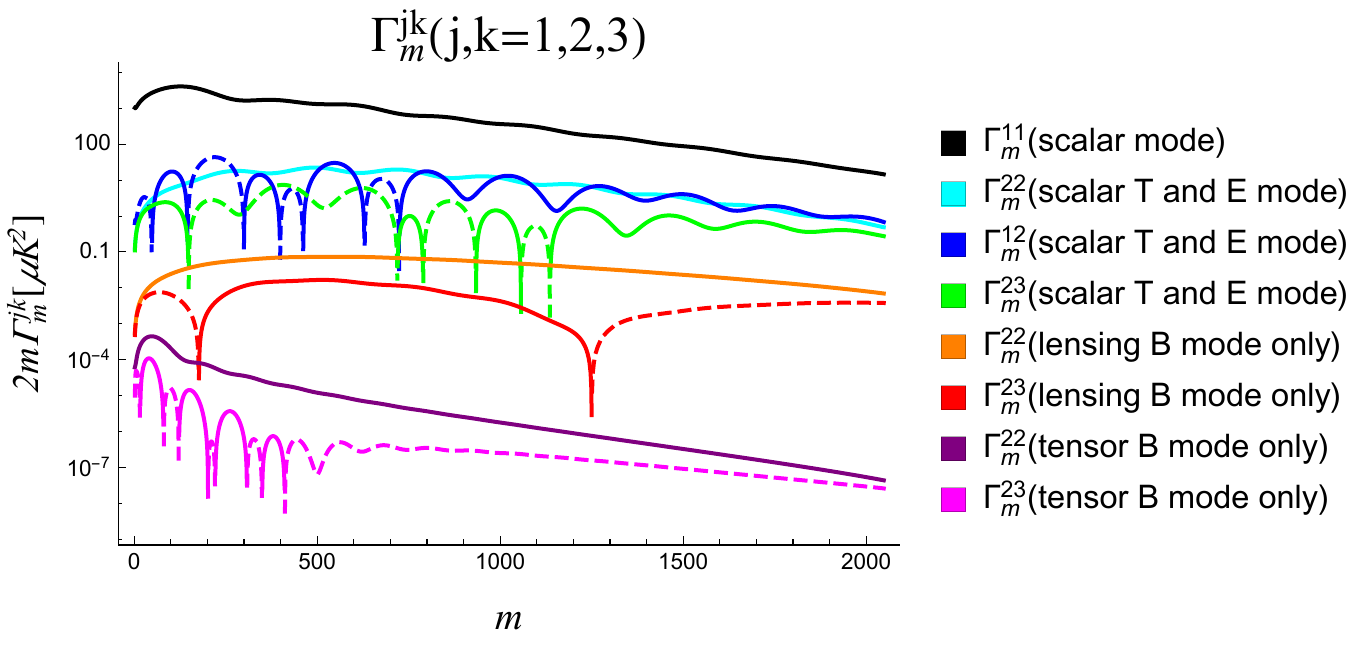}
    \includegraphics{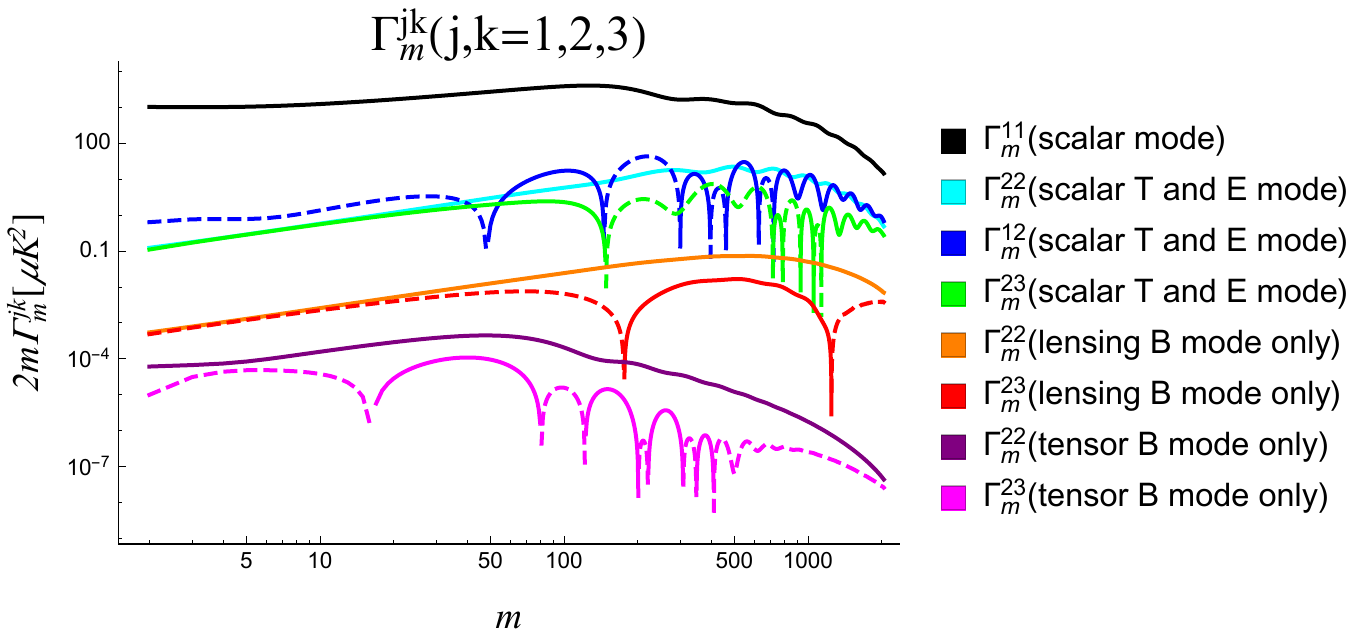}
    \caption{Ring power spectra of CMB on circular scans calculated  using Eq.~\eqref{approx relation between gamma and C}, for $\Theta=50^\circ$. The dashed lines indicate negative values. 
    $\Gamma_m^{33}$ and $\Gamma_m^{13}$ are indistinguishable from $\Gamma_m^{22}$ and $\Gamma_m^{12}$ (differences being typically of the order of 1\%), and are not plotted here.}
    \label{Fig:plotGamma}
\end{figure}

We start from some CMB spectra computed via the CLASS code \cite{2011arXiv1104.2932L}. To evaluate the relative contribution of the scalar and tensor modes to each of the $\bm{\Gamma^{jk}}$, we consider three inputs: (a) only $T$ modes and $E$ modes from primordial scalar perturbations; (b) $B$ modes due to the lensing of primordial scalar perturbations; (c) $B$ modes from primordial tensor fluctuations, with the tensor to scalar ratio $r=0.01$. Fig.~\ref{Fig:plotGamma} shows the ring power spectra computed according to Eq.~\eqref{approx relation between gamma and C}, for $\Theta=50^\circ$. An interesting feature is that while for $r=0.01$ the spectra of lensing B-modes and primordial B-modes are comparable for a substantial range of $\ell$, it is not the case for the corresponding Fourier spectra. The reason for this is the summation of the power over all values of $\ell \geq |m|$ in Eq.~\eqref{equation: relation between Gamma and C}.

	
	

\subsection{System inversion for theoretical spectra}

\begin{figure}[htbp]
	\centering
	\subfigure{\includegraphics[scale=0.37]{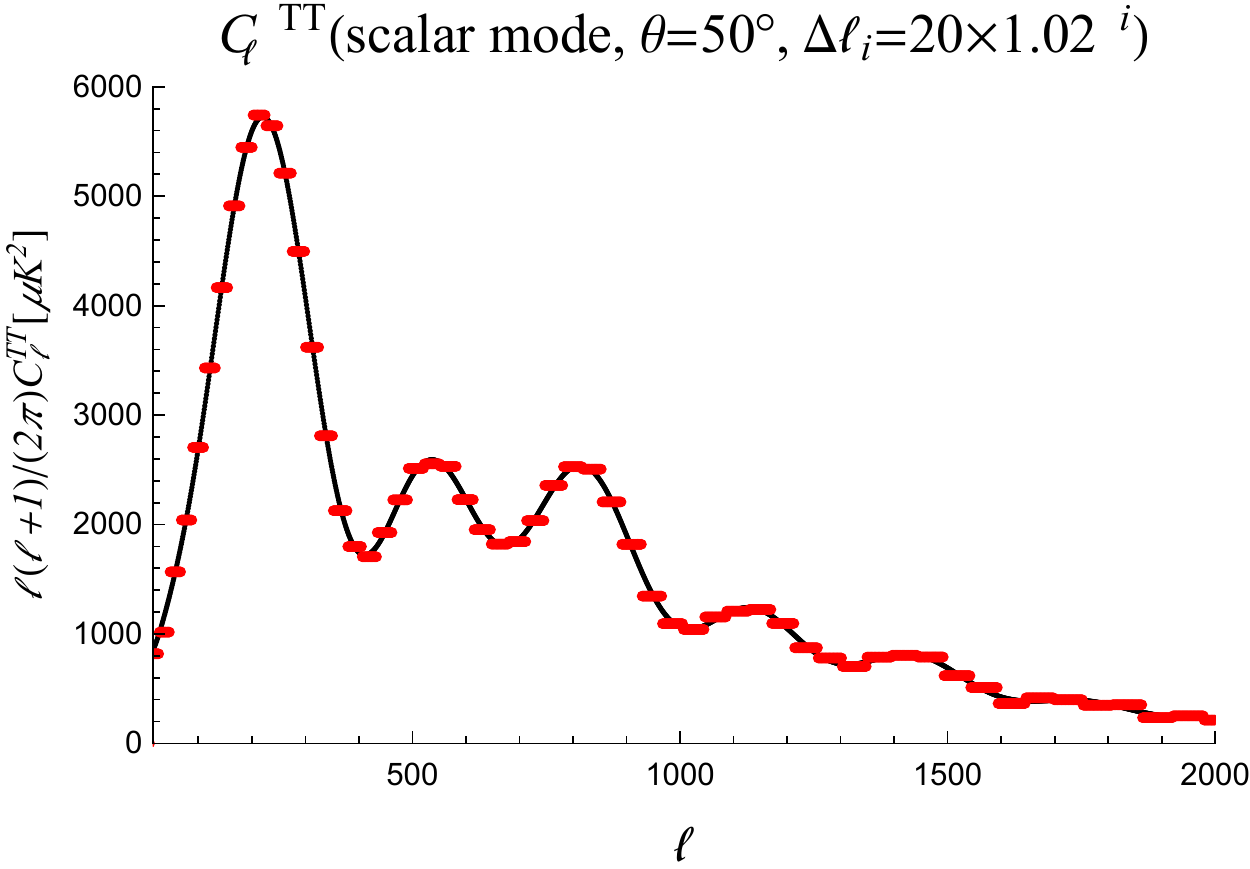}}
	\subfigure{\includegraphics[scale=0.37]{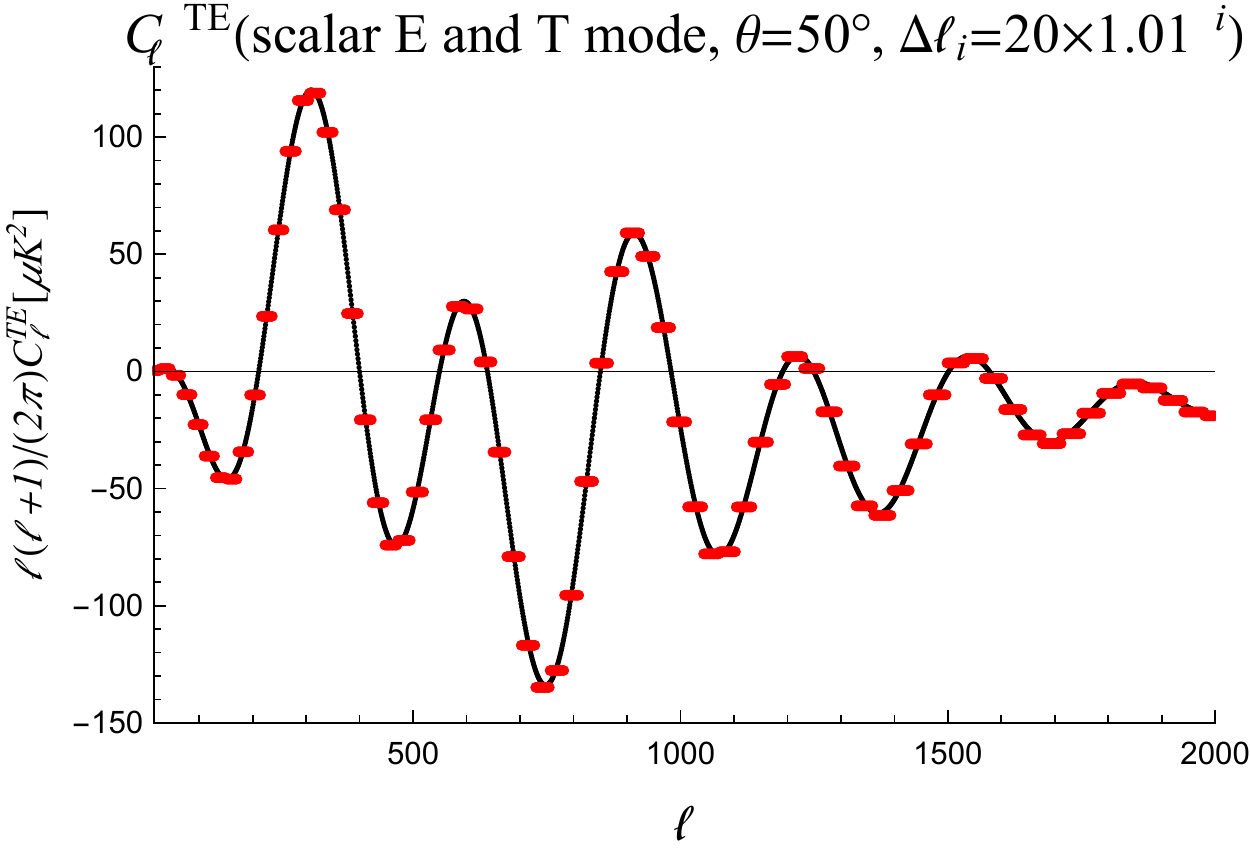}}
	\subfigure{\includegraphics[scale=0.37]{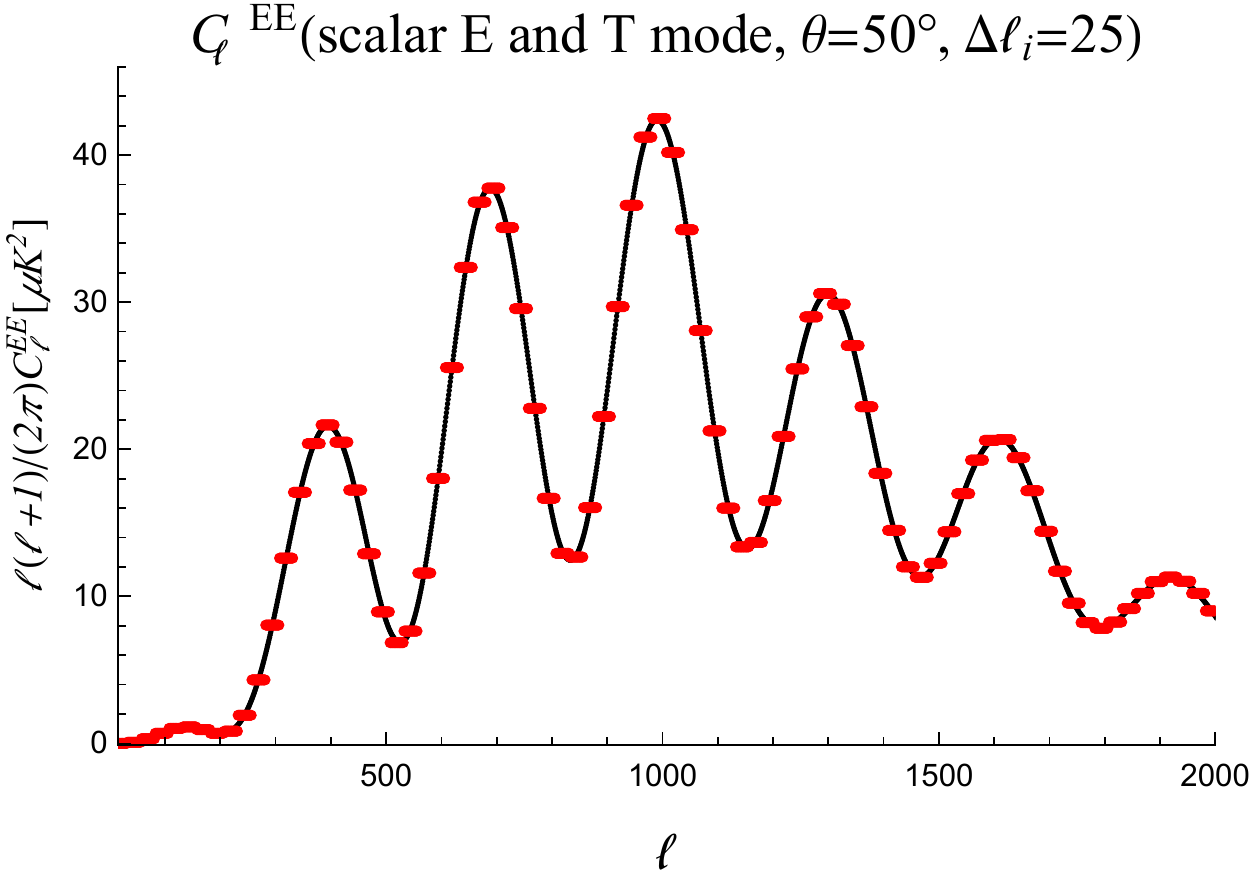}}
	
	\subfigure{\includegraphics[scale=0.37]{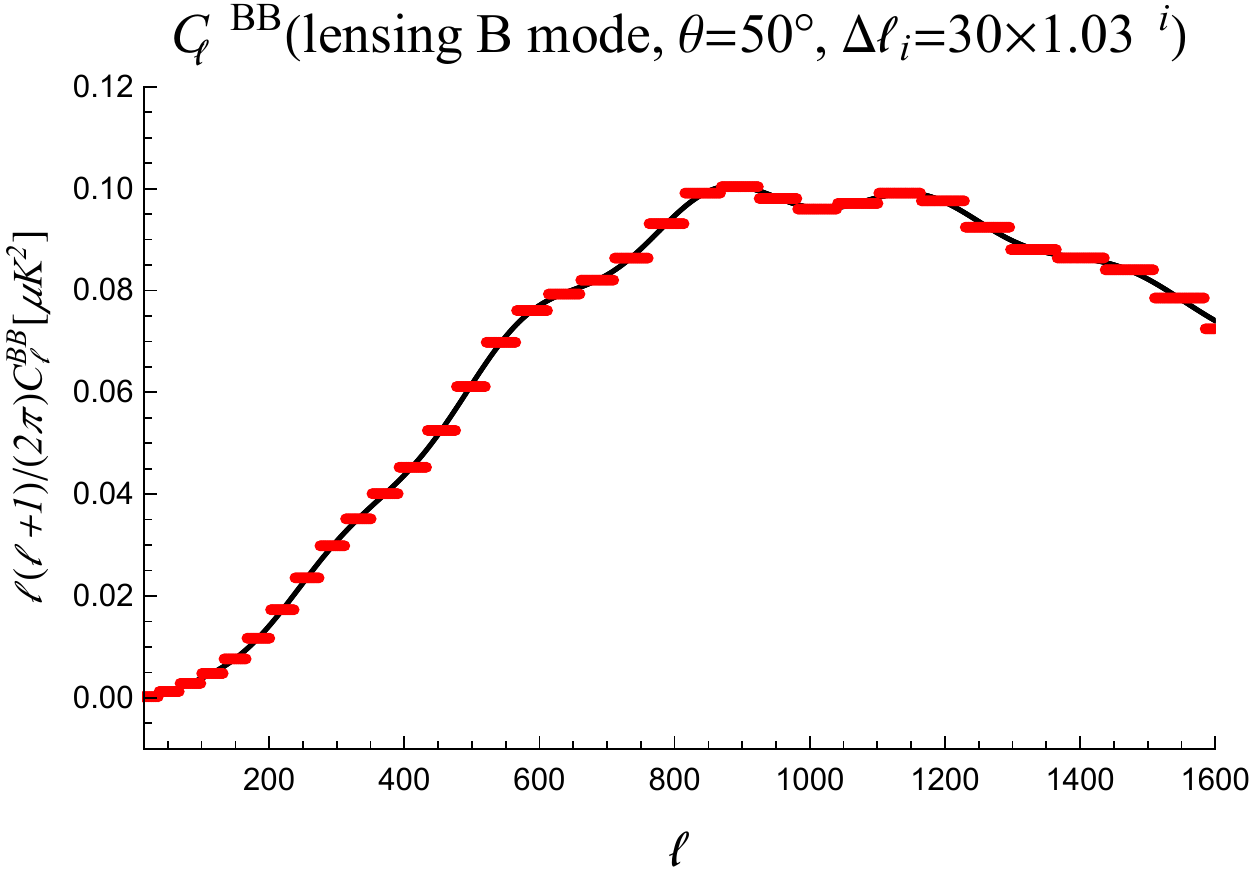}}
	\subfigure{\includegraphics[scale=0.37]{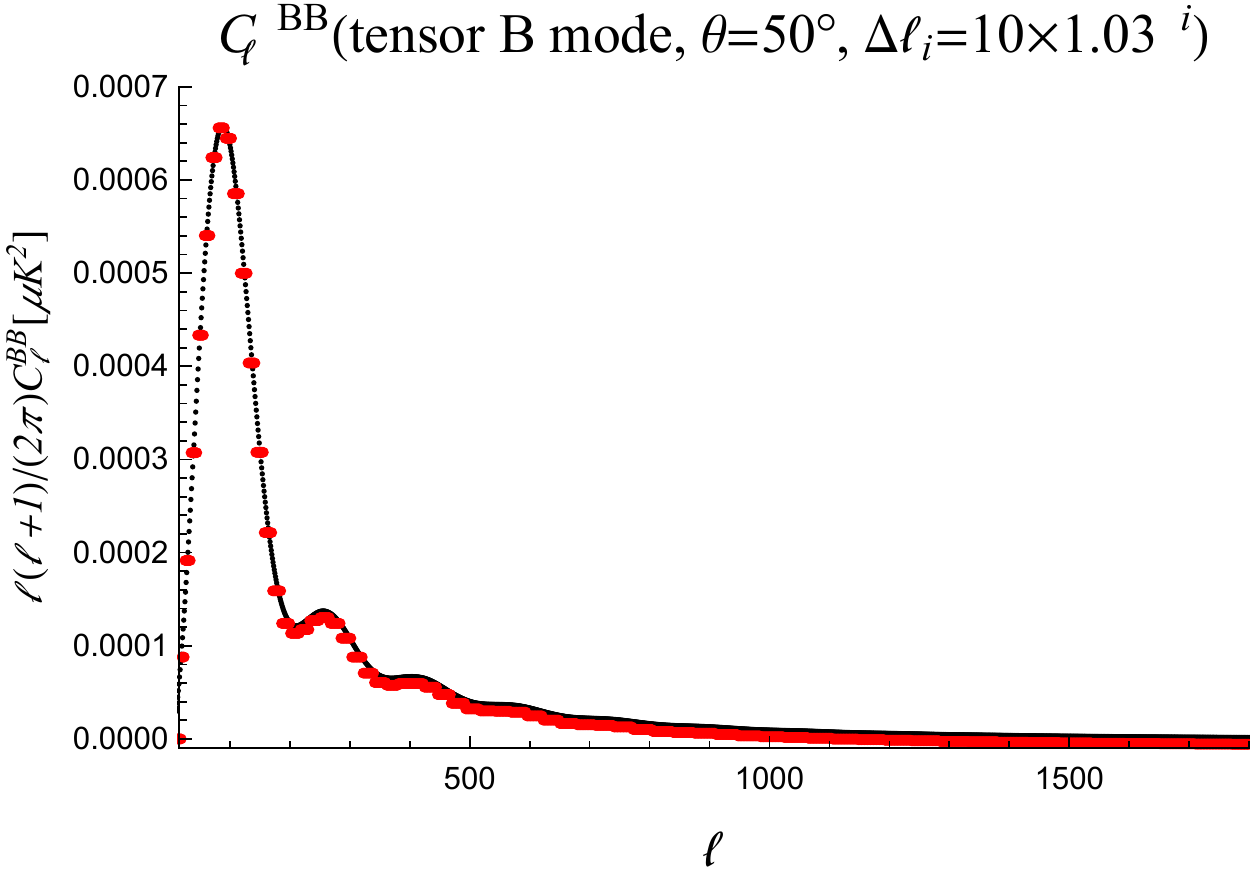}}
	\caption{Comparison between the theoretical value of CMB full sky power spectra (red points) used to calculate the theoretical ring power spectra $\Gamma_m$, and those (black lines) obtained by inverting the transformation system with theoretical $\Gamma_m$-s for $\theta=50^\circ$.}
\label{inverse_result}
\end{figure}

We now invert the transformation system to check whether the full-sky power spectra $\bm{C_\ell}$ can be recovered from the $\bm{\Gamma_m}$. Results are shown in Fig.~\ref{inverse_result}, demonstrating that the inversion works well. For this particular inversion, we considered only $\bm{\Gamma_m}$ for $|m| \geq 10$, except for the tensor B-mode spectrum, which is obtained using Fourier modes with $|m| \geq 20$. 
These values are arbitrary. For a practical experiment, one would optimize the range of $m$ being considered, as well as a possible weighting matrix in Eq.~\eqref{eq:weighted-inversion}. We postpone a possible optimization of the matrix inversion to future work, in which the noise properties and the practical scanning of a specific experiment are taken into account.

\subsection{System inversion for simulated observations}

\begin{figure}[htbp]
	\centering
	\subfigure{\includegraphics[scale=0.37]{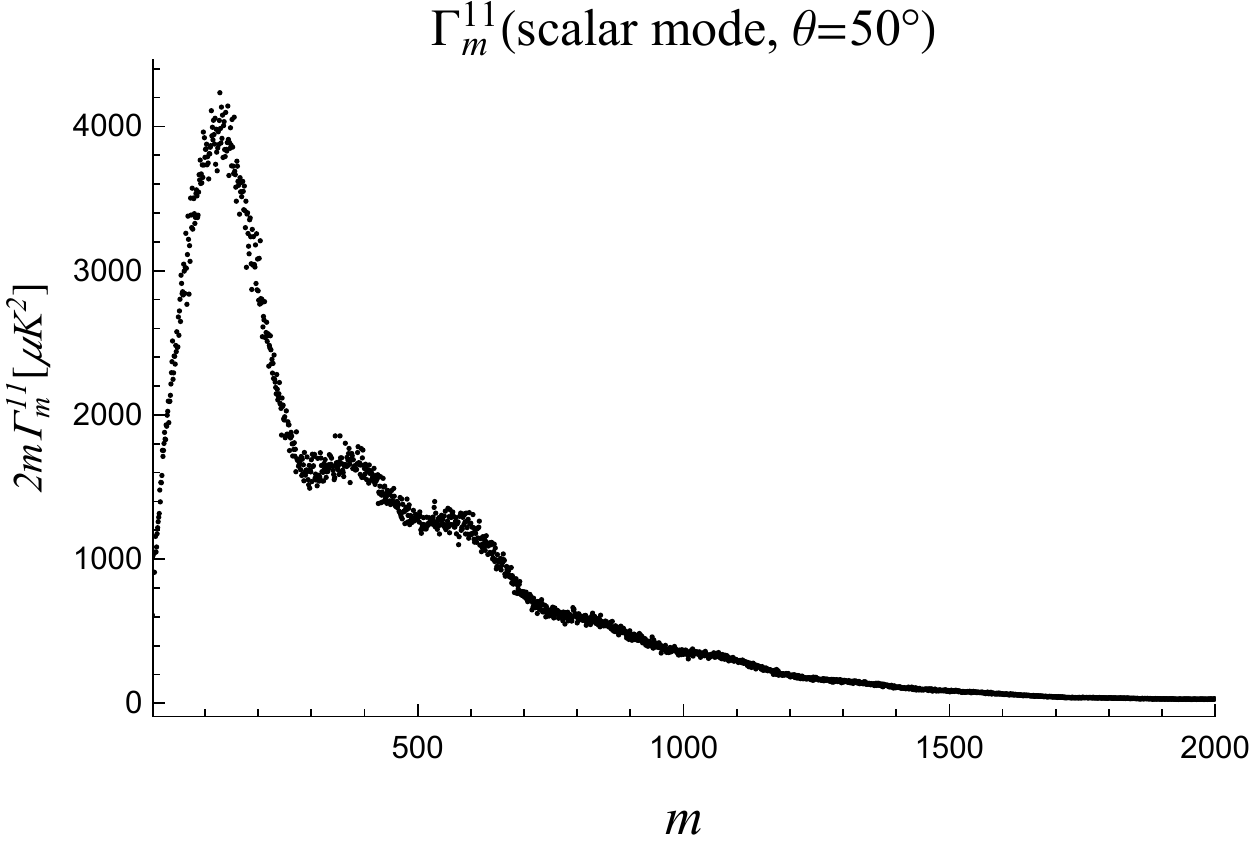}}
	\subfigure{\includegraphics[scale=0.37]{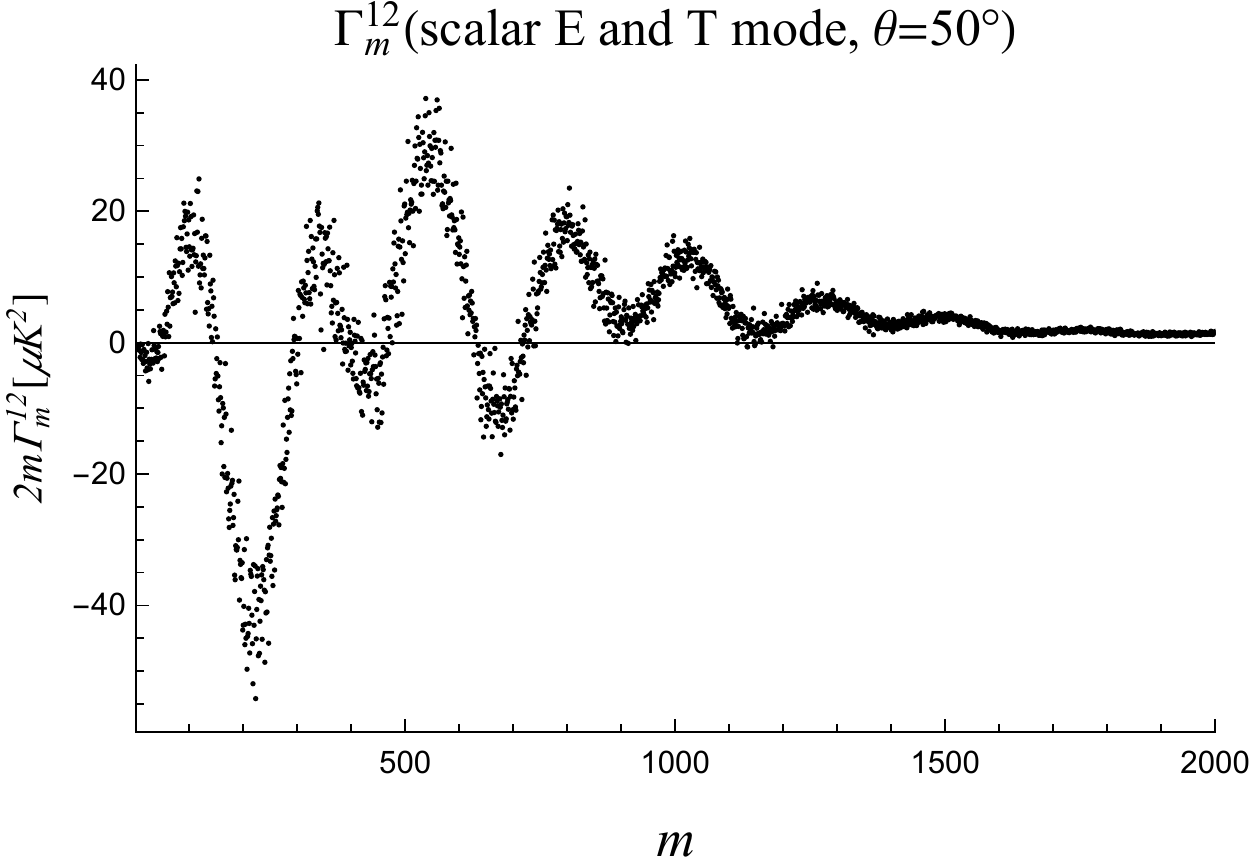}}
	\subfigure{\includegraphics[scale=0.37]{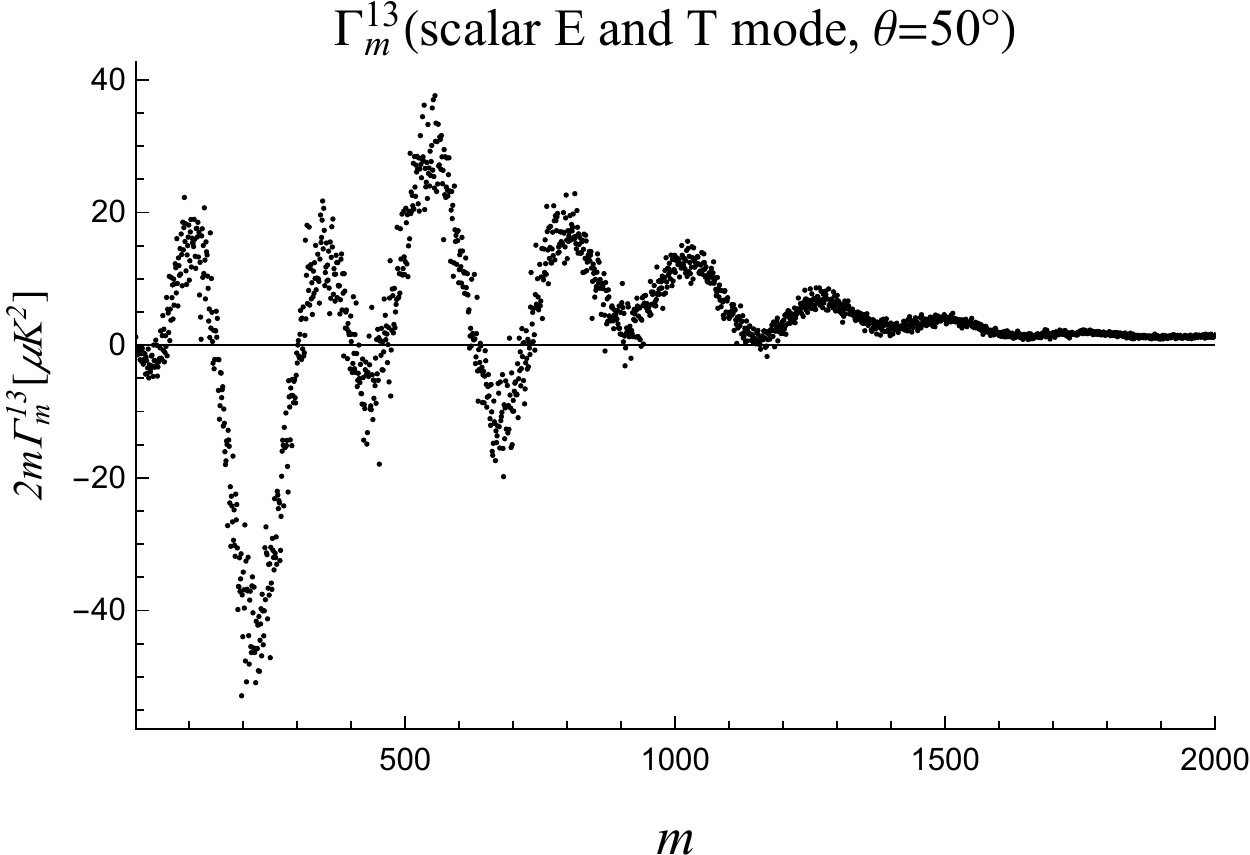}}
	
	\subfigure{\includegraphics[scale=0.37]{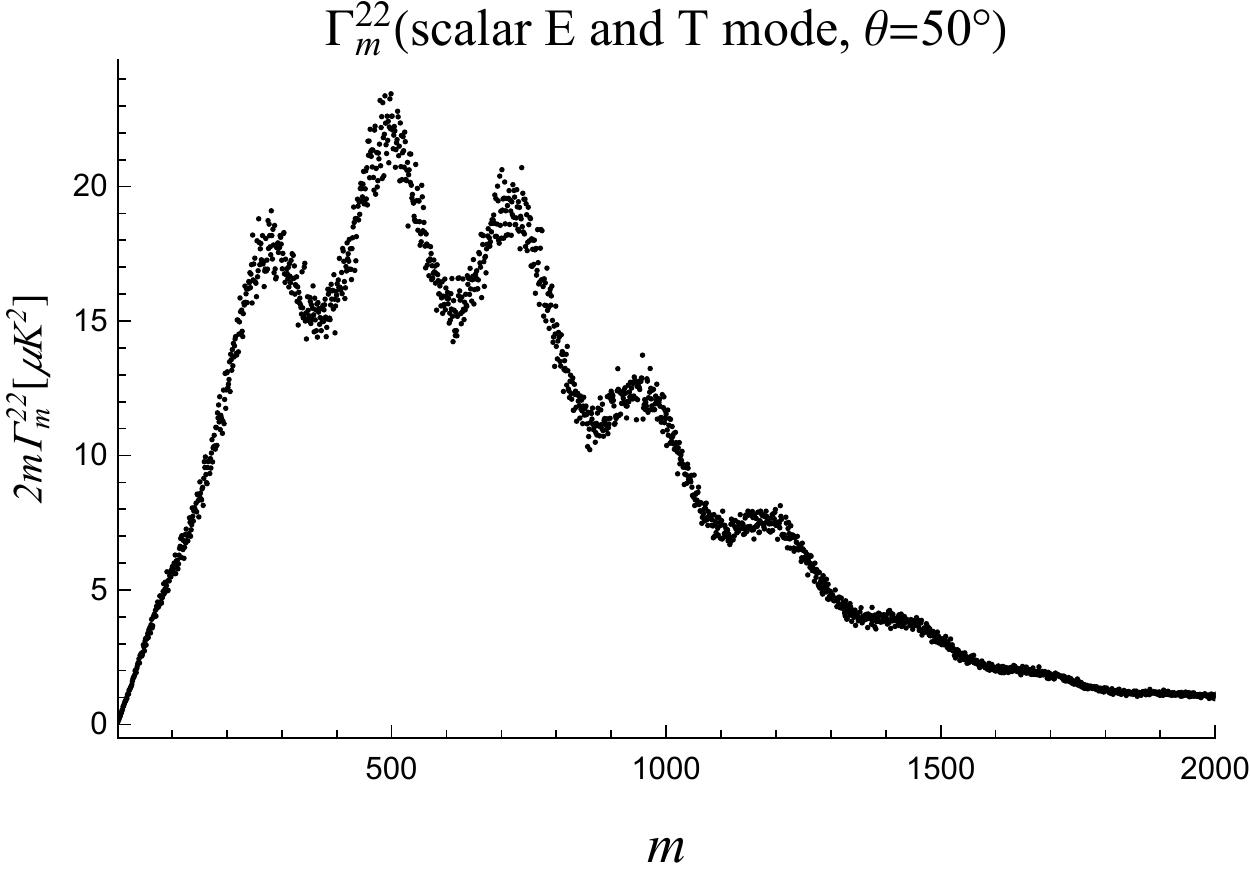}}
	\subfigure{\includegraphics[scale=0.37]{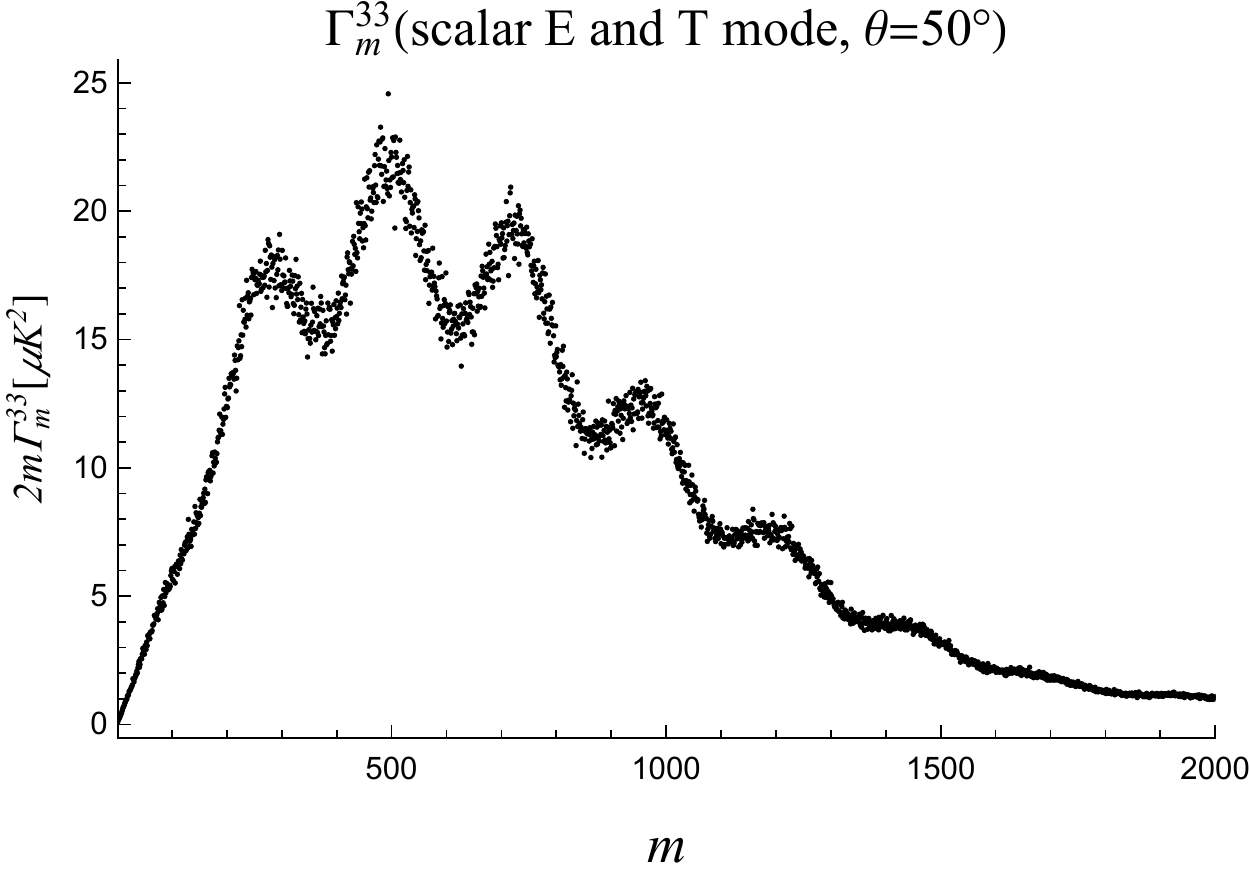}}
	\subfigure{\includegraphics[scale=0.37]{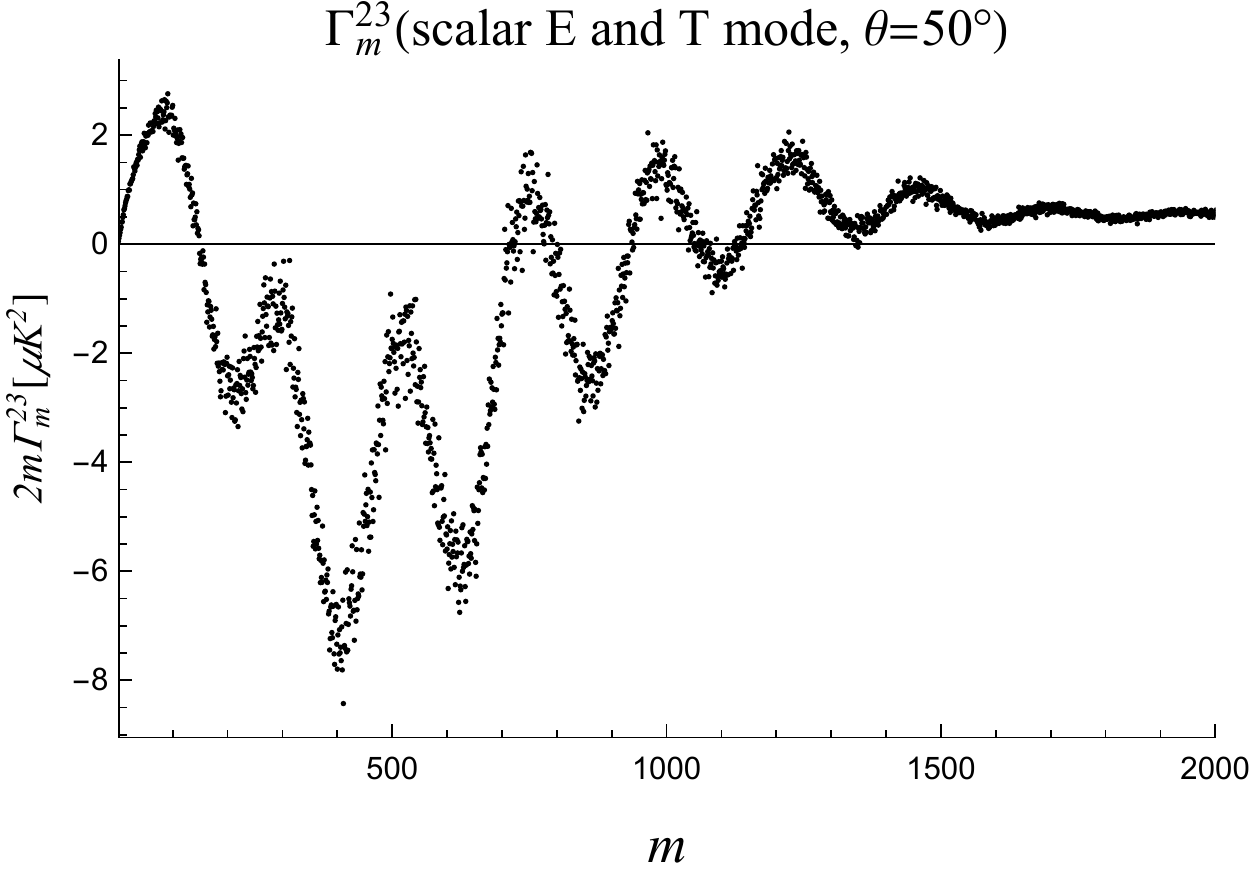}}
	
	\subfigure{\includegraphics[scale=0.37]{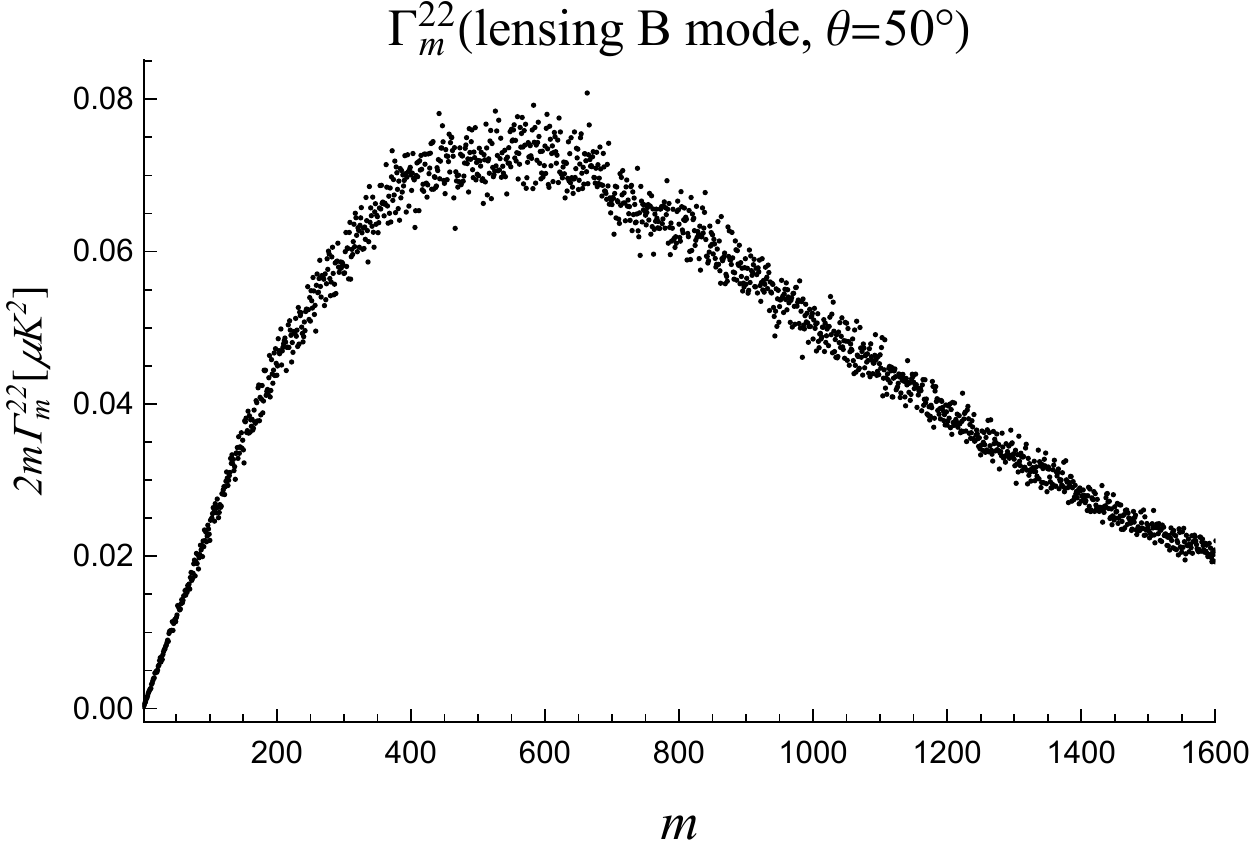}}
	\subfigure{\includegraphics[scale=0.37]{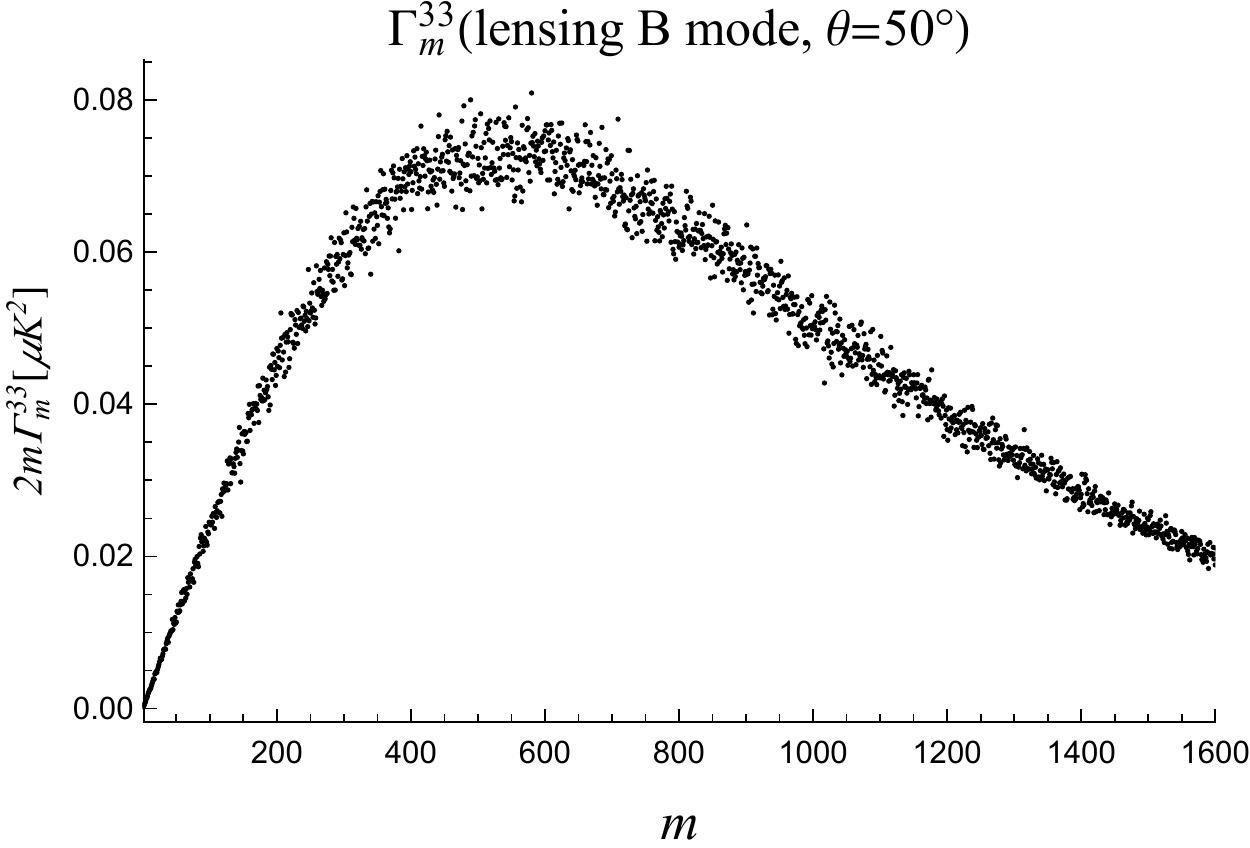}}
	\subfigure{\includegraphics[scale=0.37]{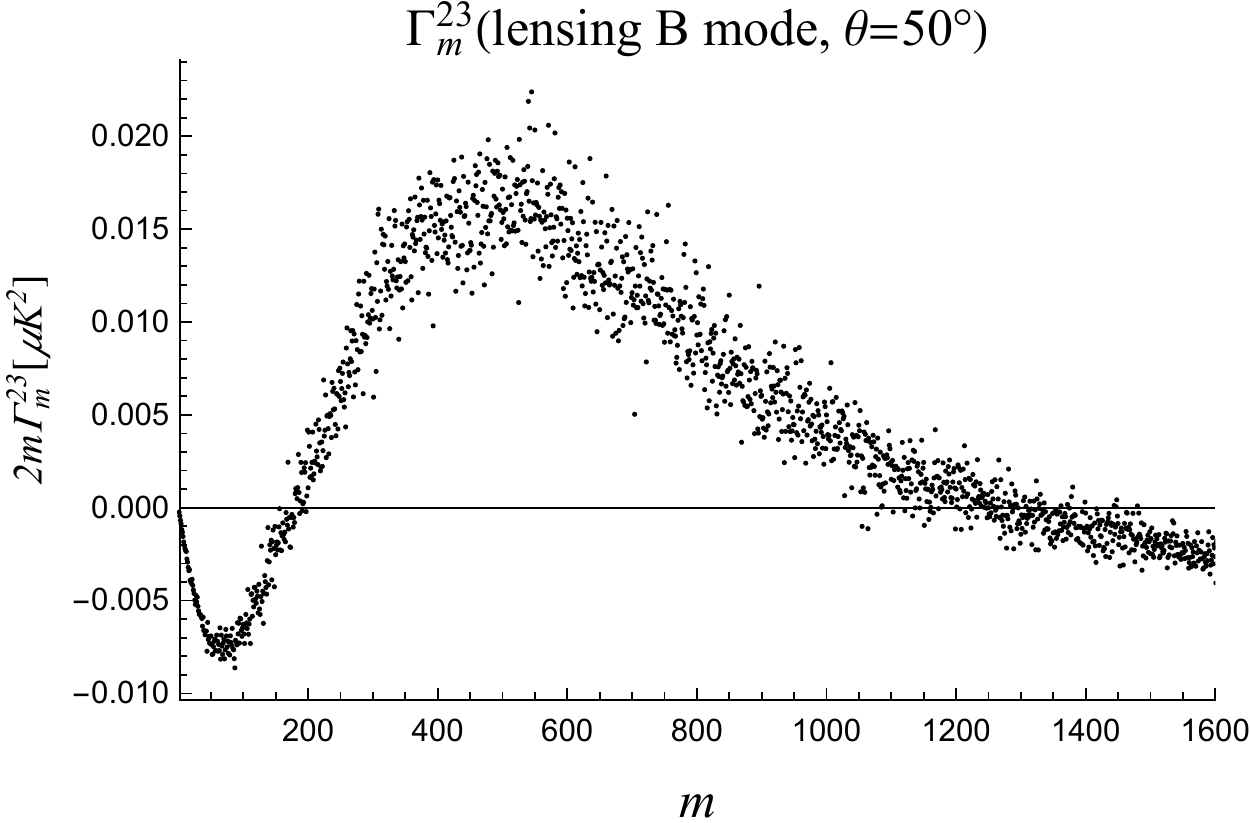}}
	
	\subfigure{\includegraphics[scale=0.37]{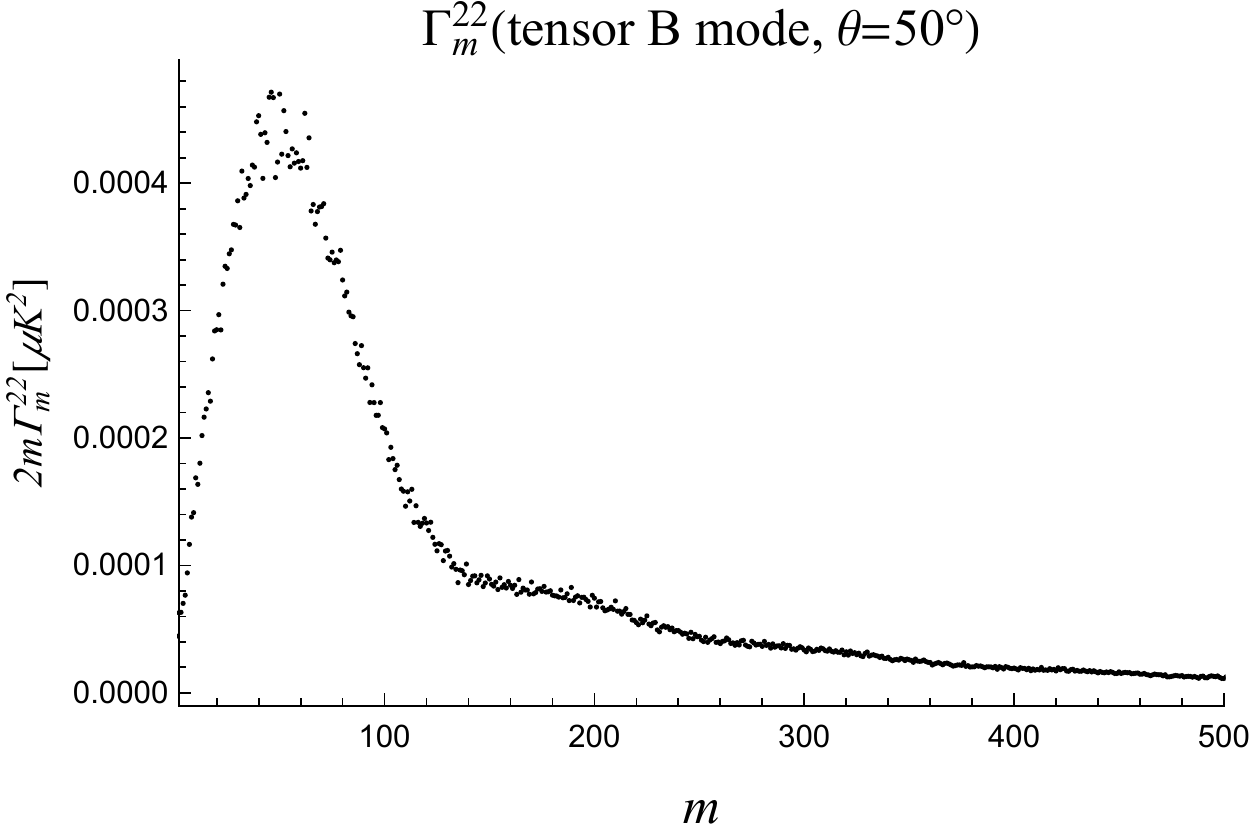}}
	\subfigure{\includegraphics[scale=0.37]{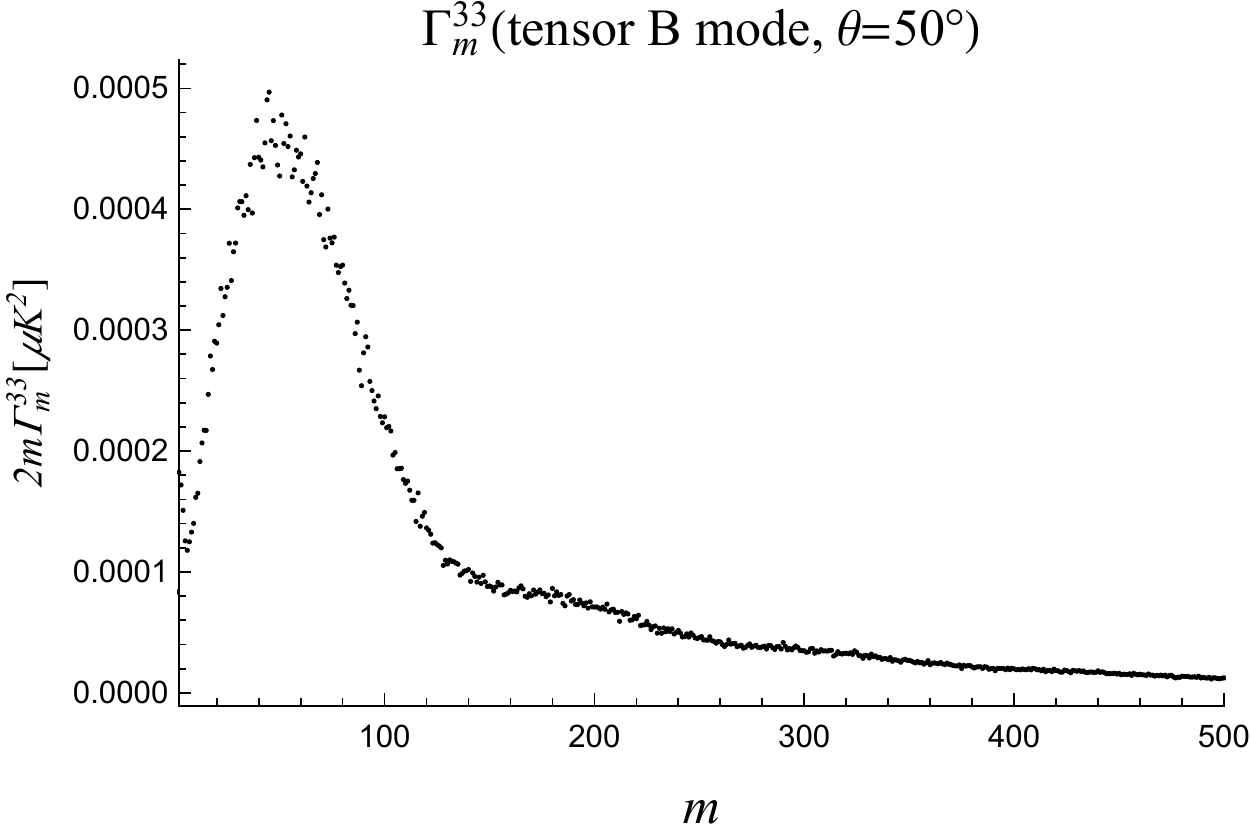}}
	\subfigure{\includegraphics[scale=0.37]{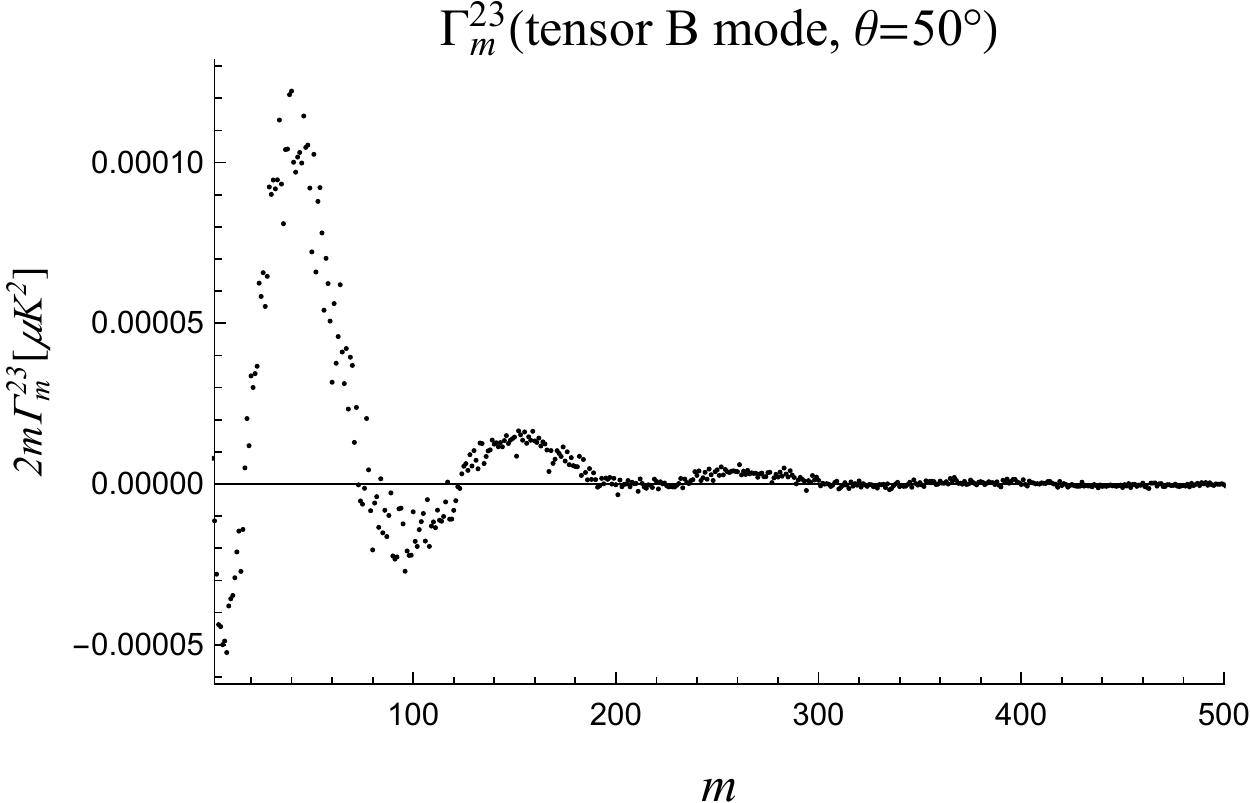}}
	\caption{Ring power spectra of CMB on circular scans from simulated observations. ($\theta =50^\circ$). }
\label{gamma}
\end{figure}

We now present results based on scans of simulated CMB maps. We generate HEALPix CMB maps independently for scalar modes and tensor modes.
Then we ``scan" the simulated sky along circular rings corresponding to a fixed colatitude $\Theta$. For each colatitude, we compute the Fourier Transform of the Stokes parameter $T$, of $Q+\mathrm{i}U$, and of $Q-\mathrm{i}U$ on the rings, and then compute the corresponding spectra $\Gamma_m$. We average $\Gamma_m$ for 700 independent simulations (i.e. 700 independent rings). Fig.~\ref{gamma} illustrates the simulated $\Gamma_m$ after averaging for $\Theta=50^\circ$. After that, we invert the system to calculate the power spectra $C_\ell$ from their ring analogues, $\Gamma_m$.

\begin{figure}[htbp]
	\centering
	\subfigure{\includegraphics[scale=0.37]{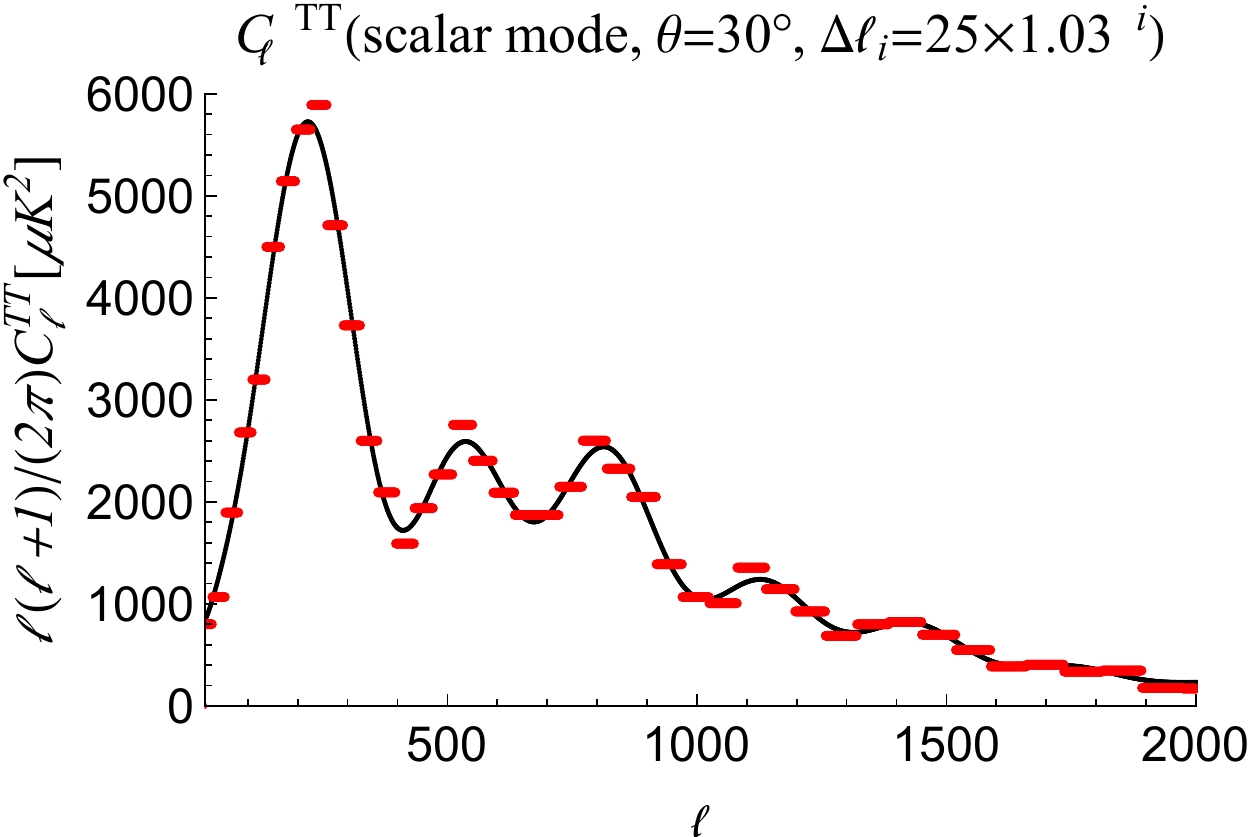}}
	\subfigure{\includegraphics[scale=0.37]{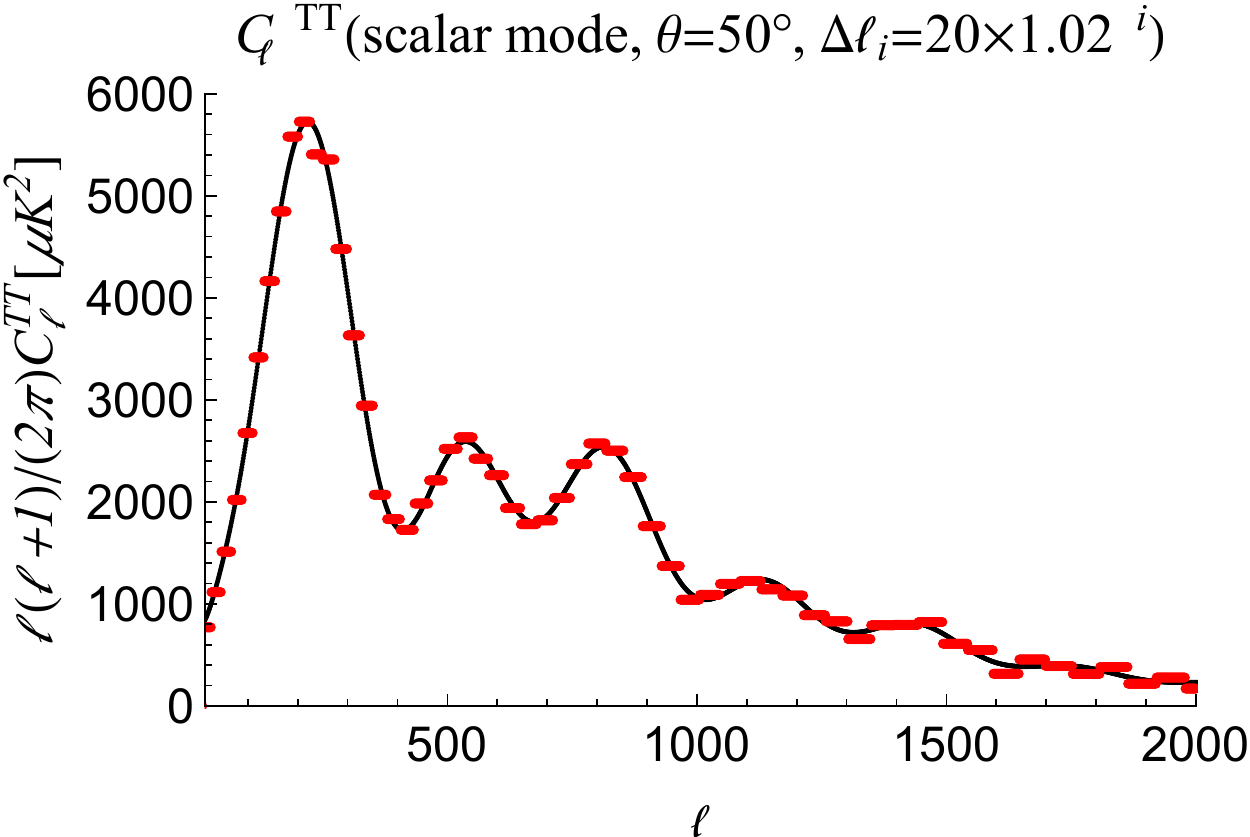}}
	\subfigure{\includegraphics[scale=0.37]{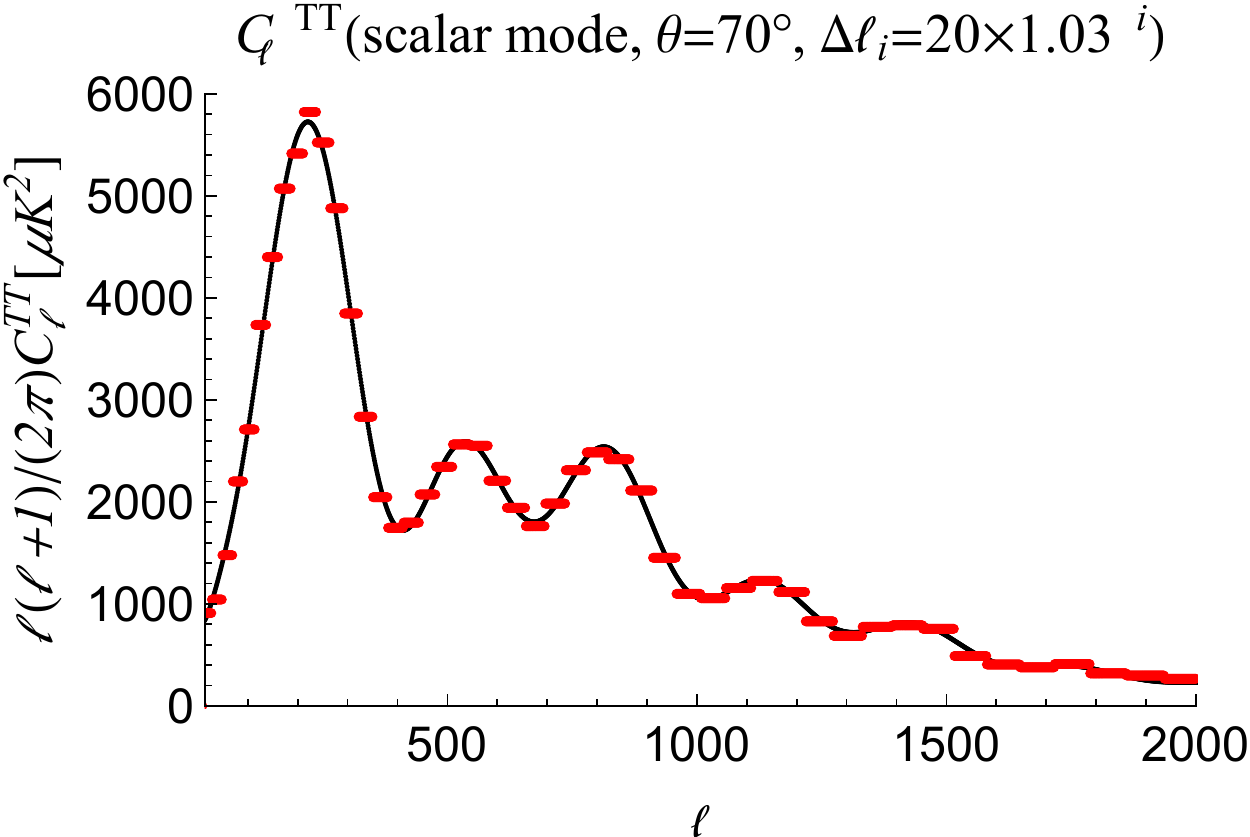}}
	\subfigure{\includegraphics[scale=0.37]{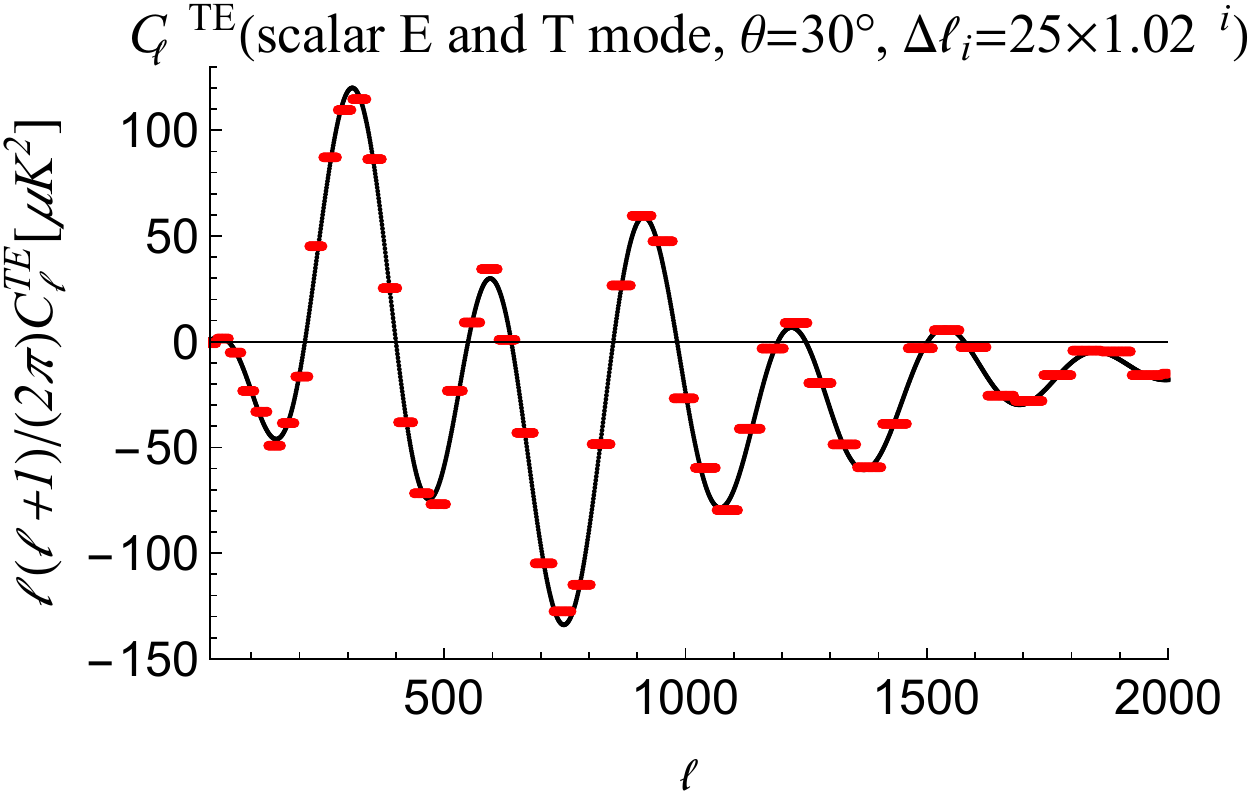}}
	\subfigure{\includegraphics[scale=0.37]{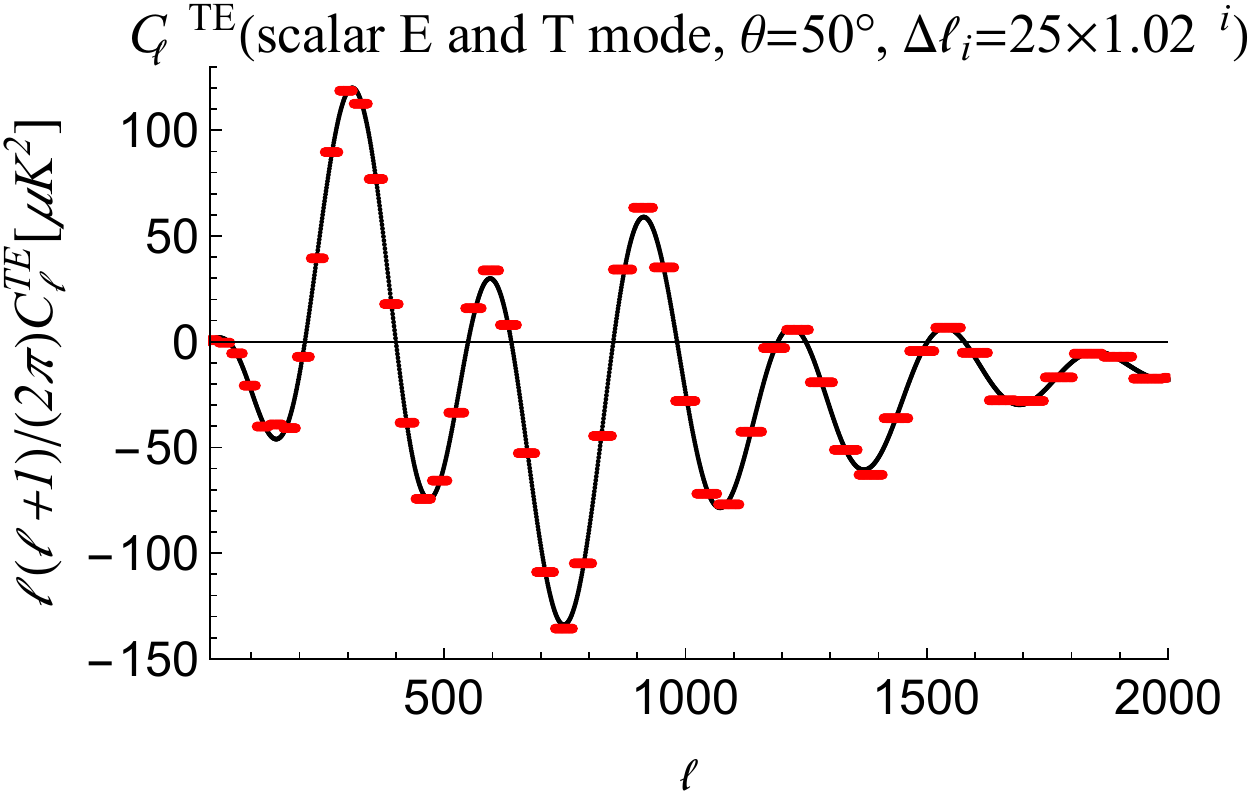}}
	\subfigure{\includegraphics[scale=0.37]{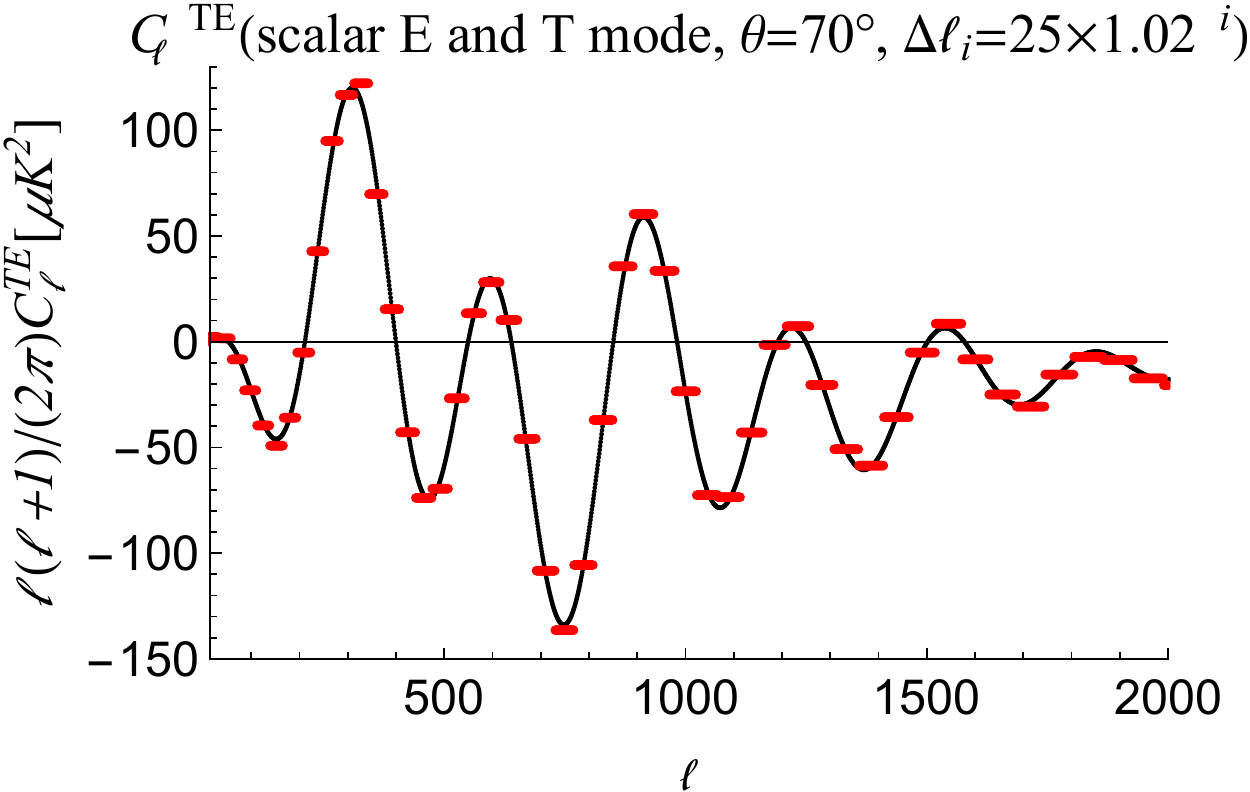}}
	\subfigure{\includegraphics[scale=0.37]{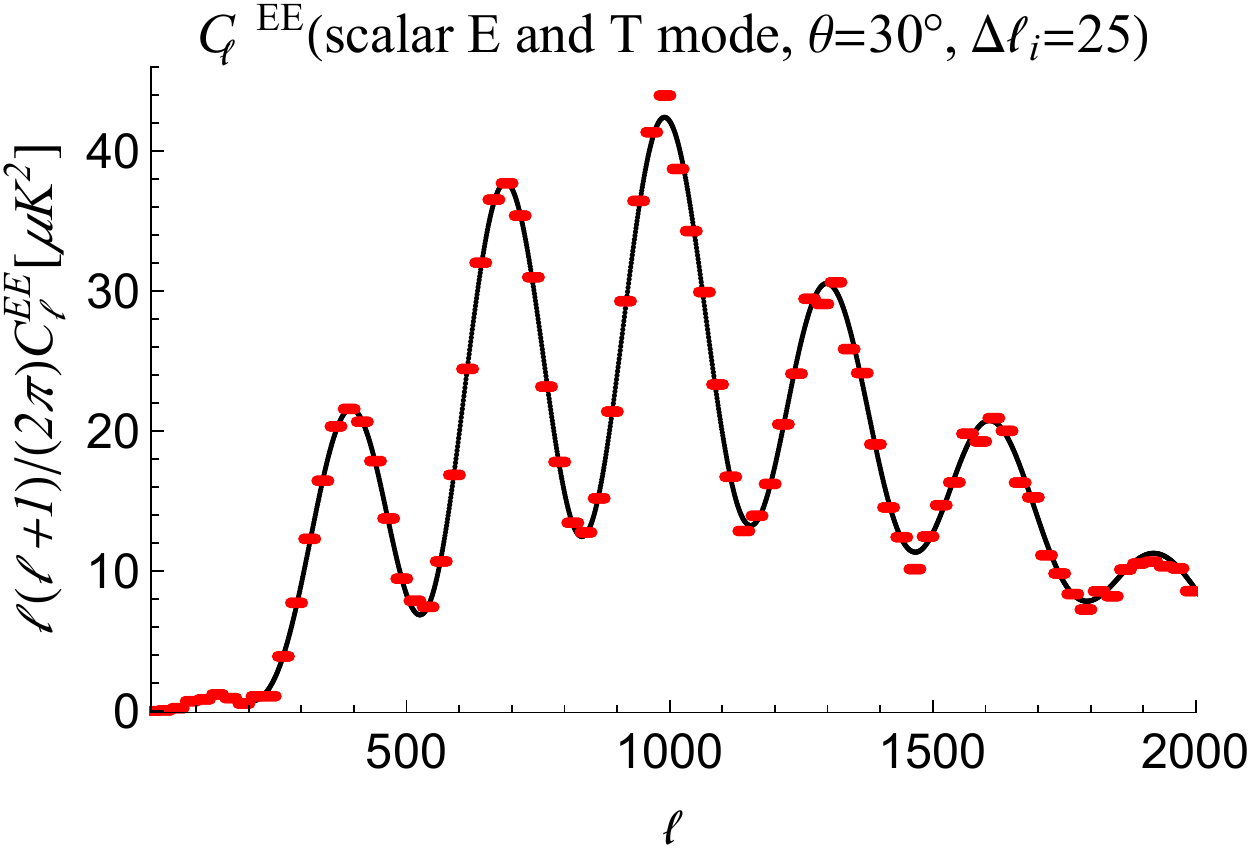}}
	\subfigure{\includegraphics[scale=0.37]{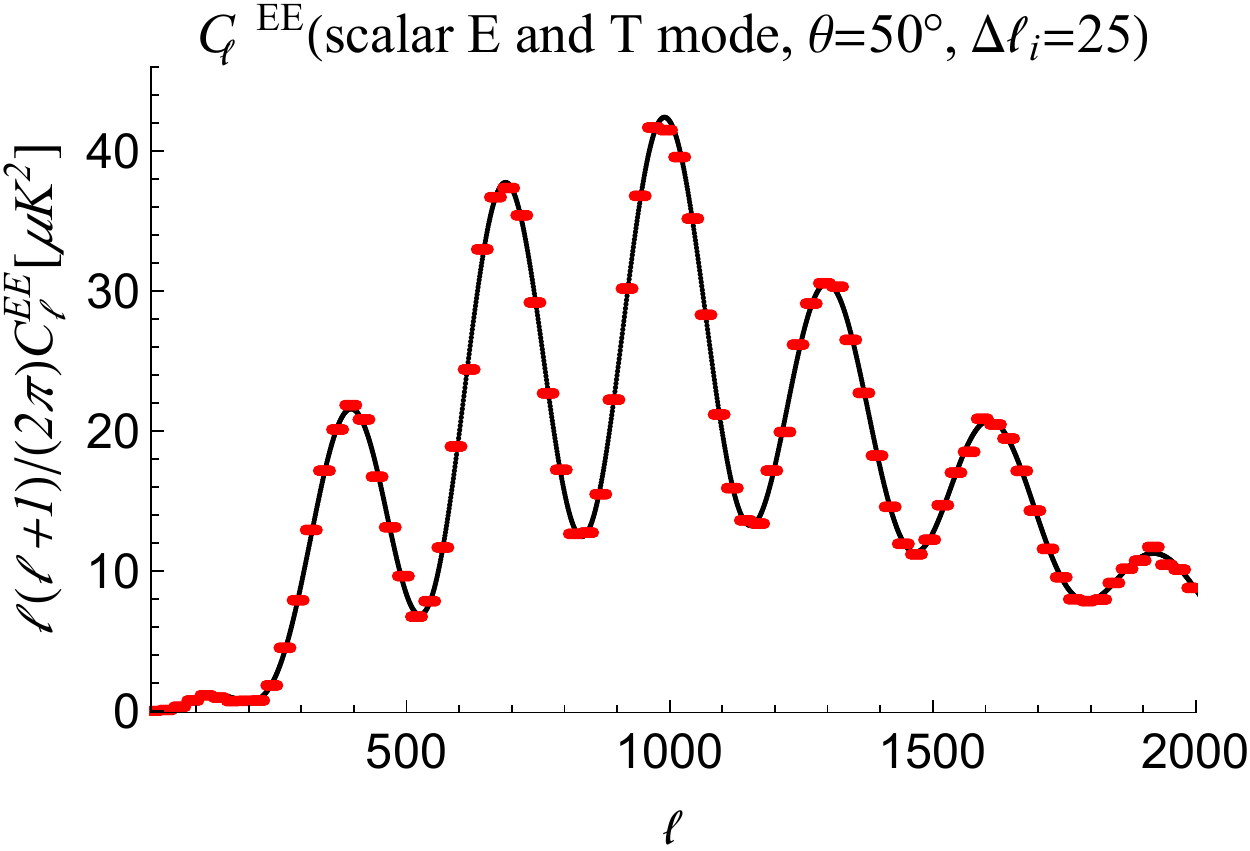}}
	\subfigure{\includegraphics[scale=0.37]{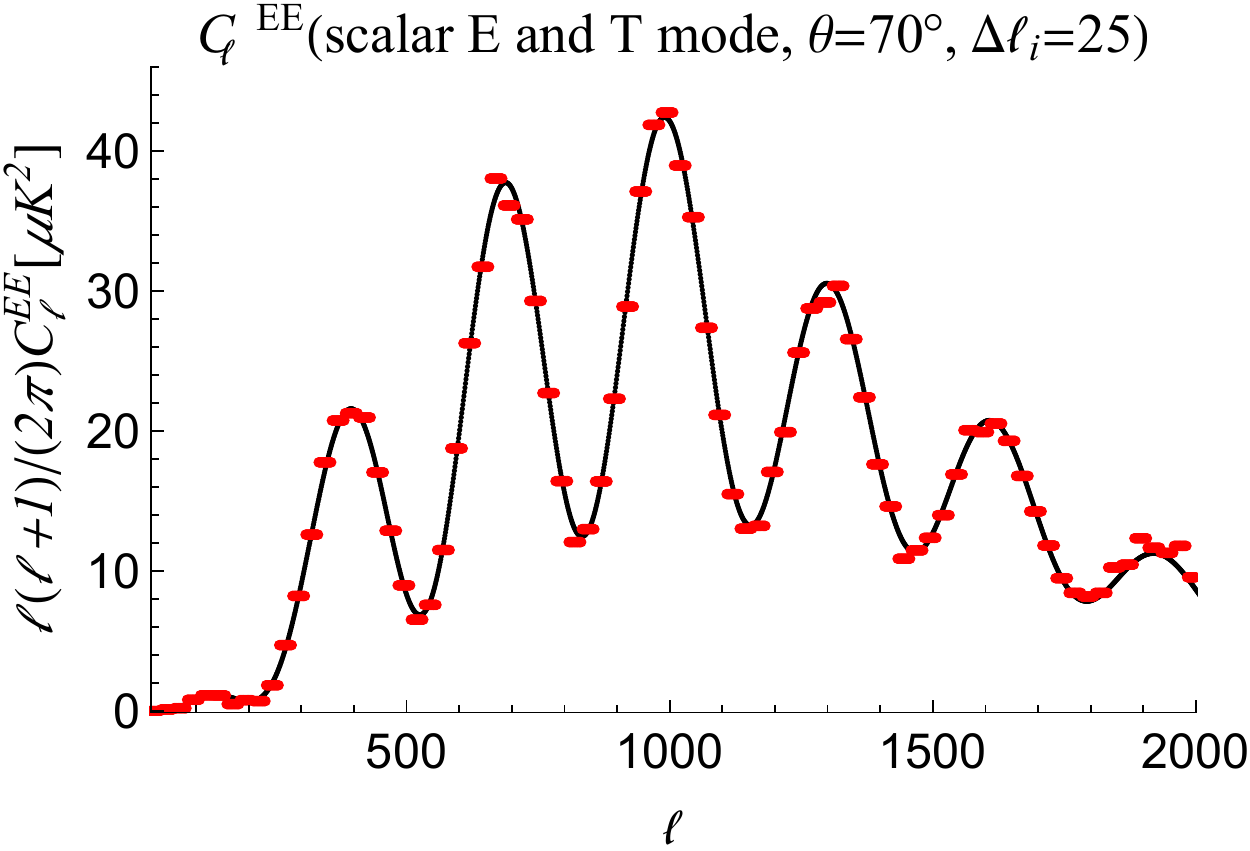}}
	\caption{Inversion of a system with 700 independent simulated rings for $C_\ell^{TT}$, $C_\ell^{TE}$, $C_\ell^{EE}$, scalar modes. Left column: $\theta =30^\circ$; Middle column: $\theta=50^\circ$; Right column: $\theta =70^\circ$.}
\label{results_all_rings_TE}
\end{figure}

\begin{figure}[htbp]
	\centering
	\subfigure{\includegraphics[scale=0.37]{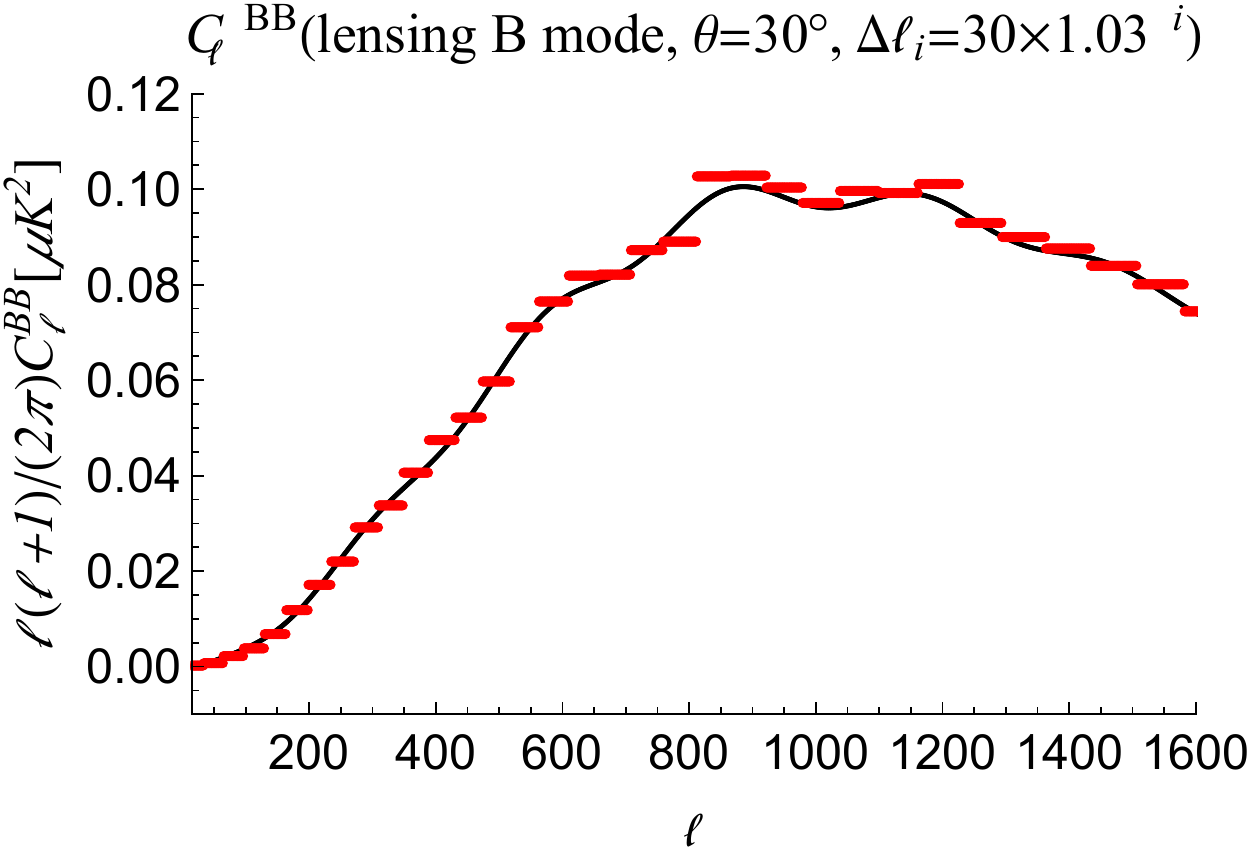}}
	\subfigure{\includegraphics[scale=0.37]{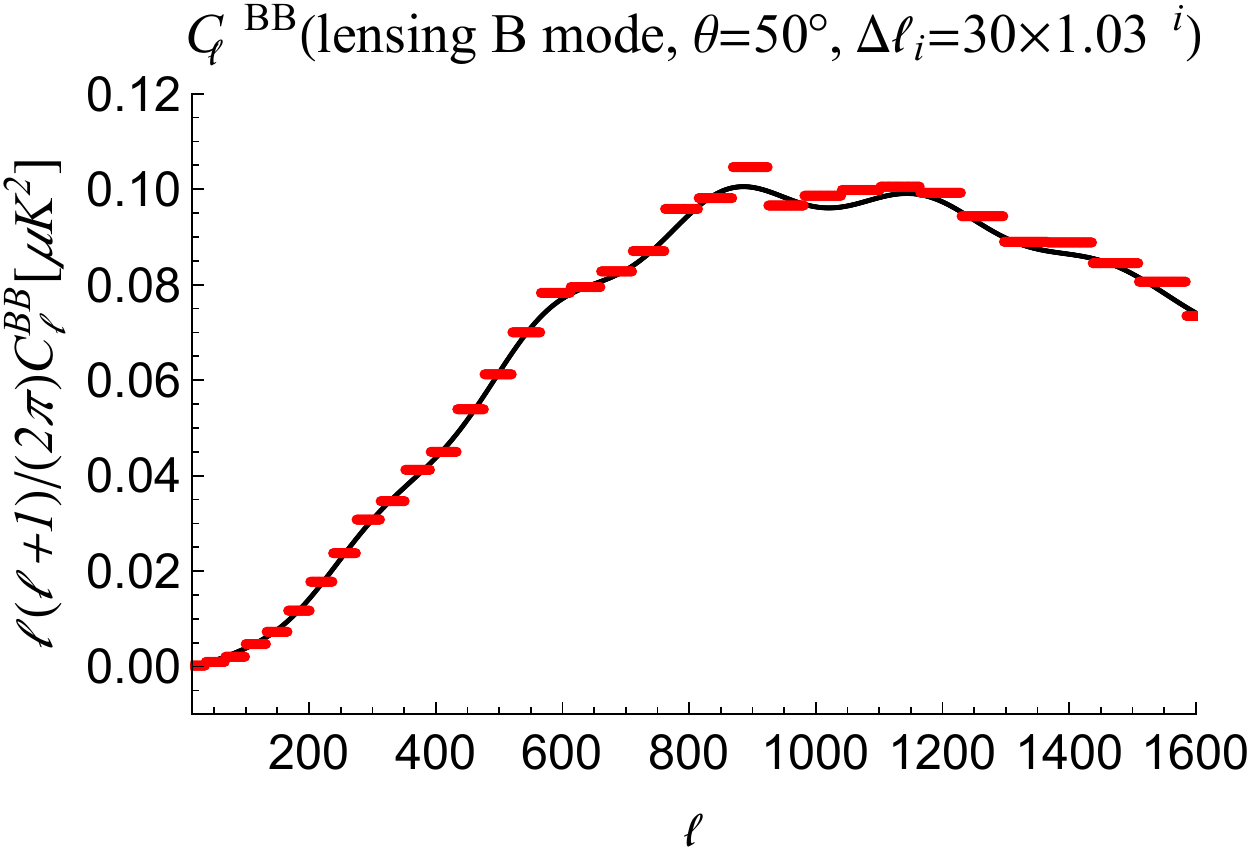}}
	\subfigure{\includegraphics[scale=0.37]{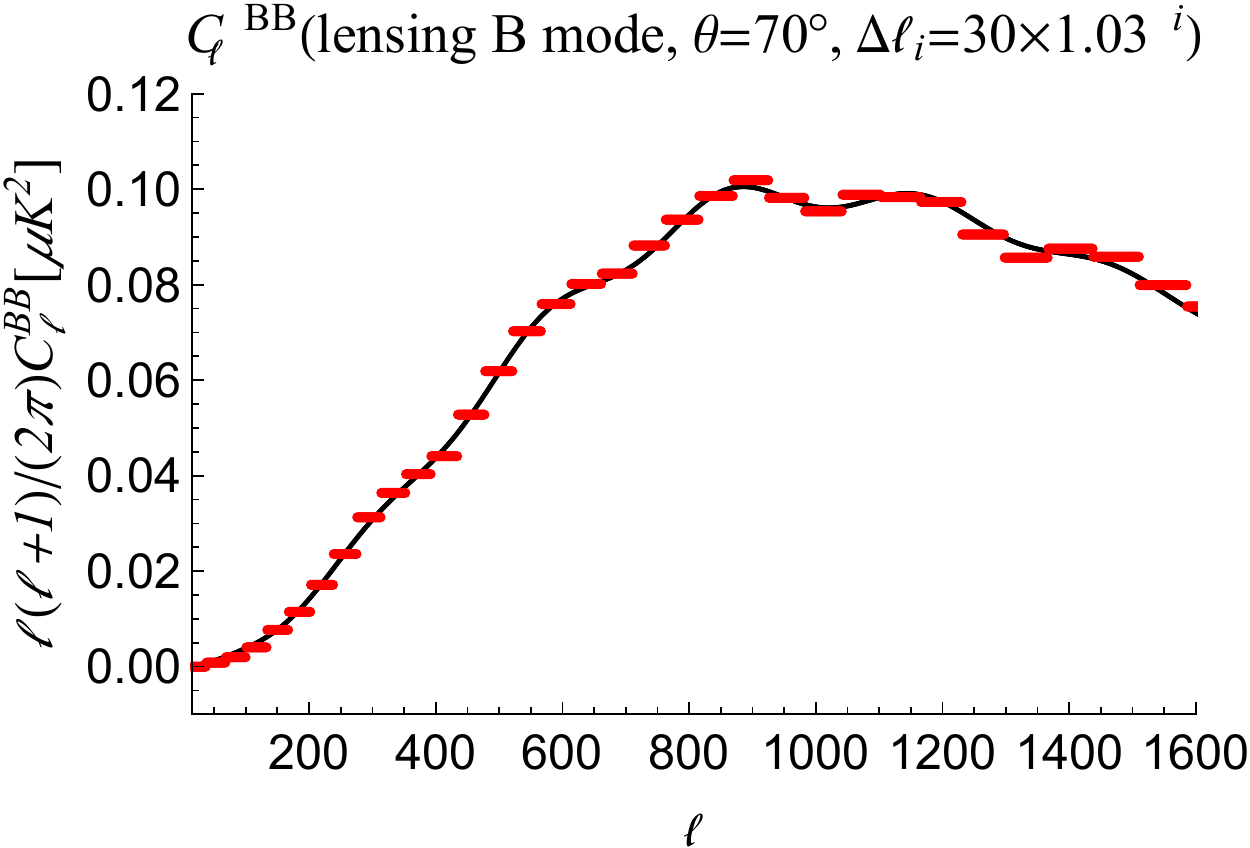}}
	\subfigure{\includegraphics[scale=0.37]{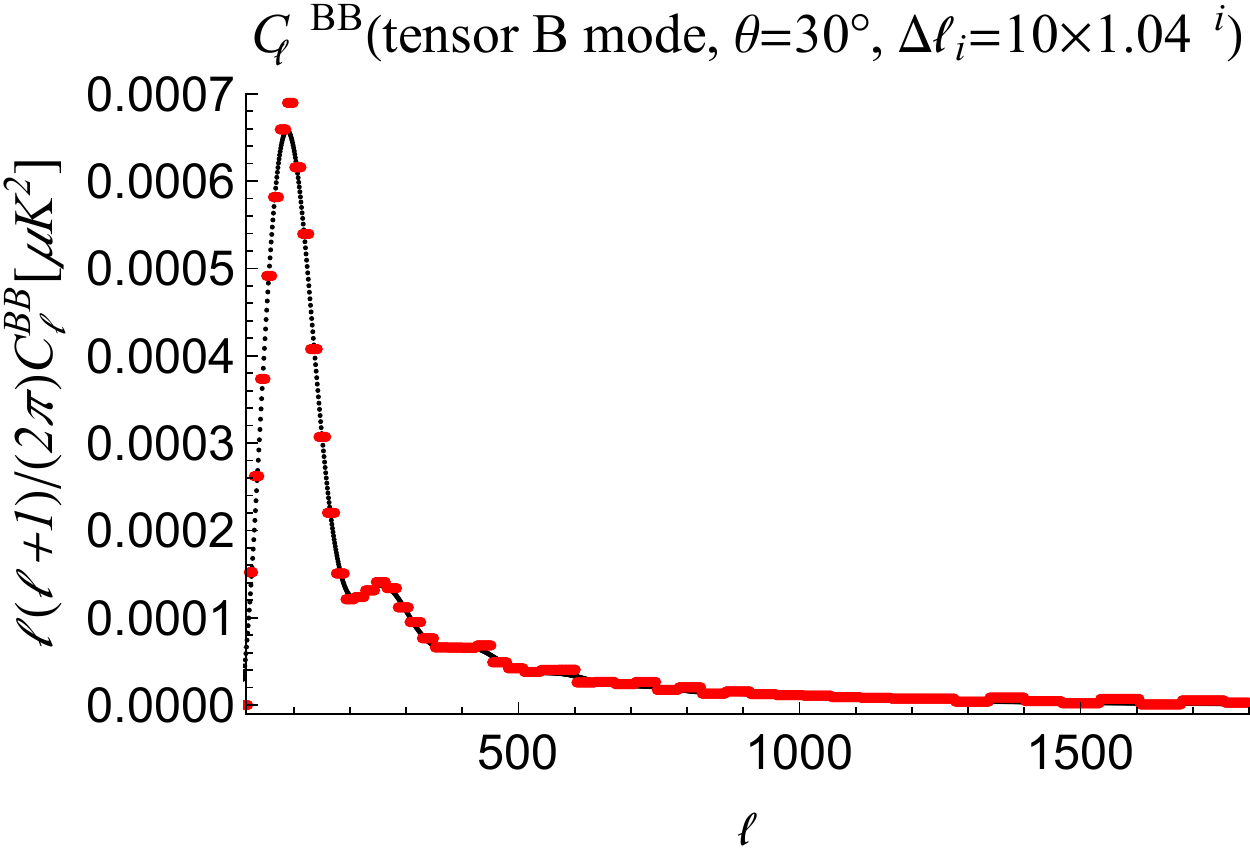}}
	\subfigure{\includegraphics[scale=0.37]{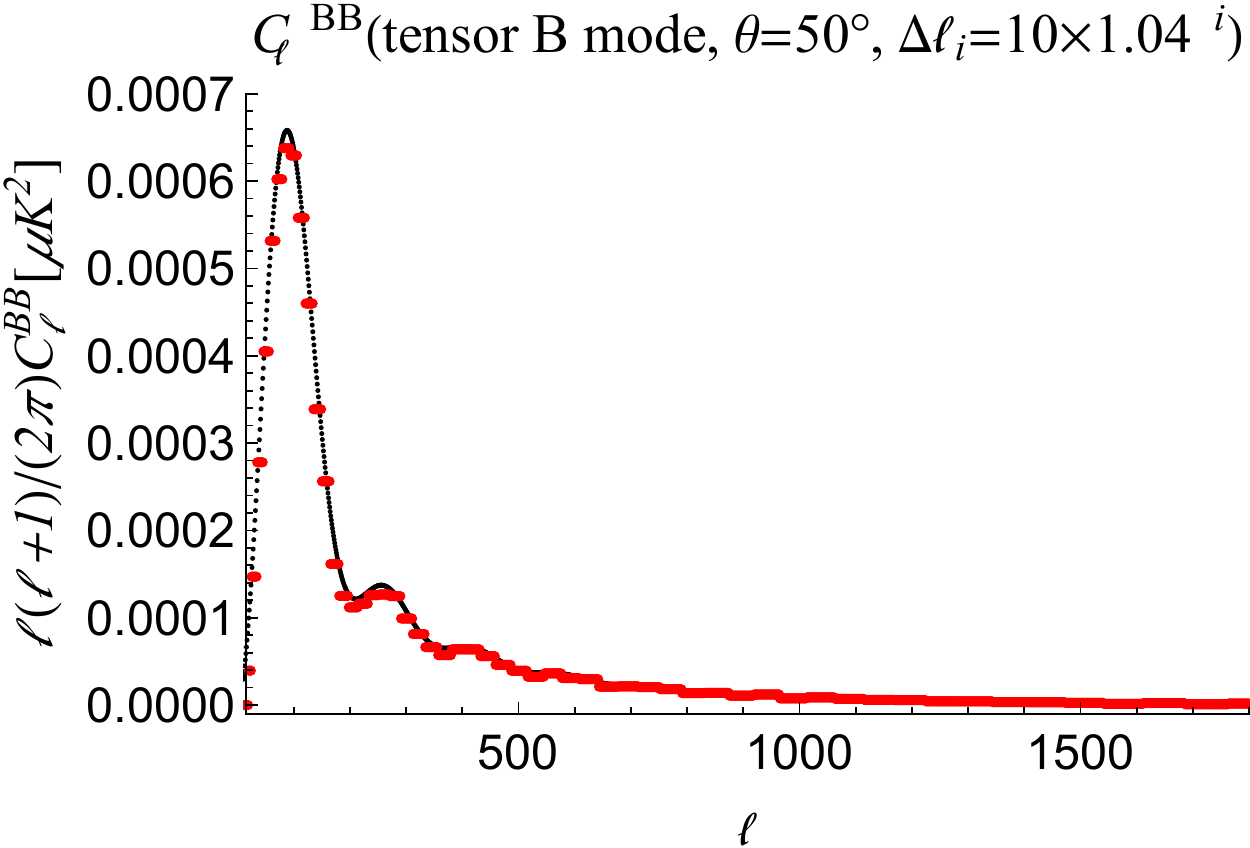}}
	\subfigure{\includegraphics[scale=0.37]{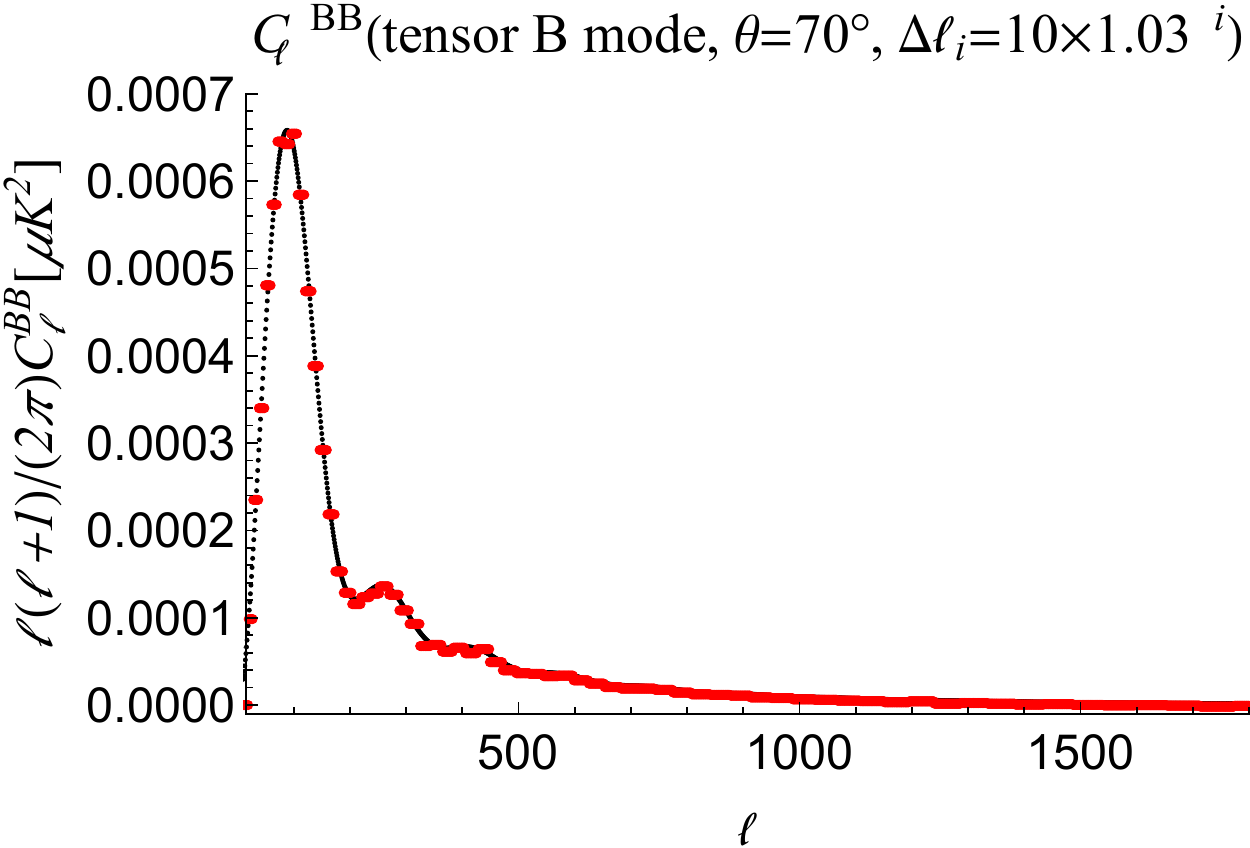}}
	\caption{Inversion of a system with 700 independent simulated rings for $C_\ell^{BB}$, lensing (top) and primordial with $r=0.01$ (bottom). Left column: $\theta =30^\circ$; Middle column: $\theta=50^\circ$; Right column: $\theta =70^\circ$.}
\label{results_all_rings_BB}
\end{figure}

We computed six auto and cross power spectra (black points): $C_\ell^{TT}$, $C_\ell^{TE}$, $C_\ell^{EE}$ (scalar T+E modes only, no B modes); $C_\ell^{BB}$ (scalar T+B modes), and $C_\ell^{TT}$, $C_\ell^{BB}$ (tensor T+B modes), compared with reference value from CLASS (red lines). Examples are shown in Fig.~\ref{results_all_rings_TE} and~\ref{results_all_rings_BB} for three values of the ring opening $\Theta$, ranging from $30^\circ$ to $70^\circ$. All of them show satisfactory estimates of the corresponding $C_\ell$, with (as expected) increasing accuracy for larger rings.

\section{Non-closing rings}

We now discuss  how the inversion of the system can be modified to take into account the fact that in practical observations, the scans can be non-perfectly closing rings.

Indeed, most CMB experiments do not measure the CMB on perfect rings. Instead, while the experiment scans the sky by spinning around a fixed spin axis in the instruments frame, the sky is slowly drifting with respect to this frame. As a consequence, rings do not exactly close.
We consider a scanning strategy such that
\begin{equation}
\Theta(\phi)=\theta+\delta(\phi)
\end{equation}
where $\theta$ is a constant number and $|\delta(\phi)| \ll \theta$, then,
\begin{gather}
\hat \alpha_{1,m}\equiv\frac{1}{2\pi}\int_0^{2\pi}d\phi T(\theta+\delta(\phi),\phi)e^{-im\phi}\\
\hat \alpha_{2,m}\equiv\frac{1}{2\pi}\int_0^{2\pi}d\phi(Q+iU)(\theta+\delta(\phi),\phi)e^{-im\phi}\\
\hat \alpha_{3,m}\equiv\frac{1}{2\pi}\int_0^{2\pi}d\phi(Q-iU)(\theta+\delta(\phi),\phi)e^{-im\phi}
\end{gather}
As shown in Appendix \ref{app:math}, to first order approximation the ring power spectra are given by:
\begin{equation}
\hat\Gamma_m^{jk}=\sum_{\ell=|m|}^{+\infty}K^\ell_m(\theta,j,k,X,Y)\cdot C^{XY}_\ell+\sum_{\ell=|m|}^{+\infty}k^\ell_m(\theta,\delta,j,k,X,Y)\cdot C^{XY}_\ell,
\end{equation}
which can be recast in matrix form as:
\begin{equation}
	\bm{\hat\Gamma}=\big(\bm{K}+\bm{k}\big) \bm{C}
\end{equation}
and the elements of $\bm{k}$ are first order small quantities. Then the inversion of this system is:
\begin{equation}
    \bm{C}\simeq\big(\bm{K^{-1}}-\bm{K^{-1}}\bm{k}\bm{K^{-1}}\big)\bm{\hat\Gamma}
\end{equation}
Of course, binning is necessary, too.

\section{Conclusion}

In this paper, we have explored the connection between the CMB temperature and polarization power spectra and their one-dimensional analogues on ring-shaped trajectories on the celestial sphere. We have shown that it is possible to estimate the CMB temperature and polarization power spectra from a set of ring-shaped scans (with no discussion, however, of the propagation of errors at this stage). Given that most future CMB experiments will scan the sky along circular or nearly circular  scans, this connection between harmonic spectra on the sphere and Fourier spectra on rings, for both temperature and polarization data, can be useful for the analysis of next generation CMB experiments, at least as intermediate steps in the analysis.

\acknowledgments
We thank Chang Feng, Yang Liu, Siyu Li, Pierre Zhang and Hao Zhai for useful discussions. This work is supported in part by the NSFC (Nos. 11653002, 11961131007, 1201101448, 11722327, 11421303), by the CAST-YESS (2016QNRC001), by the National Youth Talents Program of China, by the Fundamental Research Funds for Central Universities, by the CSC Innovation Talent Funds, and by the USTC Fellowship for International Cooperation. All numerical calculations were operated on the computer clusters {\it LINDA} \& {\it JUDY} in the particle cosmology group at USTC.

\appendix

\section{Details of theoretical arithmetic}
\label{app:math}
\subsection{From Spherical Harmonic decomposition to Fourier modes}

In this appendix, we display the connection between spherical harmonic modes of the full sky and Fourier modes of circular rings. Firstly, the spherical harmonic decompositions of $T(\theta,\phi)$, $Q+iU(\theta,\phi)$ and $Q-iU(\theta,\phi)$ are

\begin{equation}
T(\theta,\phi)=\sum_{\ell=0}^{+\infty}\sum_{m=-\ell}^{+\ell}a_{T,\ell m}Y_{\ell m}(\theta,\phi)=\sum_{m=-\infty}^{+\infty}\sum_{\ell=|m|}^{+\infty}a_{T,\ell m}Y_{\ell m}(\theta,\phi)
\end{equation}

\begin{equation}
(Q+iU)(\theta,\phi)=\sum_{\ell=2}^{+\infty}\sum_{m=-\ell}^{+\ell}a_{2,\ell m}\ \prescript{}{2}{Y}_{\ell m}(\theta,\phi)=\sum_{m=-\infty}^{+\infty}\sum_{\substack{\ell=\\min\{2,|m|\}}}^{+\infty}a_{2,\ell m}\ \prescript{}{2}{Y}_{\ell m}(\theta,\phi)
\end{equation}

\begin{equation}
(Q-iU)(\theta,\phi)=\sum_{\ell=2}^{+\infty}\sum_{m=-\ell}^{+l}a_{-2,\ell m}\ \prescript{}{-2}{Y}_{\ell m}(\theta,\phi)=\sum_{m=-\infty}^{+\infty}\sum_{\substack{\ell=\\min\{2,|m|\}}}^{+\infty}a_{-2,\ell m}\ \prescript{}{-2}{Y}_{\ell m}(\theta,\phi)
\end{equation}
According to the definition of the spin weighted spherical harmonics \cite{1967JMP.....8.2155G}:

\begin{align}
\prescript{}{s}{Y}_{\ell m}=e^{im\phi}\sqrt{\frac{(\ell+m)!(\ell-m)!}{(\ell+s)!(\ell-s)!}\frac{2\ell +1}{4\pi}}&\sin^{2\ell}\frac{\theta}{2}\nonumber\\
\times\sum_r\begin{pmatrix}\ell-s\\r\end{pmatrix}&\begin{pmatrix}\ell+s\\r+s-m\end{pmatrix}(-1)^{\ell-r-s+m}\cot^{2r+s-m}\frac{\theta}{2}
\label{expression of s_Y function}
\end{align}
%
%
and given the definition of $\mathcal{P}^k_{\ell m}$, we can rewrite the Fourier coefficients $\alpha_{k,m}$ for $(k=1,2,3)$ as:
\begin{align}
\alpha_{k,m}&=\frac{1}{2\pi}\int_0^{2\pi}d\phi\sum_{m'=-\infty}^{+\infty}\sum_{\substack{\ell=\\ min\{2,|m'|\}}}^{+\infty}a_{k,\ell m'}\Big(e^{im'\phi}\mathcal P^k_{\ell m'}(\theta)\Big)e^{-im\phi}\nonumber\\
&=\frac{1}{2\pi}\int_0^{2\pi}d\phi\sum_{m'=-\infty}^{+\infty}\sum_{\substack{\ell=\\ min\{2,|m'|\}}}^{+\infty}a_{k,\ell m'}\mathcal P^k_{\ell m'}e^{i(m'-m)\phi}\nonumber\\
&=\sum_{m'=-\infty}^{+\infty}\sum_{\substack{\ell=\\ min\{2,|m'|\}}}^{+\infty}a_{k,\ell m'}\mathcal P^k_{\ell m'}\delta_{m'm}\nonumber\\
&=\sum_{\ell=|m|}^{+\infty}a_{k,\ell m}\mathcal P_{\ell m}^{k}.
\label{alpha and a}
\end{align}
For convenience, we just consider the $\alpha_{k,m}$ for which $|m|\geqslant2$.

\subsection{Computation of \texorpdfstring{$\Gamma_m$}{} as a function of \texorpdfstring{$C_\ell$}{} }
\label{section: from C to Gamma}

By using Eq.~\eqref{alpha and a}, we connect the ring power spectrum $\Gamma_m$ to the full-sky power spectrum $C_\ell$. The $E$ mode and $B$ mode that describe the polarisation of the CMB is defined by the modes of $(Q+iU)$ and $(Q-iU)$ according to \cite{1997PhRvD..55.1830Z},
\begin{equation}
\left\{
\begin{aligned}
a_{2,\ell m}=-a_{E,\ell m}-ia_{B,\ell m}\\
a_{-2,\ell m}=-a_{E,\ell m}+ia_{B,\ell m}.
\end{aligned}
\right.
\end{equation}
The expressions of full-sky power spectrum are
\begin{equation}
\left\{
\begin{aligned}
\big\langle a_{T,\ell' m'}^*a_{T,\ell m}\big\rangle=& C_\ell^{TT}\delta_{\ell'\ell}\delta_{m'm}\\
\big\langle a_{E,\ell' m'}^*a_{E,\ell m}\big\rangle=& C_\ell^{EE}\delta_{\ell'\ell}\delta_{m'm}\\
\big\langle a_{B,\ell' m'}^*a_{B,\ell m}\big\rangle=& C_\ell^{BB}\delta_{\ell'\ell}\delta_{m'm}\\
\big\langle a_{T,\ell' m'}^*a_{E,\ell m}\big\rangle=&\big\langle a_{E,\ell' m'}^*a_{T,\ell m}\big\rangle= C_\ell^{TE}\delta_{\ell'\ell}\delta_{m'm}\\
\big\langle a_{T,\ell' m'}^*a_{B,\ell m}\big\rangle=&\big\langle a_{B,\ell' m'}^*a_{T,\ell m}\big\rangle= C_\ell^{TB}\delta_{\ell'\ell}\delta_{m'm}\\
\big\langle a_{E,\ell' m'}^*a_{B,\ell m}\big\rangle=&\big\langle a_{B,\ell' m'}^*a_{E,\ell m}\big\rangle= C_\ell^{EB}\delta_{\ell'\ell}\delta_{m'm}.
\end{aligned}
\right.\footnote{$C_\ell^{XY}=C_\ell^{YX}$, for the coefficients $C_\ell^{XY}$ are all real. }
\end{equation}
Then from Eq.~\eqref{alpha and a}, we express the ring power spectrum $\Gamma_m^{ik}$ in terms of the full-sky power spectrum $C_{\ell}^{XY}$:

\begin{align}
\big\langle\alpha_{j,m'}^*\alpha_{k,m}\big\rangle&=\Big\langle\sum_{\ell'=|m'|}^{+\infty}a_{j,\ell' m'}^*\mathcal P_{\ell' m'}^{j*}\quad\sum_{\ell=|m|}^{+\infty}a_{k,\ell m}\mathcal P_{\ell m}^k\Big\rangle\nonumber\\
&=\sum_{\ell'=|m'|}^{+\infty}\sum_{\ell=|m|}^{+\infty}\mathcal P_{\ell' m'}^j\mathcal P_{\ell m}^k\big\langle a_{j,\ell' m'}^*a_{k,\ell m}\big\rangle\nonumber\\
&=\sum_{\ell'=|m'|}^{+\infty}\sum_{\ell=|m|}^{+\infty}\mathcal P_{\ell' m'}^j\mathcal P_{\ell m}^k\sum_{X,Y}A^{XY}_{jk}C_\ell^{XY}\delta_{\ell'\ell}\delta_{m'm}\nonumber\\
&=\sum_{\ell=|m|}^{+\infty}\mathcal P_{\ell m}^j\mathcal P_{\ell m}^k\sum_{X,Y}A^{XY}_{jk}C_\ell^{XY}\delta_{m'm}\nonumber\\
&=\sum_\ell K^\ell_m(\theta,j,k,X,Y)C_\ell^{XY}\delta_{m'm}\nonumber\\
&=\Gamma_m^{jk}\delta_{m'm}
\label{main system}
\end{align}
where the elements of the matrix $A^{XY}_{ik}$ are given by the following table, equivalent to Eq.~\eqref{equation: relation between Gamma and C}:
\\
\renewcommand\arraystretch{1.5}
\begin{longtable}[c]{|c|c|c|c|c|c|c|c|c|c|}
\hline
$A_{jk}^{XY}$&$TT$&$TE$&$ET$&$TB$&$BT$&$EE$&$BB$&$EB$&$BE$\\
\hline
$jk=11$&$1$&$0$&$0$&$0$&$0$&$0$&$0$&$0$&$0$\\
\hline
$jk=12$&$0$&$-1$&$0$&$-i$&$0$&$0$&$0$&$0$&$0$\\
\hline
$jk=21$&$0$&$-1$&$0$&$+i$&$0$&$0$&$0$&$0$&$0$\\
\hline
$jk=13$&$0$&$-1$&$0$&$+i$&$0$&$0$&$0$&$0$&$0$\\
\hline
$jk=31$&$0$&$-1$&$0$&$-i$&$0$&$0$&$0$&$0$&$0$\\
\hline
$jk=22$&$0$&$0$&$0$&$0$&$0$&$1$&$1$&$0$&$0$\\
\hline
$jk=23$&$0$&$0$&$0$&$0$&$0$&$1$&$-1$&$-2i$&$0$\\
\hline
$jk=32$&$0$&$0$&$0$&$0$&$0$&$1$&$-1$&$+2i$&$0$\\
\hline
$jk=33$&$0$&$0$&$0$&$0$&$0$&$1$&$1$&$0$&$0$\\
\hline
\end{longtable}
%
%
Noting that according to the definition of $\prescript{}{s}Y_{\ell m}$ (\ref{expression of s_Y function}), $s$ and $\ell$ satisfy $\ell \geq |s|$, 
for polarization, we restrict ourselves to $l\geqslant |m|\geqslant2$.

\subsection{Getting \texorpdfstring{$C_\ell$}{} from \texorpdfstring{$\Gamma_m$}{}}


Fig.~\ref{plot_matrix_P} shows two examples of matrix $\bm{P_{jk}}$. Although $\bm{P_{jk}}$ are upper triangular matrices theoretically, the elements of $\bm{P_{jk}}$, i.e. $\mathcal{P}^j_{\ell m}\mathcal{P}^k_{\ell m}$ are extremely close to zero for $m$ larger than $M$. This is another way to explain why we choose $M\simeq L\sin\theta$. As a result, this leads to the singularity of matrix $\bm{K}$.



\subsubsection{Temperature power spectrum}
The analogue of $C^{TT}_\ell$ on the scanning ring is $\Gamma^{11}_m$:
\begin{equation}
    \bm{\Gamma^{11}}=\bm{P_{11}}\bm{C^{TT}}
\end{equation}
Then,
\begin{equation}
    \bm{\hat{C}^{TT}}=\bm{B}\Big[\big(\bm{P_{11}}\bm{B}\big)^T\big(\bm{P_{11}}\bm{B}\big)\Big]^{-1}\big(\bm{P_{11}}\bm{B}\big)^T\;\bm{\Gamma^{11}}
\end{equation}

\subsubsection{cross-power spectrum between temperature and E-mode polarization}
$C^{TE}_\ell$ is `projected' to the real part of $\Gamma^{12}_m$ and $\Gamma^{12}_m$ on the ring:
\begin{equation}
    \begin{pmatrix}
    Re(\bm{\Gamma^{12}})\\Re(\bm{\Gamma^{13}})
    \end{pmatrix}= -
    \begin{pmatrix}
    \bm{P_{12}}\\\bm{P_{13}}
    \end{pmatrix}\bm{C^{TE}}
\end{equation}

\begin{align}
    \bm{\hat{C}^{TE}}=& - \bm{B}\Big[\big(\bm{P_{12}}\bm{B}\big)^T\big(\bm{P_{12}}\bm{B}\big) + \big(\bm{P_{13}}\bm{B}\big)^T\big(\bm{P_{13}}\bm{B}\big)\Big]^{-1}\big(\bm{P_{12}}\bm{B}\big)^T\;Re(\bm{\Gamma^{12}})\nonumber\\
    & - \bm{B}\Big[\big(\bm{P_{12}}\bm{B}\big)^T\big(\bm{P_{12}}\bm{B}\big) + \big(\bm{P_{13}}\bm{B}\big)^T\big(\bm{P_{13}}\bm{B}\big)\Big]^{-1}\big(\bm{P_{13}}\bm{B}\big)^T\;Re(\bm{\Gamma^{13}})
\end{align}
\subsubsection{auto-power spectra of polarization}
\begin{equation}
    \begin{pmatrix}
    Re(\bm{\Gamma^{22}})\\Re(\bm{\Gamma^{33}})\\Re(\bm{\Gamma^{23}})
    \end{pmatrix}=
    \begin{pmatrix}
    \bm{P_{22}}&\bm{P_{22}}\\\bm{P_{33}}&\bm{P_{33}}\\\bm{P_{23}}& - \bm{P_{23}}
    \end{pmatrix}
    \begin{pmatrix}
    \bm{C^{EE}}\\\bm{C^{BB}}
    \end{pmatrix}
\end{equation}
Inverting this system, we get:
\begin{align}
    \bm{\hat{C}^{EE}}=&\quad\frac{1}{2}\bm{B}\Big[\big(\bm{P_{22}}\bm{B}\big)^T\big(\bm{P_{22}}\bm{B}\big) + \big(\bm{P_{33}}\bm{B}\big)^T\big(\bm{P_{33}}\bm{B}\big)\Big]^{-1}\big(\bm{P_{22}}\bm{B}\big)^T\;Re(\bm{\Gamma^{22}})\nonumber\\
    &+\frac{1}{2}\bm{B}\Big[\big(\bm{P_{22}}\bm{B}\big)^T\big(\bm{P_{22}}\bm{B}\big) + \big(\bm{P_{33}}\bm{B}\big)^T\big(\bm{P_{33}}\bm{B}\big)\Big]^{-1}\big(\bm{P_{33}}\bm{B}\big)^T\;Re(\bm{\Gamma^{33}})\nonumber\\
    &+\bm{B}\Big[\big(\bm{P_{23}}\bm{B}\big)^T\big(\bm{P_{23}}\bm{B}\big)\Big]^{-1}\big(\bm{P_{23}}\bm{B}\big)^T\;Re(\bm{\Gamma^{23}})
\end{align}

\begin{align}
    \bm{\hat{C}^{BB}}=&\quad\frac{1}{2}\bm{B}\Big[\big(\bm{P_{22}}\bm{B}\big)^T\big(\bm{P_{22}}\bm{B}\big) + \big(\bm{P_{33}}\bm{B}\big)^T\big(\bm{P_{33}}\bm{B}\big)\Big]^{-1}\big(\bm{P_{22}}\bm{B}\big)^T\;Re(\bm{\Gamma^{22}})\nonumber\\
    &+\frac{1}{2}\bm{B}\Big[\big(\bm{P_{22}}\bm{B}\big)^T\big(\bm{P_{22}}\bm{B}\big) + \big(\bm{P_{33}}\bm{B}\big)^T\big(\bm{P_{33}}\bm{B}\big)\Big]^{-1}\big(\bm{P_{33}}\bm{B}\big)^T\;Re(\bm{\Gamma^{33}})\nonumber\\
    &-\bm{B}\Big[\big(\bm{P_{23}}\bm{B}\big)^T\big(\bm{P_{23}}\bm{B}\big)\Big]^{-1}\big(\bm{P_{23}}\bm{B}\big)^T\;Re(\bm{\Gamma^{23}})
\end{align}

\subsubsection{Generalizing to TB and EB correlation}
Considering that B-mode polarization has the opposite behavior with T-mode and E-mode under parity transformation, correlations of TB and EB in the primordial CMB signal vanish. However, non-zero $C_\ell^{TB}$ and $C_\ell^{EB}$ can be produced by the so-called `cosmic birefringence' effect, as well as by miscalibrated polarisation angles \citep{2019PTEP.2019h3E02M, Natoli_2018}.

If $C^{TB}_\ell$ is non-zero, then $\Gamma^{12}_m$ and $\Gamma^{13}_m$ are complex, and the imaginary parts are determined by $C^{TB}_\ell$ only:

\begin{equation}
    \begin{pmatrix}
    Im(\bm{\Gamma^{12}})\\Im(\bm{\Gamma^{13}})
    \end{pmatrix}=
    \begin{pmatrix}
    -\bm{P_{12}}\\\bm{P_{13}}
    \end{pmatrix}\bm{C^{TE}}.
\end{equation}
Therefore,
\begin{align}
    \bm{\hat{C}^{TB}}=& - \bm{B}\Big[\big(\bm{P_{12}}\bm{B}\big)^T\big(\bm{P_{12}}\bm{B}\big) + \big(\bm{P_{13}}\bm{B}\big)^T\big(\bm{P_{13}}\bm{B}\big)\Big]^{-1}\big(\bm{P_{12}}\bm{B}\big)^T\;Im(\bm{\Gamma^{12}})\nonumber\\
    & + \bm{B}\Big[\big(\bm{P_{12}}\bm{B}\big)^T\big(\bm{P_{12}}\bm{B}\big) + \big(\bm{P_{13}}\bm{B}\big)^T\big(\bm{P_{13}}\bm{B}\big)\Big]^{-1}\big(\bm{P_{13}}\bm{B}\big)^T\;Im(\bm{\Gamma^{13}}).
\end{align}
Similarly, $C^{EB}_\ell$ is projected to the imaginary part of $\Gamma^{23}_m$:
\begin{equation}
    Im(\bm{\Gamma^{23}})= - 2\bm{P_{23}}\bm{C^{EB}},
\end{equation}
and
\begin{equation}
    \bm{\hat{C}^{EB}}= - \frac{1}{2}\bm{B}\Big[\big(\bm{P_{23}}\bm{B}\big)^T\big(\bm{P_{23}}\bm{B}\big)\Big]^{-1}\big(\bm{P_{23}}\bm{B}\big)^T\;Im(\bm{\Gamma^{23}})
\end{equation}

\subsection{Non-closing rings}
For $|m|\geqslant2$ and $k=1,2,3$, the Fourier coefficients on the ring are:
\begin{align}
    &\quad\hat\alpha_{k,m}\nonumber\\
    =&\quad\frac{1}{2\pi}\int_0^{2\pi}d\phi\Bigg(\sum_{n=-\infty}^{+\infty}\sum_{\substack{\ell=\\ min\{2,|n|\}}}^{+\infty}a_{k,\ell n}e^{in\phi}\mathcal P^k_{\ell n}\big(\theta+\delta(\phi)\big)\Bigg)e^{-im\phi}\nonumber\\
    \simeq&\quad\frac{1}{2\pi}\int_0^{2\pi}d\phi\sum_{n=-\infty}^{+\infty}\sum_{\substack{\ell=\\ min\{2,|n|\}}}^{+\infty}a_{k,\ell n}\Big[\mathcal P^k_{\ell m}(\theta)+\frac{d}{d\theta}\mathcal P^k_{\ell m}(\theta)\delta(\phi)\Big]e^{i(n-m)\phi}\nonumber\\
    =&\quad\frac{1}{2\pi}\sum_{\ell=|m|}^{+\infty}a_{k,\ell m}\mathcal P^k_{\ell m}(\theta)\int_0^{2\pi}d\phi+
    \frac{1}{2\pi}\sum_{\ell=|m|}^{+\infty}a_{k,\ell m}\frac{d}{d\theta}\mathcal P^k_{\ell m}(\theta)\int_0^{2\pi}\delta(\phi)d\phi\nonumber\\
    &+\frac{1}{2\pi}\sum_{\substack{n=-\infty\\ n\ne m}}^{+\infty}\sum_{\substack{\ell=\\ min\{2,|n|\}}}^{+\infty}a_{k,\ell n}\mathcal P^k_{\ell n}(\theta)\int_0^{2\pi}e^{i(n-m)\phi}d\phi\nonumber\\
    &+\frac{1}{2\pi}\sum_{\substack{n=-\infty\\ n\ne m}}^{+\infty}\sum_{\substack{\ell=\\ min\{2,|n|\}}}^{+\infty}a_{k,\ell n}\frac{d}{d\theta}\mathcal P^k_{\ell n}(\theta)\int_0^{2\pi}\delta(\phi)e^{i(n-m)\phi}d\phi\nonumber\\
    =&\sum_{\ell=|m|}^{+\infty}a_{k,\ell m}\mathcal P^k_{\ell m}(\theta)+\frac{1}{2\pi}\sum_{n=-\infty}^{+\infty}\sum_{\substack{\ell=\\ min\{2,|n|\}}}^{+\infty}a_{k,\ell n}\frac{d}{d\theta}\mathcal P^k_{\ell n}(\theta)\int_0^{2\pi}\delta(\phi)e^{i(n-m)\phi}d\phi
\end{align}

Neglecting the high order small quantities, their correlation functions are:
\begin{align}
    &\quad\Big\langle\hat\alpha_{j,m'}^*\hat\alpha_{k,m}\Big\rangle\nonumber\\
    \simeq&\quad\sum_{\ell'=|m'|}^{+\infty}\sum_{\ell=|m|}^{+\infty}\mathcal P^j_{\ell' m'}(\theta)\mathcal P^k_{\ell m}(\theta)\;\Big\langle a_{j,\ell' m'}^*a_{k,\ell m}\Big\rangle\nonumber\\
    &+\sum_{\ell'=|m'|}^{+\infty}\sum_{n=-\infty}^{+\infty}\sum_{\substack{\ell=\\ min\{2,|n|\}}}^{+\infty}\frac{1}{2\pi}\mathcal P^j_{\ell' m'}(\theta)\frac{d}{d\theta}\mathcal P_{\ell n}(\theta)\int_0^{2\pi}\delta(\phi)e^{i(n-m)\phi}d\phi\;\Big\langle a_{j,\ell' m'}^*a_{k,\ell n}\Big\rangle\nonumber\\
    &+\sum_{n'=-\infty}^{+\infty}\sum_{\substack{\ell=\\ min\{2,|n|\}}}^{+\infty}\sum_{\ell=|m|}^{+\infty}\frac{1}{2\pi}\frac{d}{d\theta}\mathcal P^j_{\ell' n'}(\theta)\mathcal P^k_{\ell m}(\theta)\int_0^{2\pi}\delta(\phi)e^{-i(n'-m')\phi}d\phi\;\Big\langle a_{j,\ell' n'}^*a_{k,\ell m}\Big\rangle\nonumber\\
    =&\quad\sum_{\ell'=|m'|}^{+\infty}\sum_{\ell=|m|}^{+\infty}\mathcal P^j_{\ell' m'}(\theta)\mathcal P^k_{\ell m}(\theta)\sum_{X,Y}A^{XY}_{jk}C_\ell^{XY}\delta_{\ell'\ell}\delta_{m'm}\nonumber\\
    &+\sum_{\ell'=|m'|}^{+\infty}\sum_{n=-\infty}^{+\infty}\sum_{\substack{\ell=\\ min\{2,|n|\}}}^{+\infty}\frac{1}{2\pi}\mathcal P^j_{\ell' m'}\frac{d}{d\theta}\mathcal P^k_{\ell n}\sum_{X,Y}A^{XY}_{jk}C_\ell^{XY}\delta_{\ell'\ell}\delta_{m'n}\int_0^{2\pi}\delta(\phi)e^{i(n-m)\phi}d\phi\nonumber\\
    &+\sum_{n'=-\infty}^{+\infty}\sum_{\substack{\ell'=\\ min\{2,|n'|\}}}^{+\infty}\sum_{\ell=|m|}^{+\infty}\frac{1}{2\pi}\frac{d}{d\theta}\mathcal P^j_{\ell' n'}\mathcal P^k_{\ell m}\sum_{X,Y}A^{XY}_{jk}C_\ell^{XY}\delta_{\ell'\ell}\delta_{n'm}\int_0^{2\pi}\delta(\phi)e^{-i(n'-m')\phi}d\phi\nonumber\\
    =&\quad\sum_{\ell=|m|}^{+\infty}\mathcal P^j_{\ell m'}(\theta)\mathcal P^k_{\ell m}(\theta)\sum_{X,Y}A^{XY}_{jk}C_\ell^{XY}\delta_{m'm}\nonumber\\
    &+\sum_{\ell=|m'|}^{+\infty}\frac{1}{2\pi}\mathcal P^j_{\ell m'}(\theta)\frac{d}{d\theta}\mathcal P^k_{\ell m'}(\theta)\sum_{X,Y}A^{XY}_{jk}C_\ell^{XY}\int_0^{2\pi}\delta(\phi)e^{i(m'-m)\phi}d\phi\nonumber\\
    &+\sum_{\ell=|m|}^{+\infty}\frac{1}{2\pi}\frac{d}{d\theta}\mathcal P^j_{\ell m}(\theta)\mathcal P^k_{\ell m}(\theta)\sum_{X,Y}A^{XY}_{jk}C_\ell^{XY}\int_0^{2\pi}\delta(\phi)e^{-i(m-m')\phi}d\phi
\end{align}

Then we get the first order approximation of ring power spectra:
\begin{align}
	\hat\Gamma^{jk}_m=&\quad\Big\langle\hat\alpha_{jm}^*\hat\alpha_{km}\Big\rangle\nonumber\\
	=&\quad\sum_{\ell=|m|}^{+\infty}\mathcal P^j_{\ell m}(\theta)\mathcal P^k_{\ell m}(\theta)\sum_{X,Y}A^{XY}_{jk}C_\ell^{XY}\nonumber\\
	&+\sum_{\ell=|m|}^{+\infty}\frac{1}{2\pi}\Big(\mathcal P^j_{\ell m}(\theta)\frac{d}{d\theta}\mathcal P^k_{\ell m}(\theta)+\frac{d}{d\theta}\mathcal P^j_{\ell m}(\theta)\mathcal P^k_{\ell m}(\theta)\Big)\sum_{X,Y}A^{XY}_{jk}C_\ell^{XY}\int_0^{2\pi}\delta(\phi)d\phi\nonumber\\
  \equiv&\quad\sum_{\ell=|m|}^{+\infty}K^\ell_m(\theta,j,k,X,Y)\cdot C^{XY}_\ell+\sum_{\ell=|m|}^{+\infty}k^\ell_m(\theta,\delta,j,k,X,Y)\cdot C^{XY}_\ell
\end{align}

\section{Details of numerical calculations}
When we use the following expression to
calculate $\mathcal P^2_{\ell m}$ and $\mathcal P^3_{\ell m}$, errors will accumulate quickly.
\begin{align}
\prescript{}{s}{Y}_{\ell m}=e^{im\phi}\sqrt{\frac{(\ell+m)!(\ell-m)!}{(\ell+s)!(\ell-s)!}\frac{2\ell +1}{4\pi}}&\sin^{2\ell}\frac{\theta}{2}\nonumber\\
\times\sum_r\begin{pmatrix}\ell-s\\r\end{pmatrix}&\begin{pmatrix}\ell+s\\r+s-m\end{pmatrix}(-1)^{\ell-r-s+m}\cot^{2r+s-m}\frac{\theta}{2}
\end{align}
Considering that scientific computing software such as Mathematica can calculate the values of associated Legendre polynomials accurately, we calculate the values of $\mathcal P^{\{s\}}_{\ell m}$ with the help of spin raising (lowering) operators. Specifically, the spin weighted spherical harmonics satisfy
\begin{equation}
\prescript{}{s}{Y}_{\ell m}(\theta,\phi)=
\begin{dcases}
\sqrt{\frac{(\ell-s)!}{(\ell+s)!}}\eth^s\ Y_{\ell m}(\theta,\phi), &0\le s\le\ell\\
\sqrt{\frac{(\ell+s)!}{(\ell-s)!}}(-1)^s\bar\eth^{-s}\ Y_{\ell m}(\theta,\phi), &-\ell\le s\le 0,
\end{dcases}
\end{equation}
and
\begin{align}
    &\mathcal{P}^2_{\ell m}(\theta)e^{im\phi}\nonumber\\
    =&\sqrt{\frac{(\ell-2)!}{(\ell+2)!}}\eth^2 Y_{\ell m}(\theta,\phi)\nonumber\\
    =&\sqrt{\frac{(\ell-2)!}{(\ell+2)!}}\left[\frac{d^2}{d\theta^2}\mathcal P^1_{\ell m}(\theta)-\frac{2m+\cos\theta}{\sin\theta}\frac{d}{d\theta}\mathcal P^1_{\ell m}(\theta)+\frac{(m+2\cos\theta)m}{\sin^2\theta}\mathcal P^1_{\ell m}(\theta)\right]e^{im\phi}
\end{align}
\begin{align}
    &\mathcal{P}^3_{\ell m}(\theta)e^{im\phi}\nonumber\\
    =&\sqrt{\frac{(\ell-2)!}{(\ell+2)!}}\bar\eth^2 Y_{\ell m}(\theta,\phi)\nonumber\\
    =&\sqrt{\frac{(\ell-2)!}{(\ell+2)!}}\left[\frac{d^2}{d\theta^2}\mathcal P^1_{\ell m}(\theta)+\frac{2m-\cos\theta}{\sin\theta}\frac{d}{d\theta}\mathcal P^1_{\ell m}(\theta)+\frac{(m-2\cos\theta)m}{\sin^2\theta}\mathcal P^1_{\ell m}(\theta)\right]e^{im\phi}
\end{align}
One the other hand, the associated Legendre polynomial satisfies the equation below
\begin{align}
\big(2\ell+1\big)\big(1-x^2\big)\frac{dP_{\ell m}(x)}{dx}&=\big(\ell+1\big)\big(\ell+m\big)P_{\ell-1,m}(x)-\ell\big(\ell-m+1\big)P_{\ell+1,m}(x)\\
\big(2\ell+1\big)xP_{\ell m}(x)&=\big(\ell+m\big)P_{\ell-1,m}(x)+\big(\ell-m+1\big)P_{\ell+1,m}(x).
\end{align}
After a series of derivations, we could obtain:
\begin{align}
\mathcal P^2_{\ell m}(\theta)&=\sqrt{\frac{2\ell +1}{4\pi}\frac{(\ell-m)!}{(\ell+m)!}\frac{(\ell-2)!}{(\ell+2)!}}\times\nonumber\\
&\Bigg\{\Big[-\big(\ell+1\big)\big(\ell+2\big)+2\big(\ell+1+m^2\big)\csc^2\theta+2m\big(\ell+2\big)\csc\theta\cot\theta\Big]P_{\ell m}(\cos\theta)\nonumber\\
&\quad-2\big(\ell-m+1\big)\big(\csc\theta\cot\theta+m\csc^2\theta\big)P_{\ell+1,m}(\cos\theta)\Bigg\},
\end{align}
\begin{align}
\mathcal P^3_{\ell m}(\theta)&=\sqrt{\frac{2\ell +1}{4\pi}\frac{(\ell-m)!}{(\ell+m)!}\frac{(\ell-2)!}{(\ell+2)!}}\times\nonumber\\
&\Bigg\{\Big[-\big(\ell+1\big)\big(\ell+2\big)+2\big(\ell+1+m^2\big)\csc^2\theta-2m\big(\ell+2\big)\csc\theta\cot\theta\Big]P_{\ell m}(\cos\theta)\nonumber\\
&\quad-2\big(\ell-m+1\big)\big(\csc\theta\cot\theta-m\csc^2\theta\big)P_{\ell+1,m}(\cos\theta)\Bigg\}.
\end{align}
We use the above formulae to calculate $\mathcal{P}^2_{\ell m}$ and $\mathcal{P}^3_{\ell m}$.

\bibliographystyle{JHEP}
\bibliography{biblio.bib}{} 

\providecommand{\href}[2]{#2}\begingroup\raggedright\begin{thebibliography}{10}

\bibitem{2011A&A...536A...1P}
{Planck Collaboration}, P.~A.~R. {Ade}, N.~{Aghanim}, M.~{Arnaud},
  M.~{Ashdown}, J.~{Aumont} et~al., \emph{{Planck early results. I. The Planck
  mission}}, \href{https://doi.org/10.1051/0004-6361/201116464}{\emph{Astron.
  Astrophys.} {\bfseries 536} (2011) A1}
  [\href{https://arxiv.org/abs/1101.2022}{{\ttfamily 1101.2022}}].

\bibitem{2019arXiv190712875P}
{Planck Collaboration}, N.~{Aghanim}, Y.~{Akrami}, M.~{Ashdown}, J.~{Aumont},
  C.~{Baccigalupi} et~al., \emph{{Planck 2018 results. V. CMB power spectra and
  likelihoods}}, {\emph{arXiv e-prints} (2019) arXiv:1907.12875}
  [\href{https://arxiv.org/abs/1907.12875}{{\ttfamily 1907.12875}}].

\bibitem{2018arXiv180706209P}
{Planck Collaboration}, N.~{Aghanim}, Y.~{Akrami}, M.~{Ashdown}, J.~{Aumont},
  C.~{Baccigalupi} et~al., \emph{{Planck 2018 results. VI. Cosmological
  parameters}}, {\emph{arXiv e-prints} (2018) arXiv:1807.06209}
  [\href{https://arxiv.org/abs/1807.06209}{{\ttfamily 1807.06209}}].

\bibitem{2014JCAP...10..007N}
S.~{Naess}, M.~{Hasselfield}, J.~{McMahon}, M.~D. {Niemack}, G.~E. {Addison},
  P.~A.~R. {Ade} et~al., \emph{{The Atacama Cosmology Telescope: CMB
  polarization at $200 < l < 9000$}},
  \href{https://doi.org/10.1088/1475-7516/2014/10/007}{\emph{JCAP} {\bfseries
  2014} (2014) 007} [\href{https://arxiv.org/abs/1405.5524}{{\ttfamily
  1405.5524}}].

\bibitem{2018JCAP...09..005K}
A.~{Kusaka}, J.~{Appel}, T.~{Essinger-Hileman}, J.~A. {Beall}, L.~E.
  {Campusano}, H.-M. {Cho} et~al., \emph{{Results from the Atacama B-mode
  Search (ABS) experiment}},
  \href{https://doi.org/10.1088/1475-7516/2018/09/005}{\emph{JCAP} {\bfseries
  2018} (2018) 005} [\href{https://arxiv.org/abs/1801.01218}{{\ttfamily
  1801.01218}}].

\bibitem{2018PhRvL.121v1301B}
{BICEP2 Collaboration}, {Keck Array Collaboration}, P.~A.~R. {Ade}, Z.~{Ahmed},
  R.~W. {Aikin}, K.~D. {Alexand er} et~al., \emph{{Constraints on Primordial
  Gravitational Waves Using Planck, WMAP, and New BICEP2/Keck Observations
  through the 2015 Season}},
  \href{https://doi.org/10.1103/PhysRevLett.121.221301}{\emph{Phys. Rev. Lett.}
  {\bfseries 121} (2018) 221301}
  [\href{https://arxiv.org/abs/1810.05216}{{\ttfamily 1810.05216}}].

\bibitem{2019arXiv190800480D}
S.~{Dahal}, M.~{Amiri}, J.~W. {Appel}, C.~L. {Bennett}, L.~{Corbett},
  R.~{Datta} et~al., \emph{{The CLASS 150/220 GHz Polarimeter Array: Design,
  Assembly, and Characterization}}, {\emph{arXiv e-prints} (2019)
  arXiv:1908.00480} [\href{https://arxiv.org/abs/1908.00480}{{\ttfamily
  1908.00480}}].

\bibitem{2018JLTP..193.1066N}
T.~{Nagasaki}, J.~{Choi}, R.~T. {G{\'e}nova-Santos}, M.~{Hattori}, M.~{Hazumi},
  H.~{Ishitsuka} et~al., \emph{{GroundBIRD: Observation of CMB Polarization
  with a Rapid Scanning and MKIDs}},
  \href{https://doi.org/10.1007/s10909-018-2077-y}{\emph{Journal of Low
  Temperature Physics} {\bfseries 193} (2018) 1066}.

\bibitem{2019arXiv191002608A}
S.~{Adachi}, M.~A.~O. {Aguilar Fa{\'u}ndez}, K.~{Arnold}, C.~{Baccigalupi},
  D.~{Barron}, D.~{Beck} et~al., \emph{{A Measurement of the Degree Scale CMB
  B-mode Angular Power Spectrum with POLARBEAR}}, {\emph{arXiv e-prints} (2019)
  arXiv:1910.02608} [\href{https://arxiv.org/abs/1910.02608}{{\ttfamily
  1910.02608}}].

\bibitem{2019Univ....5...42M}
A.~{Mennella}, P.~{Ade}, G.~{Amico}, D.~{Auguste}, J.~{Aumont}, S.~{Banfi}
  et~al., \emph{{QUBIC: Exploring the Primordial Universe with the Q $\&$U
  Bolometric Interferometer}},
  \href{https://doi.org/10.3390/universe5020042}{\emph{Universe} {\bfseries 5}
  (2019) 42}.

\bibitem{2019arXiv191005748S}
J.~T. {Sayre}, C.~L. {Reichardt}, J.~W. {Henning}, P.~A.~R. {Ade}, A.~J.
  {Anderson}, J.~E. {Austermann} et~al., \emph{{Measurements of B-mode
  Polarization of the Cosmic Microwave Background from 500 Square Degrees of
  SPTpol Data}}, {\emph{arXiv e-prints} (2019) arXiv:1910.05748}
  [\href{https://arxiv.org/abs/1910.05748}{{\ttfamily 1910.05748}}].

\bibitem{2019JCAP...02..056A}
P.~{Ade}, J.~{Aguirre}, Z.~{Ahmed}, S.~{Aiola}, A.~{Ali}, D.~{Alonso} et~al.,
  \emph{{The Simons Observatory: science goals and forecasts}},
  \href{https://doi.org/10.1088/1475-7516/2019/02/056}{\emph{JCAP} {\bfseries
  2019} (2019) 056} [\href{https://arxiv.org/abs/1808.07445}{{\ttfamily
  1808.07445}}].

\bibitem{2018ApJS..239....7E}
{EBEX Collaboration}, A.~M. {Aboobaker}, P.~{Ade}, D.~{Araujo}, F.~{Aubin},
  C.~{Baccigalupi} et~al., \emph{{The EBEX Balloon-borne
  Experiment{\textemdash}Optics, Receiver, and Polarimetry}},
  \href{https://doi.org/10.3847/1538-4365/aae434}{\emph{Astrophys. J. Suppl.}
  {\bfseries 239} (2018) 7} [\href{https://arxiv.org/abs/1703.03847}{{\ttfamily
  1703.03847}}].

\bibitem{2018JLTP..193.1112G}
R.~{Gualtieri}, J.~P. {Filippini}, P.~A.~R. {Ade}, M.~{Amiri}, S.~J. {Benton},
  A.~S. {Bergman} et~al., \emph{{SPIDER: CMB Polarimetry from the Edge of
  Space}}, \href{https://doi.org/10.1007/s10909-018-2078-x}{\emph{Journal of
  Low Temperature Physics} {\bfseries 193} (2018) 1112}
  [\href{https://arxiv.org/abs/1711.10596}{{\ttfamily 1711.10596}}].

\bibitem{2012SPIE.8452E..3FD}
P.~{de Bernardis}, S.~{Aiola}, G.~{Amico}, E.~{Battistelli}, A.~{Coppolecchia},
  A.~{Cruciani} et~al., \emph{{SWIPE: a bolometric polarimeter for the
  Large-Scale Polarization Explorer}},  in \emph{Millimeter, Submillimeter, and
  Far-Infrared Detectors and Instrumentation for Astronomy VI}, W.~S. {Holland}
  and J.~{Zmuidzinas}, eds., vol.~8452 of \emph{Society of Photo-Optical
  Instrumentation Engineers (SPIE) Conference Series}, p.~84523F, Sept., 2012,
  \href{https://doi.org/10.1117/12.926569}{DOI}
  [\href{https://arxiv.org/abs/1208.0282}{{\ttfamily 1208.0282}}].

\bibitem{2018SPIE10708E..06P}
S.~{Pawlyk}, P.~A.~R. {Ade}, D.~{Benford}, C.~L. {Bennett}, D.~T. {Chuss},
  R.~{Datta} et~al., \emph{{The primordial inflation polarization explorer
  (PIPER): current status and performance of the first flight}},  in
  \emph{Proc. SPIE Int. Soc. Opt. Eng.}, vol.~10708 of \emph{Society of
  Photo-Optical Instrumentation Engineers (SPIE) Conference Series},
  p.~1070806, Jul, 2018, \href{https://doi.org/10.1117/12.2313874}{DOI}.

\bibitem{2018JCAP...04..017D}
E.~{Di Valentino}, T.~{Brinckmann}, M.~{Gerbino}, V.~{Poulin}, F.~R. {Bouchet},
  J.~{Lesgourgues} et~al., \emph{{Exploring cosmic origins with CORE:
  Cosmological parameters}},
  \href{https://doi.org/10.1088/1475-7516/2018/04/017}{\emph{JCAP} {\bfseries
  2018} (2018) 017} [\href{https://arxiv.org/abs/1612.00021}{{\ttfamily
  1612.00021}}].

\bibitem{2006PhR...429....1L}
A.~{Lewis} and A.~{Challinor}, \emph{{Weak gravitational lensing of the CMB}},
  \href{https://doi.org/10.1016/j.physrep.2006.03.002}{\emph{Phys. Rept.}
  {\bfseries 429} (2006) 1}
  [\href{https://arxiv.org/abs/astro-ph/0601594}{{\ttfamily
  astro-ph/0601594}}].

\bibitem{2018JCAP...04..018C}
A.~{Challinor}, R.~{Allison}, J.~{Carron}, J.~{Errard}, S.~{Feeney},
  T.~{Kitching} et~al., \emph{{Exploring cosmic origins with CORE:
  Gravitational lensing of the CMB}},
  \href{https://doi.org/10.1088/1475-7516/2018/04/018}{\emph{JCAP} {\bfseries
  2018} (2018) 018} [\href{https://arxiv.org/abs/1707.02259}{{\ttfamily
  1707.02259}}].

\bibitem{2016arXiv161002743A}
K.~N. {Abazajian}, P.~{Adshead}, Z.~{Ahmed}, S.~W. {Allen}, D.~{Alonso}, K.~S.
  {Arnold} et~al., \emph{{CMB-S4 Science Book, First Edition}}, {\emph{arXiv
  e-prints} (2016) arXiv:1610.02743}
  [\href{https://arxiv.org/abs/1610.02743}{{\ttfamily 1610.02743}}].

\bibitem{2018JCAP...04..016F}
F.~{Finelli}, M.~{Bucher}, A.~{Ach{\'u}carro}, M.~{Ballardini}, N.~{Bartolo},
  D.~{Baumann} et~al., \emph{{Exploring cosmic origins with CORE: Inflation}},
  \href{https://doi.org/10.1088/1475-7516/2018/04/016}{\emph{JCAP} {\bfseries
  2018} (2018) 016} [\href{https://arxiv.org/abs/1612.08270}{{\ttfamily
  1612.08270}}].

\bibitem{2019BAAS...51c.338S}
S.~{Shandera}, P.~{Adshead}, M.~{Amin}, E.~{Dimastrogiovanni}, C.~{Dvorkin},
  R.~{Easther} et~al., \emph{{Probing the origin of our Universe through cosmic
  microwave background constraints on gravitational waves}}, {\emph{Bull. Am.
  Astron. Soc.} {\bfseries 51} (2019) 338}
  [\href{https://arxiv.org/abs/1903.04700}{{\ttfamily 1903.04700}}].

\bibitem{2010A&A...520A...1T}
J.~A. {Tauber}, N.~{Mandolesi}, J.~L. {Puget}, T.~{Banos}, M.~{Bersanelli},
  F.~R. {Bouchet} et~al., \emph{{Planck pre-launch status: The Planck
  mission}}, \href{https://doi.org/10.1051/0004-6361/200912983}{\emph{Astron.
  Astrophys.} {\bfseries 520} (2010) A1}.

\bibitem{2020JLTP..200..384L}
K.~{Lee}, J.~{Choi}, R.~T. {G{\'e}nova-Santos}, M.~{Hattori}, M.~{Hazumi},
  S.~{Honda} et~al., \emph{{GroundBIRD: A CMB Polarization Experiment with MKID
  Arrays}}, \href{https://doi.org/10.1007/s10909-020-02511-5}{\emph{Journal of
  Low Temperature Physics} {\bfseries 200} (2020) 384}.

\bibitem{1967JMP.....8.2155G}
J.~N. {Goldberg}, A.~J. {Macfarlane}, E.~T. {Newman}, F.~{Rohrlich} and
  E.~C.~G. {Sudarshan}, \emph{{Spin-s Spherical Harmonics and {$\eth$}}},
  \href{https://doi.org/10.1063/1.1705135}{\emph{Journal of Mathematical
  Physics} {\bfseries 8} (1967) 2155}.

\bibitem{2018SPIE10708E..1GF}
C.~{Franceschet}, S.~{Realini}, A.~{Mennella}, G.~{Addamo}, A.~{Ba{\'u}}, P.~M.
  {Battaglia} et~al., \emph{{The STRIP instrument of the Large Scale
  Polarization Explorer: microwave eyes to map the Galactic polarized
  foregrounds}},  in \emph{Millimeter, Submillimeter, and Far-Infrared
  Detectors and Instrumentation for Astronomy IX}, J.~{Zmuidzinas} and J.-R.
  {Gao}, eds., vol.~10708 of \emph{Society of Photo-Optical Instrumentation
  Engineers (SPIE) Conference Series}, p.~107081G, July, 2018,
  \href{https://doi.org/10.1117/12.2313558}{DOI}
  [\href{https://arxiv.org/abs/1812.03687}{{\ttfamily 1812.03687}}].

\bibitem{2018SPIE10708E..2FI}
F.~{Incardona}, M.~{Benetti}, M.~{Bersanelli}, C.~{Franceschet}, D.~{Maino},
  A.~{Mennella} et~al., \emph{{Preliminary scanning strategy analysis for the
  LSPE-STRIP instrument}},  in \emph{Millimeter, Submillimeter, and
  Far-Infrared Detectors and Instrumentation for Astronomy IX}, vol.~10708 of
  \emph{Society of Photo-Optical Instrumentation Engineers (SPIE) Conference
  Series}, p.~107082F, July, 2018,
  \href{https://doi.org/10.1117/12.2315005}{DOI}.

\bibitem{2020arXiv200811049T}
{The LSPE collaboration}, G.~{Addamo}, P.~A.~R. {Ade}, C.~{Baccigalupi}, A.~M.
  {Baldini}, P.~M. {Battaglia} et~al., \emph{{The large scale polarization
  explorer (LSPE) for CMB measurements: performance forecast}}, {\emph{arXiv
  e-prints} (2020) arXiv:2008.11049}
  [\href{https://arxiv.org/abs/2008.11049}{{\ttfamily 2008.11049}}].

\bibitem{1998astro.ph..4180D}
J.~{Delabrouille}, R.~{Gispert} and J.~L. {Puget}, \emph{{CMB anisotropies on
  circular scans}}, {\emph{arXiv e-prints} (1998) astro}
  [\href{https://arxiv.org/abs/astro-ph/9804180}{{\ttfamily
  astro-ph/9804180}}].

\bibitem{1998MNRAS.298..445D}
J.~{Delabrouille}, K.~M. {Gorski} and E.~{Hivon}, \emph{{Circular scans for
  cosmic microwave background anisotropy observation and analysis}},
  \href{https://doi.org/10.1046/j.1365-8711.1998.01646.x}{\emph{Mon. Not. Roy.
  Astron. Soc.} {\bfseries 298} (1998) 445}
  [\href{https://arxiv.org/abs/astro-ph/9710349}{{\ttfamily
  astro-ph/9710349}}].

\bibitem{2002MNRAS.331..975V}
F.~{van Leeuwen}, A.~D. {Challinor}, D.~J. {Mortlock}, M.~A.~J. {Ashdown},
  M.~P. {Hobson}, A.~N. {Lasenby} et~al., \emph{{Harmonic analysis of cosmic
  microwave background data - I. Ring reductions and point-source catalogue}},
  \href{https://doi.org/10.1046/j.1365-8711.2002.05254.x}{\emph{Mon. Not. Roy.
  Astron. Soc.} {\bfseries 331} (2002) 975}
  [\href{https://arxiv.org/abs/astro-ph/0112276}{{\ttfamily
  astro-ph/0112276}}].

\bibitem{2002MNRAS.331..994C}
A.~D. {Challinor}, D.~J. {Mortlock}, F.~{van Leeuwen}, A.~N. {Lasenby}, M.~P.
  {Hobson}, M.~A.~J. {Ashdown} et~al., \emph{{Harmonic analysis of cosmic
  microwave background data - II. From ring-sets to the sky}},
  \href{https://doi.org/10.1046/j.1365-8711.2002.05255.x}{\emph{Mon. Not. Roy.
  Astron. Soc.} {\bfseries 331} (2002) 994}
  [\href{https://arxiv.org/abs/astro-ph/0112277}{{\ttfamily
  astro-ph/0112277}}].

\bibitem{2003MNRAS.343..552A}
R.~{Ansari}, S.~{Bargot}, A.~{Bourrachot}, F.~{Couchot}, J.~{Ha{\"\i}ssinski},
  S.~{Henrot-Versill{\'e}} et~al., \emph{{Concerning the connection between the
  C$_{l}$ power spectrum of the cosmic microwave background and the
  {\ensuremath{\Gamma}}$_{m}$ Fourier spectrum of rings on the sky}},
  \href{https://doi.org/10.1046/j.1365-8711.2003.06694.x}{\emph{Mon. Not. Roy.
  Astron. Soc.} {\bfseries 343} (2003) 552}
  [\href{https://arxiv.org/abs/astro-ph/0301251}{{\ttfamily
  astro-ph/0301251}}].

\bibitem{1999A&AS..135..579C}
F.~{Couchot}, J.~{Delabrouille}, J.~{Kaplan} and B.~{Revenu}, \emph{{Optimised
  polarimeter configurations for measuring the Stokes parameters of the cosmic
  microwave background radiation}},
  \href{https://doi.org/10.1051/aas:1999191}{\emph{Astron. Astrophys. Suppl.
  Ser.} {\bfseries 135} (1999) 579}
  [\href{https://arxiv.org/abs/astro-ph/9807080}{{\ttfamily
  astro-ph/9807080}}].

\bibitem{1968ApJ...151..459S}
J.~{Silk}, \emph{{Cosmic Black-Body Radiation and Galaxy Formation}},
  \href{https://doi.org/10.1086/149449}{\emph{Astrophys. J.} {\bfseries 151}
  (1968) 459}.

\bibitem{2011arXiv1104.2932L}
J.~{Lesgourgues}, \emph{{The Cosmic Linear Anisotropy Solving System (CLASS) I:
  Overview}}, {\emph{arXiv e-prints} (2011) arXiv:1104.2932}
  [\href{https://arxiv.org/abs/1104.2932}{{\ttfamily 1104.2932}}].

\bibitem{1997PhRvD..55.1830Z}
M.~{Zaldarriaga} and U.~{Seljak}, \emph{{All-sky analysis of polarization in
  the microwave background}},
  \href{https://doi.org/10.1103/PhysRevD.55.1830}{\emph{Phys. Rev. D}
  {\bfseries 55} (1997) 1830}
  [\href{https://arxiv.org/abs/astro-ph/9609170}{{\ttfamily
  astro-ph/9609170}}].

\bibitem{2019PTEP.2019h3E02M}
Y.~{Minami}, H.~{Ochi}, K.~{Ichiki}, N.~{Katayama}, E.~{Komatsu} and
  T.~{Matsumura}, \emph{{Simultaneous determination of the cosmic birefringence
  and miscalibrated polarization angles from CMB experiments}},
  \href{https://doi.org/10.1093/ptep/ptz079}{\emph{Progress of Theoretical and
  Experimental Physics} {\bfseries 2019} (2019) 083E02}.

\bibitem{Natoli_2018}
P.~Natoli, M.~Ashdown, R.~Banerji, J.~Borrill, A.~Buzzelli, G.~de~Gasperis
  et~al., \emph{Exploring cosmic origins with {CORE}: Mitigation of systematic
  effects}, \href{https://doi.org/10.1088/1475-7516/2018/04/022}{\emph{Journal
  of Cosmology and Astroparticle Physics} {\bfseries 2018} (2018) 022}.

\end{thebibliography}\endgroup


\end{document}